 \documentclass[12pt]{article}
 \newcommand{\be}{\begin{equation}}
 \newcommand{\ee}{\end{equation}}
 \newcommand{\ba}{\begin{eqnarray}}
 \newcommand{\ea}{\end{eqnarray}}
 \newcommand{\inc}{{\it i}}
 \newcommand{\bra}{\langle}
 \newcommand{\ket}{\rangle}
 \newcommand{\efbold}{\mbox{{\boldmath $\vec f$}}}
 
 \newcommand{\erbold}{\mbox{{\boldmath $\vec r$}}}
 \newcommand{\Phibold}{\mbox{{\boldmath $\vec \Phi$}}}
 \newcommand{\mubold}{\mbox{{\boldmath $\vec \mu$}}}

 \newcommand{\gbold}{\mbox{{\boldmath $\vec{g}$}}}
 \newcommand{\pbold}{\mbox{{\boldmath $\vec p$}}}
 \newcommand{\wbold}{\mbox{{\boldmath $\vec w$}}}
 \newcommand{\dotmubold}{\dot{\bf {\mbox{\boldmath $\vec {\boldmath \mu}$}} }}
 
 \newcommand{\doterbold}{\dot{\textbf {\mbox{\boldmath $\vec {\boldmath r}$}} }}

\begin{document}
 \title{
   ~~~~~~~~~~~~~~~~~~~~~~~~~~~~~~~~~~~~~~~~~~~~~~~~~~~~~~~~~~~~~~~~~~~~~~~~~~~~~~~~~
  ${~}^{{Published~in:
  }}$~~~~~~~~~~~~~~~~~~~~~~~~~~~~~~~~~~~~~~\\
  ${~}^{\underline{``Celestial~Mechanics~and~Dynamical~Astronomy"~(2005)~\,{\bf{91}}\;:\;75\,-\,108}}$\\
 ~\\
 {\Large{\textbf{Long-term evolution of orbits about a
 precessing\footnote{~In this article, we use the word ``precession" in its
 most general sense which embraces the entire spectrum of
 changes of the spin-axis orientation -- from the long-term variations down to
 the Chandler-type wobble down to nutations and to
 the polar wonder.} oblate planet:\\
 1. The case of uniform precession.}
            }}}
 \author{
 {\Large{Michael Efroimsky}}\\
 {\small{US Naval Observatory, Washington DC 20392 USA}}\\
 {\small{e-mail: me @ usno.navy.mil~}}\\
~\\
 }

 \maketitle

 \begin{abstract}
It was believed until very recently that a near-equatorial
satellite would always keep up with the planet's equator (with
oscillations in inclination, but without a secular drift). As
explained in Efroimsky and Goldreich (2004), this misconception
originated from a wrong interpretation of a (mathematically
correct) result obtained in terms of non-osculating orbital
elements. A similar analysis carried out in the language of
osculating elements will endow the planetary equations with some
extra terms caused by the planet's obliquity change. Some of these
terms will be nontrivial, in that they will not be amendments to
the disturbing function. Due to the extra terms, the variations of
a planet's obliquity may cause a secular drift of its satellite
orbit inclination. In this article we set out the analytical
formalism for our study of this drift. We demonstrate that, in the
case of uniform precession, the drift will be extremely slow,
because the first-order terms responsible for the drift will be
short-period and, thus, will have vanishing orbital averages (as
anticipated 40 years ago by Peter Goldreich), while the secular
terms will be of the second order only. However, it turns out that
variations of the planetary precession make the first-order terms
secular. For example, the planetary nutations will resonate with
the satellite's orbital frequency and, thereby, may instigate a
secular drift. A detailed study of this process will be offered in
a subsequent publication, while here we work out the required
mathematical formalism and point out the key aspects of the
dynamics.

~\\
~\\
\end{abstract}

\section{Physical motivation and the statement of purpose}

Ward (1973, 1974) noted that the obliquity of Mars may have
suffered large-angle motions at long time scales. Later, Laskar
and Robutel (1993) and Touma and Wisdom (1994) demonstrated that
these motions may have been chaotic. This would cause severe
climate variations and have major consequences for development of
life.

It is a customary assumption that a near-equatorial satellite of
an oblate planet would always keep up with the planet's equator
(with only small oscillations of the orbit inclination) provided
the obliquity changes are sufficiently slow (Goldreich 1965,
Kinoshita 1993). As demonstrated in Efroimsky and Goldreich
(2004), this belief stems from a calculation performed in the
language of non-osculating orbital elements. A similar analysis
carried out in terms of osculating elements will contain hitherto
overlooked extra terms entailed by the planet's obliquity
variations. These terms (emerging already in the first order over
the precession-caused perturbation) will cause a secular angular
drift of the satellite orbit away from the planetary equator.

The existence of Phobos and Deimos, and the ability of Mars to
keep them close to its equatorial plane during obliquity
variations sets constraints on the obliquity variation amplitude
and rate. Our eventual goal is to establish such constraints. If
the satellites' secular inclination drifts are slow enough that
the satellites stay close to Mars' equator during its obliquity
changes through billions of years, then the rigid-planet
non-dissipative models used by Ward (1973, 1974), Laskar \&
Robutel (1993), and Touma \& Wisdom (1994) will get a totally
independent confirmation. If the obliquity-variation-caused
inclination drifts are too fast (fast enough that within a billion
or several billions of years the satellites get driven away from
Mars' equatorial plane), then the inelastic dissipation and
planetary structure must play a larger role than previously
assumed.

Having this big motivation in mind, we restrict the current
article to building the required mathematical background: we study
the obliquity-variation-caused terms in the planetary equations,
calculate their secular components and point out the resonant
coupling emerging between a satellite's orbiting frequency and
certain frequencies in the planet axis' precession. A more
thorough investigation of this interaction will be left for our
next paper.

 \section{Mathematical preliminaries:\\
 osculating elements vs orbital elements}

Whenever one embarks on integrating a satellite orbit and wants to
take into account direction variations of the planet's spin, it is
most natural to carry out this work in a co-precessing coordinate
system. This always yields orbital elements defined in the said
frame, which are ready for immediate physical interpretation by a
planet-based observer. A well camouflaged pitfall of this approach
is that these orbital elements may come out non-osculating, i.e.,
that the instantaneous ellipses (or hyperbolae) parametrised by
these elements will not be tangent to the physical trajectory as
seen in the said frame of reference.

 \subsection{The osculation condition and alternatives to it}

%

An instantaneous orbit is parametrised with six independent
Keplerian parameters. These include the three Eulerian angles
$\;\inc,\,\omega,\,\Omega\;$ which define the orientation of the
instantaneous orbital plane, the eccentricity and semimajor axis
of the instantaneous ellipse or hyperbola, and the mean anomaly at
an epoch, $\;M_o\;$, which determines the initial position of the
body. Often are employed sets of other variables. The variables
are always six in number and are some functions of the Keplerian
elements. These are, for example, the Delaunay set, the Jacobi
set, and two sets offered by Poincare. More exotic is the Hill set
which includes the true anomaly as one of the elements (Hill
1913).

Systems of planetary equations in terms of the above sets of
parameters may be derived not only through the
variation-of-parameters (VOP) method but also by means of the
Hamilton-Jacobi one. However, the latter approach, though fine and
elegant, lacks the power instilled into the direct VOP technique
and cannot account for the gauge invariance of the N-body problem
(Efroimsky 2002a,b), important feature intimately connected with
some general concepts in ODE (Newman \& Efroimsky 2003). The
Hamilton-Jacobi technique implicitly fixes the gauge, and thus
leaves the internal symmetry heavily veiled (Efroimsky \&
Goldreich 2003).
Below we shall present several relevant facts and
formulae, while a more comprehensive treatment of the matter may
be looked up in (Efroimsky \& Goldreich 2003, 2004).

As well known, a solution
  \begin{equation}
  {\bf \vec r }\;=\;{\efbold} \left(C_1, ... , C_6, \,t
\right)\,\;\;\;\,
 \label{401}
 \label{A1}
  \end{equation}
to the reduced two-body problem
  \begin{equation}
  {\bf \ddot{\vec
r}}\;+\;\frac{G\,m}{r^2}\;\frac{{\bf\vec r }}{r}\;=\;0\;\;\;\,
 \label{402}
 \label{A2}
  \end{equation}
is a Keplerian ellipse or hyperbola parameterised with some set of
six independent orbital elements $\;C_i\;$ which are constants in
the absence of disturbances.

In the N-particle setting (or, more generally, in the presence of
whatever disturbances) each body will be subject to some
perturbing force $\;\Delta{\bf{\vec
F}}(\erbold,\,{\bf{\dot{\erbold}}},\,t)\,$ acting at it from
bodies other than the primary:
  \begin{equation}
  {\bf \ddot{\vec
r}}\;+\;\frac{G\,m}{r^2}\;\frac{{\bf\vec r }}{r}\;=\;{\Delta
\bf{\vec F}}\;\;\;,
 \label{403}
 \label{A3}
  \end{equation}
 Solving the above equation of motion by the VOP
 method implies the use of (\ref{A1}) as an ansatz,
  \begin{equation}
  {\bf \vec r }\;=\;{\efbold} \left(C_1(t), \,... , \, C_6(t),
\,t \right)\,\;\;\;,
 \label{404}
 \label{A4}
  \end{equation}
 the function $\;\efbold\;$ being the same as in (\ref{A1}), and the
 ``constants" $\;C_i\;$ now being endowed with a time dependence of
 their own. Insertion of (\ref{A4}) into (\ref{A3})
 is insufficient for determining the six functions $\;C_i(t)\;$.
 because the vector equation (\ref{A3}) comprises only three scalar
 equalities. To furnish a solution three more equations are needed.
 Since the age of Lagrange it has been advised in the literature
 to employ for this purpose the conditions of osculation,
  \begin{equation}
  \sum_i \;\frac{\partial {\efbold}}{\partial C_i}\;\frac{d C_i}{d t}\;=\;0\;\;\;,
 \label{405}
 \label{A5}
  \end{equation}
 imposition whereof ensures that the physical velocity,
  \begin{equation}
   \frac{d \erbold}{dt}\;= \;\frac{ \partial {\efbold}}{\partial t}\;+\;
 \sum_i \;\frac{\partial {\efbold}}{\partial C_i}\;\frac{d C_i}{d t}\;\;\;\;,\;\;\;\;\;\;\;\;
 \label{406}
 \label{A6}
  \end{equation}
 in the disturbed case is expressed by the same function $\;{\bf
 \vec g}(C_1,\,...\,,\,C_6,\,t)\;$ as it used to in the two-body configuration:
  \begin{equation}
  {\bf
\vec g}\;=\;\frac{ \partial {\efbold}}{\partial t}\;
 \label{407}
 \label{A7}
  \end{equation}
 The conditions being essentially arbitrary, their choice
 affects only the mathematical developments but not the physical
 orbit. The orbit's invariance under the alternative choices of
 the supplementary constraints reflects the gauge freedom, i.e.,
 the internal symmetry present in this problem. In the language of
 pure mathematics this is a fiber-bundle structure. In physicists'
 terms it is an example of the gauge freedom. Essentially, this is
 merely a case of ambiguous parameterisation.

 The Lagrange constraint and
 the law of motion, (\ref{A3}), with ansatz (\ref{A4}) inserted therein,
 yields the following equation of the elements' evolution:
  \begin{equation}
  \sum_j\;[C_n\;C_j]\;\frac{dC_j}{dt}\;=\;
 \frac{\partial \efbold}{\partial C_n}\;
  {\Delta \bf{\vec F}}\;
\;\;\;\;,
 \label{408}
 \label{A8}
  \end{equation}
$[C_n\;C_j]\;$ being the matrix of Lagrange brackets introduced as
  \begin{equation}
  [C_n\;C_j]\;\equiv\;\frac{\partial \efbold }{\partial
C_n}\, \frac{\partial {\bf {{\vec g}}}}{\partial
C_j}\,-\,\frac{\partial  \efbold }{\partial C_j}\, \frac{\partial
{\bf {{\vec g}}}}{\partial C_n}\;\;\;\;.
 \label{409}
 \label{A9}
  \end{equation}
 So defined, the brackets depend neither on the time
evolution of $C_i$ nor on the choice of supplementary conditions,
but solely on the functional form of
$\,\efbold\left(C_{1,...,6}\,,\;t\right)\,$ and $\,\gbold
\equiv{\partial \efbold}/{\partial t}$.

In case we decide to relax the Lagrange constraint and to
substitute for it
  \begin{equation}
  \sum_i \;\frac{\partial \efbold}{\partial C_i}\;\frac{d
C_i}{d t}\;=\; {\bf {\vec \Phi}}(C_{1,...,6}\,,\,t)\;\;\;,\;
 \label{410}
 \label{A10}
   \end{equation}
   $\Phibold\,$ being some arbitrary function of time and
parameters $\;C_i\;$ (but, for the reasons of sheer convenience,
not of time derivatives of $\;C_i$), then instead of (\ref{A8}) we
shall get, for $\;n\,=\,1,\,...\,,\,6\;$,
  \begin{equation}
  \sum_j\;\left(\,[C_n\;C_j]\;+\;\frac{\partial \efbold }{\partial
C_n}\;\frac{\partial \bf {\vec \Phi}}{\partial C_j}\;
\,\right)\,\frac{dC_j}{dt}\;=\; \frac{\partial {\efbold
}}{\partial C_n}\; {\Delta \bf{\vec F}}\;-\; \frac{\partial{
{\efbold}}}{\partial C_n} \;\frac{\partial \bf {\vec
\Phi}}{\partial t}\;-\; \frac{\partial \bf {\vec g}}{\partial C_n}
\;{\bf {\vec \Phi}} \;\;\;\;.
 \label{411}
 \label{A11}
  \end{equation}
These gauge-invariant perturbation equations of celestial
mechanics, generated by direct application of the VOP method, were
derived in (Efroimsky 2002b). They give us an opportunity to use,
in an arbitrary gauge $\,\Phibold$, the Lagrange brackets
(\ref{A9}) defined in terms of the unperturbed functions
$\,\efbold\,$ and $\,\bf\vec g\,$. Expressions for these brackets
have long been known, and they can be employed to write down the
planetary equations in their gauge-invariant form. If our goal
were to arrive to the customary Lagrange system of planetary
equations, we would now stick to the trivial gauge (\ref{A5}) and
restrict ourselves to conservative forces solely: $\,\Delta{\bf
\vec F}={\partial R(\erbold )}/{\partial \erbold }$. We however
bear in mind the objective of exploring the interplay of two
freedoms: freedom of reference frame choice and that of gauge
fixing. For this reason not only shall we leave the gauge
arbitrary but shall also reserve for our disturbance a form
general enough to include both physical and inertial forces, the
latter showing themselves en route from an inertial coordinate
system to an accelerated one. An example at hand is a satellite
orbiting a precessing oblate planet: the oblateness will give
birth to an extra physical force, while the precession will result
in inertial forces if we decide to describe the orbit in
non-inertial axes associated with the precessing planet. What is
important about the inertial inputs is that they are not only
position but also velocity dependent. This gives us a motivation
to consider velocity-dependent disturbances. Another motivation
for this comes from the relativity where the correction to Newton'
gravity law contains velocities (Brumberg 1992)

Expression (\ref{A11}) is the most general form of the equations
for the orbital elements, in terms of the disturbing force. To get
the generic form of the equations in terms of the Lagrangian
disturbance, let us recall that the (reduced) two-body problem is
described by the unit-mass Lagrangian $\;{\cal L}_o({\bf{\vec
r}},\,{\bf \dot{\vec r}},\,t)\;=\;{\bf \dot{\vec
r}}^{\left.\,\right. 2}/2\,-\,U({\bf \vec r}\;,\;t)$, canonical
momentum $\;{\bf{\vec{p}}}\,=\,{\bf \dot{\vec r}}\;$, and
Hamiltonian $\;{\cal H}_o({\bf{\vec r}},\,{\bf{\vec
p}},\,t)\,=\,{\bf {\vec p}}^{\left.\,\right. 2}/2\,+\,U({\bf \vec
r}\;,\;t)\;$, while the disturbed setting will be described by the
perturbed functions:
\begin{eqnarray}
 {\cal L}({\bf{\vec r}},\,{\bf \dot{\vec r}},\,t)={\cal L}_o+
 \,\Delta {\cal L}\,=\,\frac{{\bf{\dot {\vec r}}}^{\left.\,\right. 2}}{2}\;-\;
  U({\bf \vec r},\,t)\,+
  \,\Delta {\cal L} ( {\bf \vec r},
 \,{ \bf { \dot { \vec {r}}}} ,\,t) \;,\;
 \label{101}
 \label{412}
 \label{1}
 \end{eqnarray}
 \begin{eqnarray}
 {\bf {\vec p}}\;= \;{\bf{\dot {\vec r}}}\;+\; \frac{\partial \,\Delta
 {\cal L}}{\partial {\bf{\dot{\vec r}}}} \;\;\;\;,\;\;
 \label{2}
 \label{413}
 \label{102}
 \end{eqnarray}
 and
 \begin{eqnarray}
 \nonumber
 {\cal H}({\bf{\vec r}},\,{\bf{\vec p}},\,t)\,=\,{\bf \vec
 p}\,{\bf {\dot {\vec r}}}\,-\,{\cal L}\,= \,\frac{{\bf \vec p}^{\left.
 \,\right. 2}}{2}\,+ \,U\,+\,\Delta {\cal H}\;\;,\;\;\\
 \label{3}
 \label{414}
 \label{103}\\
 \nonumber
 \Delta
 {\cal H}({\bf{\vec r}},\,{\bf{\vec p}},\,t)\,\equiv\,-\,\Delta
 {\cal L}\,-\,\frac{1}{2}\,\left(\frac{\partial \,\Delta {\cal L}}{\partial
 {\bf{\dot {\vec r}}}} \right)^2\;\;\;,\;\;
 \end{eqnarray}
$\Delta {\cal H}\;$ being a variation of the functional form:
$\;\Delta {\cal H}\,\equiv\,{\cal H}({\bf{\vec r}},\,{\bf{\vec
p}},\,t) - {\cal H}_o({\bf{\vec r}},\,{\bf{\vec p}},\,t) \;$. The
Euler-Lagrange equations written for the perturbed Lagrangian will
read:
\begin{equation}
 {\bf{\ddot {\vec r}}}\;=\;-\;\frac{\partial U}{\partial {\bf{\vec r}}} \;+\;
\Delta
 {\bf \vec F}\;\;\;\;,
 \label{4}
 \label{415}
 \label{104}
\end{equation}
the new term
\begin{equation}
 \Delta {\bf \vec F}\;\equiv\;\frac{\partial \,\Delta {\cal L}}{\partial
 {\bf \vec r}}\;-\;\frac{d}{dt}\,\left(\frac{\partial \,\Delta {\cal L}}{\partial
{\bf{\dot
 {\vec r}}}}\right)\;\;\;\;
 \label{5}
 \label{416}
 \label{105}
\end{equation}
being the disturbing force. We see that, in the case of
velocity-dependent perturbations, this force is equal neither to
the gradient of the Lagrangian's perturbation nor to the gradient
of negative Hamiltonian's perturbation. This should be taken into
account when comparing results obtained by different techniques.
For example, in Goldreich (1965) the word ``disturbing function"
was used for the negative perturbation of the Hamiltonian. The
gradient of so defined disturbing function was not equal to the
disturbing force.

Substitution of (\ref{5}) into (\ref{A11}) then yields the generic
form of the equations in terms of the Lagrangian
disturbance.\footnote{~Direct plugging of (\ref{5}) into
(\ref{A11}) results in
  \begin{eqnarray}
 \nonumber
\sum_j\;[C_n\;C_j]\;\frac{dC_j}{dt}\;=\;\frac{\partial
{\efbold}}{\partial C_n}\; \frac{\partial \, \Delta \cal
L}{\partial \erbold}\;-\; \frac{\partial{{\efbold}}}{\partial C_n}
\;\frac{d}{dt}\left({\bf{\vec \Phi}} \,+\,\frac{\partial\, \Delta
\cal
 L}{\partial \bf{\dot{\vec r}}}\right)\;-\; \frac{\partial \bf
{\vec g}}{\partial C_n} \;{\bf {\vec \Phi}} \;\;\;\;.
  \end{eqnarray}
 If the disturbance were to depend upon positions solely, the
 first term on the right-hand side of the above equation would be
 equal simply to $\;\partial \Delta L/\partial C_n\;$. In general, though,
 we should rely on the chain rule
  \begin{eqnarray}
  \nonumber
\frac{\partial \,\Delta \cal L}{\partial C_n} \;=\;\frac{\partial
\, \Delta \cal L }{\partial {{{\erbold}}}} \,\frac{\partial{{\bf
{\efbold}}}}{\partial C_n} \;+\;\frac{\partial \,\Delta \cal
L}{\partial {\bf {\dot {\erbold}}}^{
 \left.~\right.} } \,\frac{\partial \bf\dot{\vec r} }{\partial
 C_n}\;=\;
\frac{\partial \,\Delta \cal L}{\partial {{{\erbold}}}}
\,\frac{\partial{{\bf {\efbold}}}}{\partial C_n}
\;+\;\frac{\partial \Delta \cal L}{\partial {\bf {\dot
{\erbold}}}^{
 \left.~\right.} } \,\frac{\partial ({\bf{\vec g}}\,+\,\Phibold) }{\partial
 C_n}
\;\;\;,
  \end{eqnarray}
 insertion whereof into the preceding formula entails:
  \begin{eqnarray}
 \nonumber
\sum_j\;[C_n\;C_j]\;\frac{dC_j}{dt}\;=\;\frac{\partial \,\Delta
\cal L }{\partial C_n} \;-\;\frac{\partial \,\Delta \cal
L}{\partial {\bf {\dot {\erbold}}}^{
 \left.~\right.} } \,\;\frac{\partial \Phibold  }{\partial C_n}
\;-\; \left(\frac{\partial{\bf{\efbold}}}{\partial C_n}
 \;\frac{d}{dt}\;-\; \frac{\partial
\bf {\vec g} }{ \partial C_n}\right) \;\left({\bf {\vec \Phi }
}\;+\; \frac{\partial \,\Delta \cal L}{\partial \bf\dot{\vec r}}
\right)\;\;.\;
  \end{eqnarray}
 To arrive herefrom to (\ref{A12}), one will have to add and subtract
 $\;(1/2)\partial(\partial (\Delta {\cal L})/\partial {\bf\dot{\vec r}})/\partial C_n\;$
 on the right-hand side. Thereby one makes the gauge function $\;\Phibold\;$
 appear everywhere in the company of $\;\partial (\Delta {\cal L})/\partial
 {\bf\dot{\vec r}}\;$.} The result,
 ~\\

  \begin{eqnarray}
 \nonumber
\sum_j\;\left(\;[C_n\;C_j]\;+
 \;\frac{\partial {\efbold }}{\partial C_n}\;
  \frac{\partial }{\partial C_j}\;
  \left(\frac{\partial \,\Delta \cal
 L}{\partial {\bf\dot{\vec r}}}\;+\;{\Phibold}
  \right)\;\right)
  \frac{dC_j }{dt }\;\;=
~~~~~~~~~~~~~~~~~~~~~~~~~~~~~~~~~~~~~~~~~\\
 \label{A12}
 \label{417}
 \label{2211}\\
 \nonumber
\frac{\partial }{\partial C_n}\,\left[\Delta {\cal
L}\,+\,\frac{1}{2}\,\left(\frac{\partial \,\Delta \cal L}{\partial
\bf \dot{\vec r}} \right)^2 \right]
 \;-\;
\left( \frac{\partial \bf \vec g}{\partial C_n}\;+\;\frac{\partial
\efbold }{\partial C_n}\;\frac{\partial}{\partial
t}\;+\;\frac{\partial \,\Delta \cal L}{\partial \bf\dot{\vec
r}}\;\frac{\partial }{\partial C_n} \right)\left(
  {\bf
{\vec \Phi}}\,+\,\frac{\partial \,\Delta \cal L}{\partial
\bf\dot{\vec r }}
 \right)
 \;\;\;,\;\;
  \end{eqnarray}
not only reveals the convenience of the generalised Lagrange gauge
  \begin{eqnarray}
 \Phibold\;=\;-\;\frac{\partial\,\Delta \cal L}{\partial
 {\bf\dot{\vec r}}}\;\;\;,
 \label{A13}
 \label{2212}
  \end{eqnarray}
(which reduces to $\Phibold=0$ in the case of velocity-independent
perturbations), but also explicitly demonstrates how the
Hamiltonian variation comes into play: it is easy to notice that
the sum in square brackets is equal to $\;-\;\Delta {\cal
H}^{(cont)}$. Substitution of the general-type force $\;{\Delta
\bf{\vec{F}}}\;$ into (\ref{A11}) or of the Lagrangian
perturbation $\;\Delta L\;$ into (\ref{2211}) gives us the
gauge-invariant planetary equations, provided we know the
expressions for elements of the inverse to the Lagrange-bracket
matrix $\;[C_i\,C_j]\;$. When the elements $\;C_i\;$ are chosen as
the Kepler or Delaunay variables, (\ref{A12}) entails the
gauge-invariant versions of the Lagrange and Delaunay planetary
equations (See Appendix 1 to Efroimsky \& Goldreich (2003).)\\

 \subsection{Goldreich (1965)}

 The earliest attempts to describe satellite motion about the precessing
 and nutating Earth were undertaken by Brouwer (1959), Proskurin \&
 Batrakov (1960), and Kozai (1960).

 In 1965 Goldreich accomplished a ground-breaking work that marked the beginning
 of studies of the Martian satellite dynamics. He started out with
 two major assertions. One was that the Martian satellites had either been formed
 in the equatorial plane or been brought therein very long ago. The second was that
 Mars has experienced, though its history, a uniform precession.
 While the former proved to be almost certainly correct,\footnote{~Phobos and
 Deimos give every appearance of being captured asteroids of the carbonaceous
 chondritic type (Veverka 1977, Pang et al. 1978, Pollack et al. 1979, Tolson et
 al. 1978). Given the considerations summarised in detail by Murison (1988), it
 seems least unlikely that the Martian satellites were formerly asteroids that
 were captured by the mechanism of gas drag. If so, this must have occurred early
 in the history of the solar system while the gas disk was substantial
enough to effect a capture. At that stage of planetary formation,
the spin of the forming planet would be closely aligned with the
orbital plane of the planet's motion about the Sun – i.e., the
obliquity would be small and the gas disk would be nearly coplanar
with the planetary orbit. Energetically, a capture would be most
likely to be equatorial. This is most easily seen in the context
of the restricted three-body problem. The surfaces of zero
velocity constrain any reasonable capture to occur from directions
near the inner and outer collinear Lagrange points (Szebehely
1967, Murison 1988), which lie in the equatorial plane. Also, a
somewhat inclined capture would quickly be equatorialised by the
gas disk. If the capture inclination is too high, the orbital
energy is then too high to allow a long enough temporary capture,
and the object would hence not encounter enough drag over a long
enough time to effect a permanent capture (Murison 1988). Thus,
Phobos and Deimos were likely (in as much as we can even use that
term) to have been captured into near-equatorial orbits.} the
validity of the latter remains model-dependent. While in the
simplest approximation the planetary precession is uniform, a more
involved analysis, carried out by Ward (1973, 1974), Laskar \&
Robutel (1993), and Touma \& Wisdom (1994), offers evidence of
strongly nonuniform, perhaps even chaotic, variations of the
Martian obliquity at long time scales. It should be said, though,
that the analysis presented by these authors was model-dependent.
In particular, it was performed in the approximation of the planet
being a nondissipative rigid rotator.\footnote{~In all these
studies the planet was modelled as a rigid body. Touma \& Wisdom
(1994) emphasised in their research that they assumed Mars to be
always in its principal rotation state. A common feature of all
these models was their neglect of the inelastic relaxation of the
spin axis' precession. Whether the latter simplification is
physically justified is yet to be determined. Inelastic
dissipation is, essentially, the internal friction and results
from alternating stresses and strains emerging in a wobbling
rotator (Prendergast 1958; Burns \& Safronov 1973; Efroimsky 2001,
2002c; Molina, Moreno \& Martinez-L{\'o}pez 2003; Sharma, Burns \&
Hui 2004). It is well-known that the lower the frequency of
precession the slower dissipation of the elastic energy associated
therewith. This circumstance suggests that slow obliquity changes
will result in virtually no damping. (The same circumstance makes
one think that the fastest modes of spin precession are well
dissipated. It was for this reason that Touma \& Wisdom (1994)
accepted the model of Mars always being in its principal-rotation
state.) The real physical picture is far more complicated, mainly
due to the essentially nonlinear nature of the phenomenon. Since
Mars is not a perfectly symmetrical oblate spheroid but is a
triaxial body, its free rotation in a spin state, which is even
slightly different from the principal one, is described by the
elliptic functions of Jacobi whose decomposition over sines and
cosines contains an infinite multitude of harmonics. From here, it
is possible to demonstrate that in the course of rotation a
continuous exchange of energy among these harmonics is taking
place. Due to this phenomenon, the alternating stresses associated
with short-period precession are always getting energy from
stresses caused by long-period changes of the spin axis. This way,
dissipation of the short-time motions indirectly leads to damping
of the long-period ones. This phenomenon, called ``energy cascade"
is very well known in the theory of turbulence, but in fact it is
generic and shows itself in nonlinear situations. In the
precessing-top context it has not been explored so far, and we do
not know how effective it is in restraining the obliquity
variations caused by solar torque.} Since the orbits of both
satellites are located within less that $\,2^{{o}}\,$ from the
equator, the two said assertions come to contradiction, unless
there exists a mechanism constraining satellite orbits within the
vicinity of the primary's equator. (Otherwise, as Goldreich noted
in his paper, ``the present low inclinations of these satellites'
orbits would amount to an unbelievable coincidence.") In quest of
such a mechanism, Goldreich (1965) investigated evolution of the
Kepler elements of a satellite in a reference system co-precessing
with the planet. He followed the traditional
variation-of-parameters (VOP) scheme, i.e., assumed a two-body
setting as an undisturbed problem and then treated the inertial
forces, emerging in the co-precessing frame, as perturbation
(along with another perturbation caused by the oblateness of the
planet). Below we present a brief summary of Goldreich's results,
with only minor comments.

Goldreich began by applying formulae (\ref{412} - \ref{416}) to
motion in a coordinate system attached to the planet's centre of
mass and precessing (but not spinning) with the planet. In this
system, the equation of motion includes inertial forces and,
therefore, reads:
 \begin{eqnarray}
 \nonumber
 {\bf{\ddot{\vec {r}}}} \,=\,-\; \frac{\partial U}{\partial
 {\bf{\vec{r}}}} \,-\, 2{\mubold} \, \times \,
 {\bf{\dot{\vec{r}}}}\,-\,{\bf{\dot{\mubold}}}\,\times\,
 {\bf{\vec{r}}}\,-\,{\mubold}\times({\mubold}\times
 {\bf{\vec{r}}})\;=\\
 \label{11.3}
 \label{6}\\
 \nonumber
 -\;\frac{\partial U_o}{\partial
 {\bf{\vec{r}}}} \,-\,\frac{\partial \;\Delta U}{\partial
 {\bf{\vec{r}}}} \,-\,  2{\mubold} \, \times \,
 {\bf{\dot{\vec{r}}}}\,-\,{\bf{\dot{\mubold}}}\,\times\,
 {\bf{\vec{r}}}\,-\,{\mubold}\times({\mubold}\times
 {\bf{\vec{r}}})\;\;,
 \end{eqnarray}
 dots denoting time derivatives in the co-precessing frame
 and $\;\mubold\;$ standing for the coordinate system
 angular velocity relative to an inertial frame.\footnote{~Be
 mindful that $\;\mubold\;$, though being a precession rate
 relative to an inertial frame, is a vector defined in
 the co-precessing frame. (For details see section 8.6 in Marsden and Ratiu
 (2003) or section 27 in Arnold (1989).) In this frame,
 \ba
 \nonumber
 \mubold\;=\;{\bf{\hat x}}_1\,\frac{d\inc_p}{dt}\;+\;
 {\bf{\hat x}}_2\,\frac{dh_p}{dt}\,\sin \inc_p\;+\;{\bf{\hat
 x}}_3\,\frac{dh_p}{dt}\,\cos \inc_p\;\;\;,
 \ea
$\inc_p\;$ and $\;h_p\;$ being the inclination and the longitude
of the node of the planetary equator of date relative to that of
epoch.} Here the physical (i.e., not associated with inertial
 forces)
 potential $\,U(\erbold)\,$ consists of the (reduced) two-body part
 $\;U_o(\erbold)\,\equiv\,-\,G\,M\;\erbold/r^3
 \;$ and a term $\;\Delta U(\erbold)\;$ caused by the planet's oblateness.
 The overall disturbing force on the right-hand side of the above equation is
 generated, according to (\ref{5}), by
  \begin{eqnarray}
  \Delta {\cal L}\left({\bf {\vec r}},\,{\bf{\dot{\vec r}}},\,t
  \right) = \;-\;\Delta U( {\bf{\vec r}} ) + {\bf{\dot {\vec r} } }\,
  {\bf \cdot}\,
  ({\mubold} \times {\bf{\vec{r}}}) \,+\, \frac{1}{2}\,
  ({\mubold} \times {\bf{\vec{r}}})\,{\bf \cdot}\,({\mubold}
  \times {\bf{\vec{r}}}) \;\;\; .
  \label{11.13}
  \label{7}
  \end{eqnarray}
 Since in this case
 \begin{eqnarray}
 \nonumber
\frac{\partial \,\Delta {\cal L}}{\partial {\bf{\dot{r}}}}\;=\;
{\mubold}\times{\bf\vec{r}}\;\;\;,
 \end{eqnarray}
 then
 \begin{eqnarray}
 {\bf {\vec p}}\;= \;{\bf{\dot {\vec r}}}\;+\; \frac{\partial \,\Delta
 {\cal L}}{\partial {\bf{\dot{\vec r}}}} \;=\;{\bf{\dot {\vec r}}}\;+\;
  {\mubold}\times{\bf\vec{r}} \;\;
 \label{8}
 \end{eqnarray}
 and, therefore, the appropriate Hamiltonian variation will look:
 \begin{eqnarray}
 \nonumber
 \Delta {\cal H}\;=\;-\;\left[\Delta {\cal L}\;+\; \frac{1}{2}\left(\frac{\partial
 \,\Delta {\cal L}}{\partial {\bf{\dot{r}}}} \right)^2 \right]\;=\;\;
 \;\;\;\;\;\;\;\;\\
 \label{eq:Hpert}
 \label{9}
 \\
 \nonumber
   -\,\left[\,-\,\Delta U\,+\,{\bf\vec{p}}\cdot
 ({\mubold}\times{\bf\vec{r}})\,
 \right]\,=\,\Delta U\;-\;({\bf\vec{r}}\times{\bf\vec{p}})\cdot\mubold\;=\;
 \Delta U\;-\;{\bf\vec{J}}\cdot \mubold\;\;.\\
 \nonumber
 \end{eqnarray}
 This way, $\;\Delta {\cal H}\;$ becomes expressed through quantities
 defined in the co-precessing frame: the satellite's orbital
 momentum vector $\;{\bf\vec{J}}\;=\;{\bf\vec{r}}\times{\bf\vec{p}}\;$
 and the precession rate $\;\mubold\;$.

 Goldreich employed the above expression in the role of a disturbing
 function $\;R\;$ in the planetary equations:
  \begin{eqnarray}
   \frac{d \inc }{dt}\;=\;\frac{\cos
   \inc}{n\,a^2\,(1\,-\,e^2)^{1/2}\, \sin \inc}\;\;\frac{\partial
   \left(\,-\,\Delta {\cal H} \right) }{\partial \omega} \;-
 \;\frac{1}{n\,a^2\,(1\,-\,e^2)^{1/2}\,\sin \inc
}\;\;\frac{\partial \left(\,-\,\Delta {\cal H} \right) }{\partial
\Omega} \;\;\;{,}\;\;\;\;
 \label{10}\\
 \nonumber\\
 \nonumber\\
 \nonumber\\
\frac{d\Omega}{dt}\;=\;\frac{1}{n\,a^2\,(1\,-\,e^2)^{1/2}\,\sin
\inc }\;\;\frac{\partial \left(\,-\,\Delta {\cal H} \right)
}{\partial \inc }
\;\;\;{,}\;\;\;\;\;\;~~~~~~~~~~~~~~~~~~~~~~~~~~~~~~~~~~~~~~~~~~~~~~
 \label{11}
  \end{eqnarray}
{where}
 \be
 -\;\Delta\,{\cal H}\;\equiv\;R\;=
 \;R_{oblate}\;+\;R_{inertial}\;\;\;
 \label{12}
 \ee
 consists, according to (\ref{9}), of two
 inputs:\footnote{~Our formula (\ref{13}) slightly differs from the one
 employed by Goldreich (1965), because here we use the modern definition of $\;J_2\;$:
 \ba
 \nonumber
 U\;=\;-\;\frac{\mu}{r}\;
 \left\{1\;-\;\sum_{m=2}^{\infty}\;J_m\;\left(\frac{\rho}{r}
 \right)^m\;P_m(\sin \alpha)
 \right\}\;\;,
 \ea
 $\alpha\;$ being the satellite's latitude in the planet-related
 coordinate system. The coefficient $\;J\;$ used by Goldreich (1965) differs from
 our $\;J_2\;$ by a constant factor: $\;J\;=\;(3/2)\;J_2\;\rho^2/r^2\;$.}
  \ba
  R_{oblate}(\nu )\;\equiv\,-\,\Delta U\,=\;\frac{G\,m\;J_2}{2}\;\frac{\rho^2}{r^3}\;\left[\;1\;-
  \;3\;\sin^2 \inc\;\sin^2(\omega\;+\;\nu) \;\right]\;\;,
  \label{13}
  \ea
 and
  \be
  R_{inertial}\,\equiv\,{\bf\vec{J}}\cdot \mubold\;=
  \;\sqrt{G\;m\;a\;\left(1\;-\;e^2\right)}\;\;\;{\wbold}\cdot{\mubold}\;\;\;.
  \label{14}
  \ee
 Here
 $\;m\,\equiv\,\left(m_{primary}\,+\,m_{secondary}
 \right)\;$. The mean motion is, as ever, $\;n\;\equiv\;(G\,m)^{1/2}\,a^{-3/2}\;$,
 while $\,\rho\,$ stands for the mean radius of the primary, $\;\nu\;$ denotes the true anomaly,
 and
  \ba
  r\;=\;a\,\;\frac{1\,-\,e^2}{1\,+\,e\,\cos \nu}
  \label{15}
  \ea
 is the instantaneous orbital radius. In the right-hand side of (\ref{14}) it was assumed that
 the angular momentum is connected with the orbital elements through the well-known formula
  \ba
  {\bf\vec{J}}\;\equiv\;\erbold\,\times\,\pbold\;=\;
  \sqrt{G\,m\;a\;\left(1\;-\;e^2\right)}\;\;\;
  {\wbold}
  \label{16}
  \ea
where
 \ba
 \nonumber
 {\wbold}\;=\;{\bf\hat{x}}_1\;\sin \inc\;\sin \Omega\;-
 \;{\bf\hat{x}}_2\;\sin
 \inc\;\cos \Omega\;+\;{\bf\hat{x}}_3\;\cos
 \inc\;\;~~~~~~~~~~~~~~~~~~~
 \ea
is a unit vector normal to the instantaneous ellipse, expressed
through unit vectors
$\;{\bf\hat{x}}_1,\,{\bf\hat{x}}_2,\,{\bf\hat{x}}_3\;$ associated
with the co-precessing frame $\,\it x_1,\,x_2,\,x_3\,$ (the axes
$\,\it x_1\,$ and $\,\it x_2\,$ lying in the planet's equatorial
plane). The afore-written expression for $\;\wbold\;$ evidently
yields:
 \be
 R_{inertial}\,=\;\sqrt{G\,m\;a\;\left(1\;-\;e^2\right)}\;\;\;\left(\;
 \mu_1\;\sin i\;\sin \Omega\;-\;\mu_2\;\sin i\;\cos \Omega
 \;+\;\mu_3\;\cos i \;\right)\;\;,
 \;\;\;
 \label{17}
 \ee
 while (\ref{13}) may be, in the first approximation, substituted
 with its secular part, i.e., with its average over the orbital period:
 \be
 \bra R_{oblate} \ket \,=\,\frac{n^2\;J_2}{4}\;\,{\rho}^2\;\;
 \frac{3\;\cos^2i\;-\;1}{\left(1\;-\;e^2\right)^{3/2}}\;\;,
 \label{18}
 \ee
 the averaging having been carried out through the medium of formula
 (\ref{A14}) from the Appendix, with (\ref{15}) inserted. With the aid of (\ref{17})
 and (\ref{18}), the planetary equations (\ref{11}) and (\ref{12}) will simplify to:
 \begin{eqnarray}
 \frac{d \inc }{dt}\;=\;-\;\mu_1\;\cos \Omega  \;-\;
 \mu_2\;\sin \Omega \;\;\;{,}
 \label{19}
 \end{eqnarray}
 ~\\
 \begin{eqnarray}
 \frac{d\Omega}{dt}\,=\;-\;\frac{3}{2}\;{n\;J_2}\;\left(\frac{\rho}{a}\right)^2\;\frac{\cos
 {i}}{\left(1\;-\;e^2
 \right)^2}\;+\;O\left(\;\frac{|\mubold|}{J_2\;n}\;\right)
 \;\;\;{.}
 \label{20}
 \end{eqnarray}
 The latter results in the well-known node-precession formula,
 \be
 \Omega\;=\;\Omega_o\;-\;\frac{3}{2}\;{n\;J_2}\;\left(\frac{\rho}{a}\right)^2\;\;
 \frac{\cos {i}}{\left(1\;-\;e^2
 \right)^2}\;\;\;\left(t\;-\;t_o\right)\;\;\;\;.
 \label{21}
 \ee
 Its insertion into the former entails
  \be
 i\;=\;-\;\frac{\mu_1}{\chi}\;\cos \left[\;-\;\chi\;\left(t\;-\;t_o\right)\;+
 \;\Omega_o
 \right]\;+\;\frac{\mu_2}{\nu}\;\sin \left[\;-\;\chi\;\left(t\;-\;t_o\right)\;+\;\Omega_o
 \right]\;+\;i_o \;\;\;\;{,}
 \label{22}
 \ee
 where
 \be
 \;\;\;\chi\;\equiv\;
 \frac{3}{2}\;{n\;J_2}\;\left(\frac{\rho}{a}\right)^2\;\;\frac{\cos {i}}{\left(1\;-\;e^2
 \right)^2}\;\;.
 \label{23}
 \ee
 In Goldreich (1965), equation (\ref{22}) was the main result, its derivation
 being valid for wobble which is slow ($\;|\mubold\,|\,\ll\,J_2\;n\;$) and close
 to uniform ($\;|{ \dotmubold } \,|/| { \mubold } |\,\ll\,J_2\;n\,$){.}

 Despite a warning issued by Goldreich in his paper, this result has often been
 misinterpreted and, therefore, misused in publications devoted to satellites and
 rings of wobbling planets, as well as in the literature on orbits about tumbling
 galaxies.

 In (\ref{22}) $\,\it i\,$ stands for the inclination
 defined in co-precessing axes associated with the planet's
 equator, and therefore (\ref{22}) clearly demonstrates that, in the
 course of
 obliquity changes, this inclination oscillates about zero, with
 no secular shift accumulated. Does this necessarily mean that the satellite
 \textbf{orbit}, too, oscillates about the equatorial
 plane, without a secular deviation therefrom? Most surprisingly,
 the answer to this question is negative. The reason for this is
 that so calculated orbital elements, though defined in the co-precessing
 frame, are \textbf{not osculating} therein. In other words, in the frame
 where
 the elements are introduced, the instantaneous ellipses
 parameterised by these elements are not tangent to the physical
 orbit as seen in this frame.

 This circumstance was emphasised by Goldreich, who noticed
 that formula (\ref{16}) normally (i.e., when employed in an
 inertial frame) connects the osculating elements defined in
 that frame with the angular momentum $\;\erbold\times\pbold\;$
 defined in the same, inertial, frame (i.e., with
 $\;\erbold\times\doterbold\;$). Since in the above calculation
 the frame is not inertial (and, therefore, the angular momentum is
 different
 from $\;\erbold\times\doterbold\;$ but is equal to
 $\;\erbold\times\pbold\,=\,\erbold\times(\doterbold\,+\,
 \mubold\times\erbold)
 \;$), then the orbital elements returned by (\ref{16}) cannot
 be osculating in
 this frame.\footnote{~Were these elements osculating in the frame
 wherein they had been defined, then
 formula (\ref{16}) would read:
 $\;\erbold\,\times\,\doterbold\;=\;\sqrt{G\,m\;a\;
 \left(1\;-\;e^2\right)}\;{\wbold}\;$,
 i.e., would connect the elements with the velocity in that frame.
 In reality, though, it reads:
$\;\erbold\,\times\,\pbold\;=\;\sqrt{G\,m\;a\;\left(1\;-\;
e^2\right)}\;{\wbold}\;$, i.e., connects the elements with the
momentum $\;\pbold\,=\,\doterbold\,+\,\mubold\times\erbold\;$
which happens to coincide with the satellite's velocity relative
to the inertial axes. This situation was formulated by Goldreich
in the following terms: the orbital elements emerging in the above
derivation are defined in the co-precessing frame, but are
osculating in the inertial one. This illustrative metaphor should
not, though, be overplayed: the fact that the elements emerging in
Goldreich's computation return the inertial-frame-related velocity
does not mean that this inclination may be interpreted as that
relative to the invariable plane. (The elements were introduced in
the co-precessing frame!)} On these grounds Goldreich warned the
reader of the peculiar nature of the elements used in his
integration.

To this we would add that it is not at all evident that the
inertial-forces-caused alteration of the planetary equations
should be achieved through amending the disturbing function with
the momentum-dependent variation of the negative Hamiltonian,
$\;{\bf\vec{J}}\cdot \mubold\;$. While the common fallacy
identifies the disturbing function with the negative Hamiltonian
perturbation, in reality this rule-of-thumb works (and yields
elements that are osculating) only for disturbances dependent
solely upon positions, not upon velocities (i.e., for Hamiltonian
perturbations dependent only upon coordinates, but not upon
momenta). Ours is not that case and, therefore, more alterations
in the planetary equations are needed to account for the frame
precession, if we wish these equations to render osculating
elements. However, if one neglects this circumstance and simply
amends the disturbing function with $\;{\bf\vec{J}}\cdot
\mubold\;$, then the planetary equations will give some elements
different from the osculating ones. It will then become an
interesting question as to whether such elements will or will not
coincide with those rendered by (\ref{15}) when this formula is
used in non-inertial frames.

All these subtle issues get untangled in the framework of the
gauge formalism. Application of this formalism to motions in
non-inertial frames of references was presented in Efroimsky \&
Goldreich (2004). The main results proven there are the
following.\\

1. If one attempts to account for the inertial forces by
   simply adding the term $\;{\bf{\vec{J}}}\cdot\mubold\;$ to
   the disturbing function, with no other alterations made in
   the planetary equations, then these equations indeed do
   render quantities that may be interpreted as some orbital elements
   (i.e., as parameters of some instantaneous conics).
   These elements are NOT
   osculating and, therefore, the instantaneous conics
   parameterised by there elements are not tangent to the physical
   orbit. Hence, these elements cannot, generally, be attributed
   a direct physical interpretation,\footnote{When the orbit
   evolution is
   sufficiently slow, the observer can attribute some physical
   meaning to elements of the osculating conic. For example,
   whenever an observer talks about the inclination or the
   eccentricity of a perturbed orbit, he naturally implies those
   of the osculating
   ellipse or hyperbola.} except in situations where their
   deviation from the osculating elements remains sufficiently
   small.\\

2. By a remarkable coincidence, these non-osculating
   elements turned out to be identical with those emerging in
   formula (\ref{16}). This coincidence was implicitly taken
   for granted by Goldreich (1965), which reveals his truly
   incredible scientific intuition.\\

3. To build up a system of planetary equations that render
   osculating elements of the orbit as seen in the co-precessing
   coordinate system, one has not only to add
   $\;{\bf{\vec{J}}}\cdot\mubold\;$ to the disturbing function,
   but also to amend each of these equations with several extra
   terms. Some of those terms are of order
   $\;(\,|\mubold\,|/(J_2\;n)\,)^2\;$; some others are of order
   $\;|{ \dotmubold } \,|/ ( | { \mubold } |\,J_2\;n)\,$.
   Most importantly, some terms are of first order in the
   precession-caused perturbation $\;|\mubold\,|/(J_2\;n)\;$,
   which means right away that the non-osculating elements
   used by Goldreich (1965) differ from the osculating ones
   already in the first order.
   While a more comprehensive account on this topic, with the
   resulting equations, will be offered at the end of this
   section, here we shall touch upon only one question which
   gets immediately raised by the presence of such first-order
   differences. This question is: what are the averages of these
   differences? Stated alternatively, do the secular components
   of the said non-osculating elements differ considerably
   from those of the
   osculating ones? Goldreich (1965) stated, without proof, that
   the secular components
   differ
   only in high orders over the velocity-dependent part of the
   perturbation. In our
   paper we shall probe the limits for this assumption.\\

 \subsection{Brumberg and Kinoshita}

 A development, part of which was similar to that of Goldreich
(1965), was independently carried out by Brumberg, Evdokimova \&
Kochina (1971), who studied orbits of artificial lunar satellites
in a coordinate system co-precessing with the Moon. In that
article, too, the non-osculating nature of the resulting orbital
variables did not go unnoticed. The authors called these variables
``contact elements" and stated (though never proved) that these
variables return not the correct value of the velocity but that of
the momentum. Later, one of these authors rightly noted in his
book (Brumberg 1992) that the contact elements differ from the
osculating ones already in the first order over the
velocity-dependent part of the perturbation. In subsection 1.1.3
of that book, he unsuccessfully tried to derive analytical
transformations interconnecting these sets of
variables.\footnote{~Contrary to the author's statement, formula
(1.1.41) in Brumberg (1992) is not rigorous, but is valid only to
first order. (To make it rigorous, one should substitute
everywhere, except in the denominators, $\;\doterbold\;$ with
$\;\doterbold\;-\;\partial R/\partial \doterbold\;$.) Besides, the
author did not demonstrate his derivation of formula (1.1.43), for
$\;M_o\;$, from (1.1.42). (In Brumberg's book the mean anomaly is
denoted with $\,\it{l}\,$, not with $\,M\,$. )

Most importantly, the qualitative reasoning presented by the
author in the paragraph preceding formula (1.1.43) is unrigorous
and essentially incorrect. The cause of this is that the author
compares the planetary equations for contact elements, written in
a precessing frame, with the equations for osculating ones,
written in an inertial frame, instead of comparing two such
systems (for contact and for osculating elements), both of which
are written in a precessing frame. This makes a big difference
because, as we already explained above, transition to a precessing
frame does not simply mean addition of an extra term to the
disturbing function.

Despite all these mathematical irregularities, the averaged system
of planetary equations (1.1.44), ``derived" by Brumberg for the
first-order secular perturbations, turns out to be correct in the
limit of uniform precession. Just as in the preceding subsection
we had a reason to praise the unusual intuition of Goldreich, so
here we have to pay tribute to the excellent intuition of
Brumberg, intuition which superseded his flawed mathematics.}

A similar attempt was undertaken in a very interesting article by
Ashby \& Allison (1993). Though the authors succeeded in many
other points, their attempt to derive formulae for such a gauge
transformation was not successful.\footnote{~To carry out the
gauge transformation, the authors used a set of intermediate
variables $\;\{Q^k_{(o)}\,,\;P_{k\,(o)}\}\;$, which were canonical
and, at the same time, osculating. As follows from the theorem
proven by Efroimsky \& Goldreich (2003), these variables are
nonexistent when the perturbation depends upon velocities.}

 The setting, considered by Goldreich (1965) in the context of
 Martian satellites and by Brumberg et al (1971) in the context
 of circumlunar orbits, later emerged in the article by Kinoshita
 (1993), who addressed the satellites of Uranus.

 Kinoshita's treatment of the problem was based on the following
 mathematical construction. Denote satellite
 positions and velocities in the inertial and in the co-precessing axes with
 {$\;\left\{\,\erbold\;'\,,\;{\bf{\vec v}}\;'\,\right\}$} and with
 {$\;\left\{\,\erbold\,,\;{\bf{\vec v}}\,\right\}\;$},
 correspondingly.\footnote{~We use notations opposite to those in
 Kinoshita (1993), in order to conform with the notations of
 Goldreich (1965).} Interconnection between them will be implemented by
 an orthogonal matrix $\;\hat A\;$,
 \ba
 \erbold\,=\,\hat{A}\,\erbold~'~~,~~~~~
 {\bf\vec{v}}\,\equiv\,\doterbold\,=\,\dot{\hat A}\,\erbold~'\,+\,
 \hat{A}\,{\dot{\erbold}}~'\,=\,\dot{\hat A}\,
 {\hat{A}}^{-1}\,\erbold\,+\,\hat{A}\,{\bf\vec v}~'\,=
 \;-\;\mubold\,\times\,\erbold
 \,+\,\hat{A}\,{\bf\vec p}~'\;\;,\;\;\;
 \ea
 $\mubold\;$ being the precession rate as seen in the
 co-precessing coordinate system, and the inertial velocity
 $\;{\bf\vec{v}}~'\;$ being identical to the inertial momentum
 $\;{\bf\vec{p}}~'\;$. Kinoshita suggested interpreting
 this interconnection as a canonical transformation between variables
 {$\;\{\,\erbold~'\,,\;\pbold~'\,\}\;$ }
 and  {$\;\{\,\erbold\,,\;\pbold\,\}\;$}, implemented by
 generating function
 \begin{eqnarray}
 F_2\;=\;\pbold\,\cdot\,\hat{A}\erbold~'\;=\;
 \left(\,\hat{A}^T\,\pbold\;\right)\;\cdot\; \erbold~' \;\;\;.\;
 \end{eqnarray}
 This choice of generating function rightly yields
 \begin{eqnarray}
 \erbold\,=\,\frac{\partial F_2 }{\partial \pbold
 }\,=\,\hat{A}\,\erbold~'\;\;,
 \end{eqnarray}
 while the interconnection between momenta will look:
 \ba
 \pbold~'\,=
 \,\frac{\partial F_2}{\partial
 \erbold~'}\,=\,\hat{A}^T\,\pbold\;\;,\;\;\;
 \mbox{i.e.,}\;\;\;\;
 \pbold \,=\,\left(\,\hat{A}^T\,\right)^{-1}\,\pbold~'\,=\,
 \hat{A}\,\pbold~'\;\;\;,\;\;\;
 \end{eqnarray}
whence $\;\pbold\;=\;{\bf\vec{v}}\;+\;\mubold\,\times\,\erbold\;$.
The Hamiltonian in precessing axes will read
 \ba
 \nonumber
 H\left(\,\erbold\,,\;\pbold\,\right)\;=\;
 H^{(inert)}\left(\,\erbold~'\,,\;\pbold~'\,\right)\,+\,
 \frac{\partial F_2 }{\partial t }\,=\,
 H^{(inert)}\left(\erbold~'\,,\;\pbold~'\,\right)\,+
 \,\pbold\,\cdot\,\dot{\hat{A}}\,
 \erbold~'
 \end{eqnarray}
 \begin{eqnarray}
 =\;H^{(inert)}\left(\erbold~',\,\pbold~'\right)\;-\;
 \pbold \;\cdot\; \left(\mubold\,\times\,\erbold \right) \;=\;
  H^{(inert)}\left(\erbold~',\,\pbold~'\right) \;-\;
 \left(\erbold\,\times\,\pbold
 \right)\;\cdot\;\mubold\;\;\;\;.~~~~~
 \end{eqnarray}
 The Hamiltonian perturbation, caused by the inertial
 forces, is $\,-\,\left(\erbold\,\times\,\pbold
\right)~\cdot~\mubold\,=\,-\,{\bf\vec{J}}~\times~\mubold\,$,
vector $\,\bf\vec J\,$ being the orbital angular momentum
 as seen in the co-precessing frame. Comparing this with
 (\ref{9}), we see that employment of the above canonical
 transformation is but another method of stepping on the
 same rake. In distinction from Goldreich (1965) and
 Brumberg et al (1971), Kinoshita in his paper did not
 notice that he was working with non-osculating elements.

 The problem with Kinoshita's treatment is that the
 condition of canonicity in some situations comes into
 contradiction with the osculation condition. In other words,
 canonicity sometimes implicitly contains a constraint that
 sometimes is different from the Lagrange constraint (\ref{A5}).
 This issue was comprehensively elucidated in the work
 Efroimsky \& Goldreich (2003). The authors began with the
 reduced two-body setting and thoroughly re-examined the
 Hamilton-Jacobi procedure, which leads one from the spherical
 coordinates and the corresponding canonical momenta to the
 set of Delaunay variables. While in the undisturbed two-body
 case this procedure yields the Delaunay variables which are
 trivially osculating (and parameterise a fixed Keplerian ellipse
 or hyperbola), in the perturbed case the situation becomes more
 involved. According to the theorem proven in that paper, the
 resulting Delaunay elements are osculating (and parameterise a
 conic tangent to the perturbed trajectory) if the Hamiltonian
 perturbation depends solely upon positions, not upon momenta
 (or, the same: if the Lagrangian perturbation depends upon
 positions but not upon velocities). Otherwise, the Delaunay
 elements turn out to be non-osculating (and parametrise the
 physical trajectory with a sequence of non-tangent conics).
 As one can see from the above equation, the Hamiltonian
 perturbation, caused by the inertial forces, depends upon
 the momentum, and this circumstance indicates the problem.
 This trap, in which many have fallen, is of special importance
 in General Relativity, because the relativistic corrections to
 the equations of motion are velocity-dependent.\footnote{~In an
 interesting article (Chernoivan \& Mamaev 1999), the authors
 addressed the two-body problem on a curved background. The
 curvature entailed a velocity-dependent relativist correction,
 which was treated as a perturbation. After carrying out the
 Hamilton-Jacobi development, the authors arrived at canonical
 variables analogous to the Delaunay elements. Orbit integration
 in terms of these variables would be as correct as in
 terms of any others. The problems began when the authors used
 these elements to come to some conclusions regarding
 the perihelion precession. Those conclusions need
 to be reconsidered, because they were rendered on the basis
 of Delaunay elements that were non-osculating.
 Similar comments may be made about the work by Richardson \&
 Kelly (1988) who addressed, using the Hamiltonian formalism, the
 relativist two-body problem in the post-Newtonian approximation.}

 \section{Planetary equations}

 In this section we shall briefly spell out some results
obtained in Efroimsky \& Goldreich (2004) and shall use these
results to derive the Lagrange-type planetary equations (\ref{41}
- \ref{46}) for osculating elements in a coordinate system
co-precessing with an oblate primary.

 \subsection{Planetary equations for contact elements}

Above, in subsection 2.1, we provided a very short account of the
gauge formalism. Expression (\ref{A12}), presented there, is the
most general form of the planetary equations for an arbitrary set
of six independent orbital elements, written in terms of an
arbitrary disturbance of the Lagrangian.

When the elements $\;C_i\;$ are chosen to be the Keplerian or
Delaunay sets of variables, we arrive at the gauge-invariant
versions of the Lagrange or Delaunay planetary equations,
correspondingly. They are written down in the Appendix to
Efroimsky \& Goldreich (2003). Interplay between the gauge freedom
and the freedom of frame choice is explained at length in Section
3 of the article Efroimsky \& Goldreich (2004), which addresses
orbits about a precessing planet. It is demonstrated in that work
that, if one chooses to describe the motion in terms of the
non-osculating elements that were introduced in a co-precessing
frame and defined in the generalised Lagrange gauge\footnote{~It
is an absolutely crucial point that choice of a gauge and choice
of a reference frame are two totally independent procedures. In
each frame one has an opportunity to choose among an infinite
variety of gauges.} (\ref{A13}), then the corresponding
Hamiltonian perturbation will read:
 \begin{equation}
 \Delta {\cal H}^{(cont)}\;=\;-\;\left[\;R_{oblate}(\efbold)+
 \mubold\cdot(\efbold\times {\bf {\vec{g}}})\;\right]\;\;\;,
 \label{24}
 \end{equation}
while the planetary equations (\ref{A12}) acquire the form
 \ba
 \nonumber
 [C_r\;C_i]\;\frac{dC_i}{dt}\;=\;\frac{\partial\;\left(\;-\;\Delta {\cal H}^{(cont)}\;\right)}{\partial C_r}\;
 \;\;\;,
 \ea
 or:
 \begin{equation}
 [C_r\;C_i]\;\frac{dC_i}{dt}\;=\;\frac{\partial}{\partial
 C_r}\;
 \left[\;R_{oblate}( \efbold )\;+\;\mubold\cdot(\efbold\times {\bf{\vec g}})\;\right]\;\;\;,
 \label{25}
 \end{equation}
 where ${{\efbold}}$ and ${\vec{\bf{g}}}$ stand for the
undisturbed (two-body) functional expressions of the position and
velocity via the time and the chosen set of orbital elements:
 \begin{eqnarray}
 \nonumber
 {\bf \vec r }\;=\;{\efbold} \left(C_1, ... , C_6, \,t \right)~~~~~~~~~~~~~~~~~~~~~~~~~~~~~~~\\
 \label{26}\\
 \nonumber
 {\bf \vec v}\;=\;{\bf \vec g} \left(C_1, ... , C_6, \,t \right)\;\equiv\;
 \frac{\partial }{\partial t}\,{\efbold} \left(C_1, ... , C_6, \,t \right)\;\;\,
 \end{eqnarray}
(so that $\;{\efbold} \left(C_1(t), ... , C_6(t), \,t \right)\;$
and $\;{\bf \vec g} \left(C_1(t), ... , C_6(t), \,t \right)\;$
become the ansatz for solving the disturbed problem). For
$\;R_{oblate}\;$ in (\ref{24}) - (\ref{25}), one can employ,
dependent upon the desired degree of rigour, either the exact
expression (\ref{13}) or its orbital average (\ref{18}).

In the generalised Lagrange gauge (\ref{A13}) the canonical
momentum becomes:
 \ba
{\bf\vec p}\;=\;{\bf\dot{\vec{r}}}\;+\;\frac{\partial\, \Delta
{\cal L}}{\partial {\bf\dot{\vec{r}}}}\;=\;{\bf\vec
g}\;+\;\Phibold\;+\;\frac{\partial\, \Delta {\cal L}}{\partial
{\bf\dot{\vec{r}}}} \;=\;{\bf\vec g} \;\;,
 \label{27}
 \ea
which means that its functional dependence upon the time and the
chosen set of orbital elements is the same as it used to be in the
unperturbed case where both the velocity and the momentum were
simply equal to $\;{\bf \vec g} \left(C_1, ... , C_6, \,t
\right)\;$. This also means that $\;\Delta {\cal H}^{(cont)}\;$
coincides with Goldreich's $\;\Delta {\cal H}\;$ given by
(\ref{13}).

We see that the generalised Lagrange gauge (\ref{A13}) singles out
the same set of non-osculating elements which showed up in the
studies by Goldreich (1965) and Brumberg, Evdokimova \& Kochina
(1971), -- the set of ``contact elements." This is why in
(\ref{24}) the Hamiltonian perturbation was written with the
superscript ``{\it{cont}}."

In (\ref{25}) the Lagrange-bracket matrix is defined in the
unperturbed two-body fashion (\ref{A9}) and can, therefore, be
trivially inverted. Hence, when the elements are chosen as the
Keplerian ones, the appropriate equations look like the customary
Lagrange-type equations (i.e., like (\ref{11}) and (\ref{12})
above), with the disturbing function given by (\ref{13}) or,
equivalently, by (\ref{24}):
 \begin{equation}
 \frac{da}{dt}\;=\;\frac{2}{n\,a}\;\;\frac{\partial\left(\,-\,\Delta {\cal H}^{(cont)}\right)
 }{\partial M_o}\;\;\;\;\;,\;\;\;\;\;\;\;\;
 \label{28}
 \label{446}
  \end{equation}
 ~\\
  \begin{eqnarray}
 \frac{de}{dt}\,=\,\frac{1-e^2}{n\,a^2\,e}\;\;\frac{\partial
 \left(\,-\,\Delta {\cal H}^{(cont)} \right)  }{\partial M_o}\;-\;
\frac{(1\,-\,e^2)^{1/2}}{n\,a^2\,e} \;\frac{\partial
\left(\,-\,\Delta {\cal H}^{(cont)} \right) }{\partial \omega}
\;\;\;,\;
 \label{29}
 \label{447}
  \end{eqnarray}
 ~\\
  \begin{eqnarray}
 \frac{d\omega}{dt}\,=\,\frac{\;-\,\cos \inc
}{n\,a^2\,(1\,-\,e^2)^{1/2}\, \sin \inc }\;\frac{\partial
\left(\,-\,\Delta {\cal H}^{(cont)} \right) }{\partial \inc }
\,+\,
 \frac{(1-e^2)^{1/2}}{n\,a^2\,e}\;\frac{\partial \left(\,-\,\Delta {\cal H}^{(cont)} \right) }{\partial
e}
 \label{30}
 \label{448}
  \end{eqnarray}
 ~\\
  \begin{eqnarray}
 \frac{d \inc }{dt}\;=\;\frac{\cos
\inc}{n\,a^2\,(1\,-\,e^2)^{1/2}\, \sin \inc}\;\;\frac{\partial
\left(\,-\,\Delta {\cal H}^{(cont)} \right) }{\partial \omega} \;-
 \;\frac{1}{n\,a^2\,(1\,-\,e^2)^{1/2}\,\sin \inc
}\;\;\frac{\partial \left(\,-\,\Delta {\cal H}^{(cont)} \right)
}{\partial \Omega} \;\;\;,\;\;
 \label{31}
 \label{449}
  \end{eqnarray}
 ~\\
 ~\\
  \begin{eqnarray}
\frac{d\Omega}{dt}\;=\;\frac{1}{n\,a^2\,(1\,-\,e^2)^{1/2}\,\sin
\inc }\;\;\frac{\partial \left(\,-\,\Delta {\cal H}^{(cont)}
\right) }{\partial \inc } \;\;\;,\;\;\;\;\;\;
 \label{32}
 \label{450}
  \end{eqnarray}
 ~\\
 ~\\
  \begin{eqnarray}
\frac{dM_o}{dt}\,=\,\;-\,\frac{1\,-\,e^2}{n\,a^2\,e}\,\;
\frac{\partial \left(\,-\,\Delta {\cal H}^{(cont)} \right)
 }{\partial e } \;-\;
 \frac{2}{n\,a}\,\frac{\partial \left(\,-\,\Delta {\cal H}^{(cont)} \right)
 }{\partial a }
\;\;\;\;,\;\;
 \label{33}
 \label{451}
  \end{eqnarray}

~\\

 \subsection{Planetary equations for osculating elements}

When one introduces elements in the precessing frame and also
demands that they osculate in this frame (i.e., makes them obey
the Lagrange gauge $\,\Phibold\,=\,0\,$) then the Hamiltonian
variation reads:\footnote{~ Just as $\;\Delta {\cal H}^{(cont)}\;$
in (\ref{24}), this Hamiltonian variation is still
 equal to $\;\;-\,\left[R({{\efbold}},\,t)\,+\,
 \mubold\cdot{\bf{\vec J}}\right]\,=\,-\,\left[R({{\efbold}},\,t)\,+\,
 \mubold\cdot(\efbold\times{\bf{\vec p}})\right]\;$. However, the
 canonical momentum now is different from $\;\bf\vec g\;$ and reads as:
 $\;{\bf{\vec p}}\;=\;{\bf{\vec g}}\;+\;(\mubold\times\efbold)\;$.}
 \begin{equation}
 \Delta {\cal H}^{(osc)}\;=\;-\,\left[\;R_{oblate}(\nu)\,+\,
 \mubold\cdot(\efbold\times{\bf{\vec g}})\;+\;
 (\mubold\times\efbold)\cdot(\mubold\times\efbold)\;\right]\;\;\;,
 \label{eq:DelHosc}
 \label{34}
 \end{equation}
 while equation (\ref{A12}) becomes:
 \begin{eqnarray}
 \nonumber [C_n\;C_i]\;\frac{dC_i}{dt}\;=\;-\;
 \frac{\partial\,\Delta {\cal H}^{(osc)}}{\partial
 C_n}\,~~~~~~~~~~~~~~~~~~~~~~~~~
 \nonumber \\ &&
 \label{35}\\
 \nonumber +\;\mubold\cdot \left(\frac{\partial{\efbold}}{\partial
 C_n}\times {\bf{\vec g}}\;-\;{\efbold}\times \frac{\partial{\vec{\bf
 g}}}{\partial C_n}\right)\;-\; {\bf{\dot{\mubold
 }}}\cdot\left(\efbold\times \frac{\partial \efbold }{\partial
 C_n}\right)\;-\;\left(\mubold\times\efbold\right) \;\frac{\partial
 }{\partial C_n}\left(\mubold\times\efbold\right) \;\;\;.
 \end{eqnarray}
To ease the comparison of this equation with (\ref{25}), it is
convenient to split the expression for $\;\Delta {\cal
H}^{(osc)}\;$ in (\ref{34}) into two parts:
 \ba
\Delta {\cal H}^{(cont)}\;=\;-\,\left[\;R_{oblate}({{\efbold
}},\,t)\;+\;\mubold\cdot(\efbold \times {\bf{\vec g}})\;\right]
 \label{36}
 \ea
 and
 \ba
 -\,(\mubold\times\efbold)\cdot(\mubold\times\efbold)\;\;\;,
 \label{37}
 \ea
 and then to group the latter part with the last term on the right-hand side of (\ref{35}):
 \begin{eqnarray}
 \nonumber [C_n\;C_i]\;\frac{dC_i}{dt}\;=\;-\;
 \frac{\partial\;\Delta {\cal H}^{(cont)}}{\partial C_n}~~~~~~~~~~~~~~~~
 \nonumber \\ &&
 \label{eq:avoposc}
 \label{38}\\
 \nonumber
 +\;\mubold\cdot
 \left(\frac{\partial{\efbold}}{\partial C_n}\times
 {\bf{\vec g}}\;-\;{\efbold}\times
 \frac{\partial{\vec{\bf g}}}{\partial C_n}\right)\;-\;
 {\bf{\dot{\mubold }}}\cdot\left(\efbold\times
 \frac{\partial \efbold }{\partial C_n}\right)\;+\;\left(\mubold\times
\efbold\right)
 \;\frac{\partial }{\partial C_n}\left(\mubold\times\efbold\right) \;\;\;.
 \end{eqnarray}
 One other option is to fully absorb the $\;O(|\mubold|^2)\;$ term
 into $\;\Delta {\cal H}\;$, i.e., to introduce the effective ``Hamiltonian"
 \begin{equation}
 \Delta {\cal H}^{(eff)}\;=\;-\,\left[\;R_{oblate}(\nu)\,+\,
 \mubold\cdot(\efbold\times{\bf{\vec g}})\;+\;\frac{1}{2}\;
 (\mubold\times\efbold)\cdot(\mubold\times\efbold)\;\right]\;\;\;
 \label{39}
 \end{equation}
and to write down the equations like this:
 \begin{eqnarray}
 [C_n\;C_i]\;\frac{dC_i}{dt}\;=\;-\;
 \frac{\partial\;\Delta {\cal H}^{(eff)}}{\partial C_n}~
 +\;\mubold\cdot
 \left(\frac{\partial{\efbold}}{\partial C_n}\times
 {\bf{\vec g}}\;-\;{\efbold}\times
 \frac{\partial{\vec{\bf g}}}{\partial C_n}\right)\;-\;
 {\bf{\dot{\mubold }}}\cdot\left(\efbold\times
 \frac{\partial \efbold }{\partial C_n}\right)\;\;\;.\;\;\;\;
 \label{40}
 \end{eqnarray}

For $\;{C_i}\;$ being chosen as the Keplerian elements, inversion
of the Lagrange brackets will yield the following Lagrange-type
system:
  ~\\
 \ba
 \frac{da}{dt}\,=\,
 \frac{2}{n a}\,\left[\,\;\frac{\partial\left(\,-\,\Delta {\cal H}^{(eff)}\right)
 }{\partial M_o}\,-\,
 {\bf{\dot{\mubold }}}\cdot\left(\efbold\times
 \frac{\partial \efbold }{\partial
 M_o}\right)\;\right]\;\;,\;\;\;~~~~~~~~~~~~~~~~~~~~~~~~~~~~~~~~~~~~~~~~~~~
 \label{41}
 \label{459}
  \ea
  ~\\
 ~\\
  \begin{eqnarray}
  \nonumber
 \frac{de}{dt}\,=\,\frac{1-e^2}{n\,a^2\,e}\;\;\left[\;\frac{\partial
 \left(\,-\,\Delta {\cal H}^{(eff)} \right)  }{\partial M_o}
 \,-\,
 {\bf{\dot{\mubold }}}\cdot\left(\efbold\times
 \frac{\partial \efbold }{\partial M_o}\right)
 \;\right]~~\,~~~~~~~~~~~~~~~~~~~~~~~~~~~~~~~~~~~~~~~~~\\
 \nonumber\\
 \label{460}
 \label{42}\\
 \nonumber\\
 \nonumber
 -\;
 \frac{(1\,-\,e^2)^{1/2}}{n\,a^2\,e} \;\left[\;\frac{\partial
 \left(\,-\,\Delta {\cal H}^{(eff)} \right) }{\partial \omega} \,~ +\, \mubold\cdot
 \left(\frac{\partial{\efbold}}{\partial \omega}\times
 {\bf{\vec g}}\,-\,{\efbold}\times
 \frac{\partial{\vec{\bf g}}}{\partial \omega}\right)\,-\,
 {\bf{\dot{\mubold }}}\cdot\left(\efbold\times
 \frac{\partial \efbold }{\partial \omega}\right)
 \;\right]~~,~~~
  \end{eqnarray}
 ~\\
 ~\\
  \begin{eqnarray}
 \nonumber
 \frac{d\omega}{dt}\,=~
 \frac{\;-\,\cos \inc
}{n a^2(1-e^2)^{1/2}\, \sin \inc }\left[ \frac{\partial
\left(\,-\,\Delta {\cal H}^{(eff)} \right) }{\partial \inc } \,~ +
~ \mubold\cdot
 \left(\frac{\partial{\efbold}}{\partial \inc}\times
 {\bf{\vec g}}\,-\,{\efbold}\times
 \frac{\partial{\vec{\bf g}}}{\partial \inc}\right)\,-\,
 {\bf{\dot{\mubold }}}\cdot\left(\efbold\times
 \frac{\partial \efbold }{\partial \inc}\right)\,
  \right]~~~~\\
 \nonumber\\
 \label{43}\\
 \nonumber\\
 \nonumber
 +\,~\frac{(1-e^2)^{1/2}}{n\,a^2\,e}\;
 \left[\;\frac{\partial \left(\,-\,\Delta {\cal H}^{(eff)} \right) }{\partial e}\,
+ ~ \mubold\cdot
 \left(\frac{\partial{\efbold}}{\partial e}\times
 {\bf{\vec g}}\,-\,{\efbold}\times
 \frac{\partial{\vec{\bf g}}}{\partial e}\right)\,-\,
 {\bf{\dot{\mubold }}}\cdot\left(\efbold\times
 \frac{\partial \efbold }{\partial e}\right)
 \;\right]\;\;,~~~~~~~~~~~
  \end{eqnarray}
 ~\\
 ~\\
 ~\\
 \ba
 \nonumber
 \frac{d \inc
 }{dt}\;=~ \frac{\cos \inc}{n a^2\,(1 - e^2)^{1/2} \sin
 \inc}
 \left[\frac{\partial \left(\,-\,\Delta {\cal H}^{(eff)} \right)}{\partial
 \omega}\;+ \; \mubold\cdot
 \left(\frac{\partial{\efbold}}{\partial \omega}\times
 {\bf{\vec g}}\,-\,{\efbold}\times
 \frac{\partial{\vec{\bf g}}}{\partial \omega}\right)\,-\,
 {\bf{\dot{\mubold }}}\cdot\left(\efbold\times
 \frac{\partial \efbold }{\partial \omega}\right)
 \,\right]\,-\;\;\;\\
 \nonumber\\
 \label{44}\\
 \nonumber\\
 \nonumber
 \frac{1}{na^2\,(1-e^2)^{1/2}\,\sin \inc
 }\,\left[\,\frac{\partial \left(\,-\,\Delta {\cal H}^{(eff)} \right) }{\partial
 \Omega}\;+ \; \mubold\cdot
 \left(\frac{\partial{\efbold}}{\partial \Omega}\times
 {\bf{\vec g}}\,-\,{\efbold}\times
 \frac{\partial{\vec{\bf g}}}{\partial \Omega}\right)\,-\,
 {\bf{\dot{\mubold }}}\cdot\left(\efbold\times
 \frac{\partial \efbold }{\partial \Omega}\right)
 \,\right]\;{,}\;\;\;
  \ea
 ~\\
 ~\\
  \ba
\frac{d\Omega}{dt}\,=~\frac{1}{n a^2\,(1-e^2)^{1/2}\,\sin \inc
}\,\left[\,\frac{\partial \left(\,-\,\Delta {\cal H}^{(eff)}
\,\right) }{\partial \inc }\,~ + ~\mubold\cdot
 \left(\frac{\partial{\efbold}}{\partial \inc}\times
 {\bf{\vec g}}\,-\,{\efbold}\times
 \frac{\partial{\vec{\bf g}}}{\partial \inc}\right)\,-\,
 {\bf{\dot{\mubold }}}\cdot\left(\efbold\times
 \frac{\partial \efbold }{\partial \inc}\right)\,
 \right]\;,\;\;\;
 \label{45}
  \ea
 ~\\
 ~\\
  \begin{eqnarray}
  \nonumber
\frac{dM_o}{dt}\,=\,\;-\,\;\frac{1\,-\,e^2}{n\,a^2\,e}\,\;
\left[\;\frac{\partial \left(\,-\,\Delta {\cal H}^{(eff)} \right)
 }{\partial e }~ + ~
 \mubold\cdot
 \left(\frac{\partial{\efbold}}{\partial e}\times
 {\bf{\vec g}}\,-\,{\efbold}\times
 \frac{\partial{\vec{\bf g}}}{\partial e}\right)\,-\,
 {\bf{\dot{\mubold }}}\cdot\left(\efbold\times
 \frac{\partial \efbold }{\partial e}\right)
 \;\right] ~~~~~\\
 \nonumber\\
 \label{46}\\
 \nonumber\\
 \nonumber
 -\;~\frac{2}{n\,a}\,\left[\;\frac{\partial \left(\,-\,\Delta {\cal H}^{(eff)} \right)
 }{\partial a }~ + ~ \mubold\cdot
 \left(\frac{\partial{\efbold}}{\partial a}\times
 {\bf{\vec g}}\,-\,{\efbold}\times
 \frac{\partial{\vec{\bf g}}}{\partial a}\right)\,-\,
 {\bf{\dot{\mubold }}}\cdot\left(\efbold\times
 \frac{\partial \efbold }{\partial a}\right)
 \;\right]
 \;\;,~~~~~
  \end{eqnarray}
  ~\\
  terms $\;\mubold\cdot\left(\;(\partial \efbold/\partial M_o)\times {\bf\vec g}\,-\,
  (\partial {\bf \vec g}/\partial M_o)\times\efbold\;\right)\;$ being
  omitted in (\ref{41} - \ref{42}), because these terms vanish
  identically (see the Appendix).
In equations (\ref{40} - \ref{46}), $\;\Delta {\cal H}^{(eff)}\;$
is given by (\ref{39}). In a rigorous analysis, with the
true-anomaly dependence of all terms in (\ref{41} - \ref{46})
taken into account, the Hamiltonian will look, according to
(\ref{13} - \ref{14}) and (\ref{17}):
 \begin{eqnarray}
 \nonumber
 \Delta {\cal H}^{(eff)}\;=\;-\;\left[\;R_{oblate}(\nu)+
 \mubold\cdot(\efbold\times {\bf {\vec{g}}})\,+\,\frac{1}{2}\,\left(\mubold\times
 \efbold\right)
 \,\cdot\,\left(\mubold\times\efbold\right)\;\right]\;=~~~~~~~~~~~~~~~~~
 ~~~~~~~~~~\\
 \nonumber\\
 \nonumber\\
 \nonumber
 -\;\frac{G\,m\;J_2}{2}\;\frac{\rho^2}{r^3}\;\left[\;1\;-
 \;3\;\sin^2 \inc\;\sin^2(\omega\;+\;\nu) \;\right]\;\;-\;\;
 \sqrt{G\;m\;a\;\left(1\;-\;e^2\right)}\;\;\;{\wbold}\cdot{\mubold}\;-\,\frac{1}{2}\,\left(\mubold\times
 \efbold\right)
 \,\cdot\,\left(\mubold\times\efbold\right)
 \ea
 \ba
 \nonumber\\
 \nonumber
 =\;-\;\frac{G\;m\;J_2}{2}\;\frac{\rho^2}{a^3}\;
 \left(\;\frac{1\;+\;e\;\cos\nu}{1\;-\;e^2}\;\right)^3\;
 \left[\;1\;-\;3\;\sin^2 \inc\;\sin^2(\omega\;+\;\nu)\;\right]\\
 \nonumber\\
 \label{47}\\
 \nonumber
 -\;\sqrt{G\,m\;a\;\left(1\;-\;e^2\right)}\;\;\;\left(\;
 \mu_1\;\sin i\;\sin \Omega\;-\;\mu_2\;\sin i\;\cos \Omega
 \;+\;\mu_3\;\cos i \;\right)\\
 \nonumber\\
 \nonumber\\
 \nonumber
 -\,\;\frac{1}{2}\;\left(\mubold\times
 \efbold\right) \,\cdot\,\left(\mubold\times\efbold\right)
 ~~~~.~~~~~~~~~~~~~~~~~~~~~~~~~~~~~~~~~
 \end{eqnarray}
Otherwise, when all terms on the right-hand side of (\ref{41} -
\ref{46}) are substituted with their orbital averages (denoted
with the $\;\langle\;.\,.\,.\;\rangle\;$ symbol), the Hamiltonian
will be, due to (\ref{18}):
 \begin{eqnarray}
 \nonumber
 \Delta {\cal H}^{(eff)}\;=\;-\;\left[\;\bra R_{oblate} \ket\;+\;
 \mubold\cdot(\efbold\times {\bf {\vec{g}}})\;+\;\frac{1}{2}\;\;{\langle}
 \;\left(\mubold\times\efbold\right)\cdot\left(\mubold\times\efbold\right)\;
 \rangle\;\right]\;~~~~~~~~~~~~~~~~~~~~~~~~~~~~~\\
 \label{48}\\
 \nonumber
 =\;-\;\frac{G\,m\,J_2}{4}\;\;\frac{\rho^2}{a^3}\;\;
 \frac{3\,\cos^2i\,-\,1}{\left(1\,-\,e^2\right)^{3/2}}\,-\,
 \sqrt{G\,m\,a\,\left(1\,-\,e^2\right)}\;\;\left(\,
 \mu_1\;\sin i\;\sin \Omega\,-\,\mu_2\;\sin i\;\cos \Omega
 \,+\,\mu_3\;\cos i \,\right)\\
 \nonumber\\
 \nonumber\\
 \nonumber
 -\;\frac{1}{2}\;\;\langle\;\left(\mubold\times
\efbold\right) \cdot \left(\mubold\times\efbold\right)\;\rangle
 ~~~~.~~~~~~~~~~~~~~~~~~~~~~~~~~~~~~~~~~~~~~~~~~~~~~~~~~~~~~~~~~~~~~~~~~~~~~~~~~~~~
 \end{eqnarray}
 The expression for $\;\efbold\times\bf\vec g\;$ is true-anomaly-independent and,
 therefore, does not need to be bracketed with the averaging symbols
 $\;\langle\;.\,.\,.\;\rangle\;$. The expression for $\;\left(\mubold\times
 \efbold\right)\cdot\left(\mubold\times\efbold\right)\;$ through the orbital
 elements is too cumbersome, and here we do not write it down explicitly. (See
 formula (\ref{A93}) in the Appendix.) When we permit ourselves to neglect the
 $\;O(|\mubold|^2)\;$ inputs, all three Hamiltonians, $\;\Delta {\cal H}^{(osc)}\;$,
 $\;\Delta {\cal H}^{(cont)}\;$, and $\;\Delta {\cal H}^{(eff)}\;$ coincide:
 \ba
 \nonumber
 \Delta
{\cal H}^{(eff)}\;\approx\;\Delta {\cal
H}^{(osc)}\;\approx\;\Delta {\cal H}^{(cont)}\;=\;-\;\left[\;\bra
R_{oblate} \ket\;+\;
 \mubold\cdot(\efbold\times {\bf {\vec{g}}})\;\right]\;~~~~~~~~~~~~~~~~~~~~~~~~~~~\\
 \label{49}\\
 \nonumber\\
 \nonumber
 =\;-\;\frac{G\,m\,J_2}{4}\;\;\frac{\rho^2}{a^3}\;\;
 \frac{3\,\cos^2i\,-\,1}{\left(1\,-\,e^2\right)^{3/2}}\,-\,
 \sqrt{G\,m\,a\,\left(1\,-\,e^2\right)}\;\;\left(\,
 \mu_1\;\sin i\;\sin \Omega\,-\,\mu_2\;\sin i\;\cos \Omega
 \,+\,\mu_3\;\cos i \,\right)\;\;.
 \ea

Two important issues should be dwelt upon at this point.

First, we would remind the reader that the function $\;\Delta
{\cal H}^{(cont)}\;$, given by expression (\ref{24}), yields the
correct functional form of the Hamiltonian only in case we express
the Hamiltonian through the contact elements (and calculate these
through (\ref{25}) or (\ref{28} - \ref{33})). In the currently
considered case of osculating elements, this $\;\Delta {\cal
H}^{(cont)}\;$ is NOT the correct expression for the Hamiltonian.
The correct functional dependence of the Hamiltonian upon the
osculating elements, $\;\Delta {\cal H}^{(osc)}\;$, is given by
formula (\ref{34}). Though in this dynamical problem the
Hamiltonian is unique, its functional dependencies through the
contact and through the osculating elements differ from one
another, one being $\;\Delta {\cal H}^{(cont)}\;$ as in
(\ref{24}), another being $\;\Delta {\cal H}^{(osc)}\;$ as in
(\ref{34}). As for the function $\;\Delta {\cal H}^{(eff)}\;$
rendered by (\ref{39}), it is not really a Hamiltonian, but is
simply a convenient mathematical entity. In the approximation,
where $\;O(|\mubold|^2)\;$ terms are neglected, there is no
difference between these three functions. Despite this, the
$\;O(|\mubold |)\;$ and $\;O(|{\bf{\dot{\mubold}}}|)\;$ terms do
stay in equations (\ref{40} - \ref{46}) for osculating elements.

Second, we would comment on our use of expressions (\ref{16} -
\ref{17}) in our derivation of (\ref{47} - \ref{49}). Above, in
subsections 2.2 and 3.1, the use of (\ref{14}) and (\ref{17}) was
based on formula (\ref{16}) which interconnected
 $\;{\bf\vec J}\;=\;\erbold\;\times\;\pbold\;=\;
 \efbold\;\times\;\left(\doterbold\;+\;\mubold\;\times\;\efbold \right)\;$
 with contact elements $\;a\;$, $\;e\;$, $\;\inc\;$, and $\;\Omega\;$.
 As demonstrated in Efroimsky \& Goldreich (2004), in
 that frame $\;\doterbold\;=\;\gbold\;-\;\mubold\;\times\;\efbold\;$.
 Hence, formula (\ref{16}) interconnects the contact elements with
 $\;{\bf\vec J}\;=\;\efbold\;\times\;\gbold\;$. In the present subsection,
 we use formula (\ref{16}) in its usual
capacity, i.e., to interconnect $\;{\bf\vec
J}\;=\;\erbold\;\times\;\doterbold\;=\;\efbold\;\times\;\gbold\;$
with osculating elements $\;a\;$, $\;e\;$, $\;\inc\;$, and
$\;\Omega\;$. It may seem confusing that, though in both cases
this formula can be written down in the same way,
 \ba
 \efbold\;\times\;\gbold\;=\;\sqrt{G\,m\;a\;\left(1\;-\;e^2\right)}\;\;\;
 {\wbold}\;\;\;,
 \label{50}
 \ea
its meaning is so different. The clue to understanding this
difference lies in the fact that in one case
$\;\doterbold\;=\;\gbold\;-\;\mubold\;\times\;\efbold\;$ (and,
therefore, the elements are contact), while in the other case
$\;\doterbold\;=\;\gbold\;$ (which makes the elements osculating).
For more details see Efroimsky \& Goldreich (2004).

\section{Comparison of the orbital calculations, performed in
terms of the contact elements, with those performed in terms of
the osculating elements. The simplest approximation.}

As explained in Section 2, it follows from equations (\ref{10} -
\ref{11}) that the initially small inclination remains so in the
course of the oblate primary's precession. Whether this famous
result may be interpreted as keeping the satellites in the
near-equatorial zone of a precessing planet will depend upon how
well the non-osculating (contact) inclination emerging in
(\ref{10} - \ref{23}) approximates the physical, osculating,
inclination rendered by (\ref{41} - \ref{46}).

\subsection{The averaged planetary equations}

Comparing equations (\ref{41} - \ref{46}) for osculating elements
with equations (\ref{28} - \ref{33}) for contact elements, we
immediately see that they differ already in the first order over
the precession rate $\;\mubold\;$ and, therefore, the values of
the contact elements will differ from those of their osculating
counterparts in the first order, too. A thorough investigation of
this difference would demand numerical implementation of both
systems and would be extremely time-consuming. Meanwhile, we can
get some preliminary estimates by asking the following,
simplified, question: how will the secular, i.e., averaged over an
orbital period, components of the contact and orbital elements
differ from one another? To answer this question, we perform the
following approximations:\\

{\underline{\bf{1.}}}~~~~ In equations (\ref{41} - \ref{46}), we
substitute both the $\;R_{oblate}\;$ term in the Hamiltonian and
the $\;\mubold$-dependent terms with their averages (so that, for
example, the $\;R_{oblate}\;$ term will be now substituted with
$\;\bra R_{oblate}\ket\;$ expressed via (\ref{18})~).\\

{\underline{\bf{2.}}}~~~~ We neglect the terms of order
$\;{{\mubold\,}^2}\,$. This way, we restrict the length of time
scales involved. (Over sufficiently long times even small terms
may accumulate to a noticeable secular
correction.) However, we can now benefit from the approximate equality
 (\ref{49}).\\


So truncated and averaged system of Lagrange-type equations will
read:
  ~\\
 \ba
 \frac{da}{dt}\,=\,
 \frac{2}{n a}\,\left[
 \;-\;\langle \;\dotmubold\; \left(\efbold\,\times\,
 \frac{\partial \efbold}{\partial M_o}  \right)\; \rangle\;\right]
 \;\;\;\;\;~~~~~~~~~~~~~~~~~~~~~~~~~~~~~~~~~~~~~~~~~~~~~~~~~~~~~~~~
 \label{51}
  \ea
  ~\\
 ~\\
  \begin{eqnarray}
  \nonumber
 \frac{de}{dt}\,=\,
 \frac{1-e^2}{n\,a^2\,e}\;\;\left[\;
 \;-\;\langle \;\dotmubold\; \left(\efbold\,\times\,
 \frac{\partial \efbold}{\partial M_o}  \right)\; \rangle \;\right]~~
 \,~~~~~~~~~~~~~~~~~~~~~~~~~~~~~~~~~~~~~~~~~~~~~~~~~~~
 \\
 \nonumber\\
 \label{52}
 \\
 \nonumber\\
 \nonumber
 -\;\frac{(1\,-\,e^2)^{1/2}}{n\,a^2\,e} \;\left[\;
 \langle~ \mubold\cdot
 \left(\frac{\partial{\efbold}}{\partial \omega}\times
 {\bf{\vec g}}\,-\,{\efbold}\times
 \frac{\partial{\vec{\bf g}}}{\partial \omega}\right)~\rangle
 \;-\;\langle \; \dotmubold\; \left(\efbold\,\times\,
 \frac{\partial \efbold}{\partial \omega}  \right)\;
 \rangle\;\right]~~,~~~~~~~~
  \end{eqnarray}
 ~\\
 ~\\
  \begin{eqnarray}
 \nonumber
 \frac{d\omega}{dt}\,=~
 \frac{\;-\,\cos \inc
}{n a^2(1-e^2)^{1/2}\, \sin \inc }\left[ \frac{\partial
\left(\,-\,\Delta {\cal H}^{(eff)} \right) }{\partial \inc } \,~ +
~ \langle~\mubold\cdot
 \left(\frac{\partial{\efbold}}{\partial \inc}\times
 {\bf{\vec g}}\,-\,{\efbold}\times
 \frac{\partial{\vec{\bf g}}}{\partial \inc}\right)\,~\rangle
 \;-\;\langle \; \dotmubold\; \left(\efbold\,\times\,
 \frac{\partial \efbold}{\partial \inc}  \right)\;
 \rangle\;
  \right]~~~~
  \ea
  \ba
 \nonumber\\
 \nonumber\\
 \label{53}\\
 \nonumber
 +\,~\frac{(1-e^2)^{1/2}}{n\,a^2\,e}\;
 \left[\;\frac{\partial \left(\,-\,\Delta {\cal H}^{(eff)} \right) }{\partial e}\,
+ ~ \langle~\mubold\cdot
 \left(\frac{\partial{\efbold}}{\partial e}\times
 {\bf{\vec g}}\,-\,{\efbold}\times
 \frac{\partial{\vec{\bf g}}}{\partial e}\right)~\rangle
 \;-\;\langle \; \dotmubold\; \left(\efbold\,\times\,
 \frac{\partial \efbold}{\partial e}  \right)\;
 \rangle\;
 \;\right]\;\;,~~~~~~
  \end{eqnarray}
 ~\\
 ~\\
 ~\\
 \ba
 \nonumber
 \frac{d \inc
 }{dt}\;=~ \frac{\cos \inc}{n a^2\,(1 - e^2)^{1/2} \sin
 \inc}
 \left[
 \; \langle~\mubold\cdot
 \left(\frac{\partial{\efbold}}{\partial \omega}\times
 {\bf{\vec g}}\,-\,{\efbold}\times
 \frac{\partial{\vec{\bf g}}}{\partial \omega}\right)~\rangle
 \;-\;\langle \; \dotmubold\; \left(\efbold\,\times\,
 \frac{\partial \efbold}{\partial \omega}  \right)\;
 \rangle\;
 \right]\,-\;\;\;
 ~~~~~~~~~~~~~~~\\
 \nonumber\\
 \label{54}\\
 \nonumber\\
 \nonumber
 \frac{1}{na^2\,(1-e^2)^{1/2}\,\sin \inc
 }\,\left[\,\frac{\partial \left(\,-\,\Delta {\cal H}^{(eff)} \right) }{\partial
 \Omega}\;+ \; \langle~\mubold\cdot
 \left(\frac{\partial{\efbold}}{\partial \Omega}\times
 {\bf{\vec g}}\,-\,{\efbold}\times
 \frac{\partial{\vec{\bf g}}}{\partial \Omega}\right)~\rangle
 \;-\;\langle \; \dotmubold\; \left(\efbold\,\times\,
 \frac{\partial \efbold}{\partial \Omega}  \right)\;
 \rangle\;
 \,\right]\;{,}\;\;\;
  \ea
 ~\\
 ~\\
  \ba
\frac{d\Omega}{dt} = \frac{1}{n a^2\,(1-e^2)^{1/2}\,\sin \inc
}\,\left[\,\frac{\partial \left(\,-\,\Delta {\cal H}^{(eff)}
\,\right) }{\partial \inc }\, + \,\langle \,\mubold\cdot
 \left(\frac{\partial{\efbold}}{\partial \inc}\times
 {\bf{\vec g}}\,-\,{\efbold}\times
 \frac{\partial{\vec{\bf g}}}{\partial \inc}\right)\,\rangle
 \,-\,\langle \, \dotmubold\, \left(\efbold\,\times\,
 \frac{\partial \efbold}{\partial \inc}  \right)\,
 \rangle\,
 \right]\,,\;\;
 \label{55}
  \ea
 ~\\
 ~\\
  \begin{eqnarray}
  \nonumber
\frac{dM_o}{dt}\,=\,\;-\,\;\frac{1\,-\,e^2}{n\,a^2\,e}\,\;
\left[\;\frac{\partial \left(\,-\,\Delta {\cal H}^{(eff)} \right)
 }{\partial e }~ + ~
 \langle~\mubold\cdot
 \left(\frac{\partial{\efbold}}{\partial e}\times
 {\bf{\vec g}}\,-\,{\efbold}\times
 \frac{\partial{\vec{\bf g}}}{\partial e}\right)~\rangle
 \;-\;\langle \; \dotmubold\; \left(\efbold\,\times\,
 \frac{\partial \efbold}{\partial e}  \right)\;
 \rangle
 \;\right]\\
 \nonumber\\
 \label{56}\\
 \nonumber\\
 \nonumber
 -\;~\frac{2}{n\,a}\,\left[\;\frac{\partial \left(\,-\,\Delta {\cal H}^{(eff)} \right)
 }{\partial a }~ + ~ \langle ~\mubold\cdot
 \left(\frac{\partial{\efbold}}{\partial a}\times
 {\bf{\vec g}}\,-\,{\efbold}\times
 \frac{\partial{\vec{\bf g}}}{\partial a}\right)~\rangle
 \;-\;\langle \; \dotmubold\; \left(\efbold\,\times\,
 \frac{\partial \efbold}{\partial a}  \right)\;
 \rangle\;
 \right]
 \;\;,~~~~~
  \end{eqnarray}
  ~\\
the Hamiltonian here being approximated by (\ref{49}), and the
angular brackets signifying orbital averaging. In equations
(\ref{51}), (\ref{52}) and (\ref{54}) we took into account that
the averaged Hamiltonian (\ref{49}) bears no dependence upon
$\;M_o\;$ and $\;\omega\;$. (This, though, will not be the case
for the exact, $\;\nu$-dependent, Hamiltonian given by (\ref{34})
and (\ref{13}) ! )

Calculation of $\;\left(\;(\partial \efbold/\partial C_j)\times
{\bf\vec g}\,-\,(\partial {\bf \vec g}/\partial
C_j)\times\efbold\;\right)\;$ and $\;-\;\dotmubold\;\left(\;
\efbold\,\times\,{\partial \bf\vec g}/{\partial C_j} \;\right)\;$
takes pages of algebra. A short synopsis of this calculation is
offered in the Appendix. Here follows the outcome:
  ~\\
 \ba
 \mubold\;\cdot\;\left(\;\frac{\partial \efbold}{\partial a}\;\times\;{\bf \vec
g}\;-\;\efbold\;\times\;\frac{\partial {\bf \vec g}}{\partial
a}\;\right)\;=\;\frac{3}{2}\;\;
\mu_{\perp}\;\;\sqrt{\frac{G\;m\;\left(1\;-\;e^2\right)}{a}}\;\;,~~~~~~~~~~~~~~~~~~~~
 \label{57}
 \label{475}
 \ea
~\\
  \ba
 \mubold\;\cdot\;\left(\; \frac{\partial \efbold}{\partial e}\;\times\;{\bf \vec
g}\;-\;\efbold\;\times\;\frac{\partial {\bf \vec g}}{\partial
e}\;\right)\;=\;
 -\;\,\mu_{\perp}\;\,\frac{n\,a^2\;\left(3\,e\,+\,2\,\cos \nu\,+\,e^2\,\cos \nu  \right)}{\left(1\,+\,e\,\cos
 \nu\right)\;\sqrt{1\;-\;e^2}}
 ~~~,~~
 \label{58}
 \label{476}
 \ea
~\\
 \ba
 \mubold\;\cdot\;\left(\; \frac{\partial \efbold}{\partial \omega}\;\times\;{\bf
\vec g}\;-\;\efbold\;\times\;\frac{\partial {\bf \vec g}}{\partial
\omega}\;
\right)\;=\;-\;2\;\;\mu_{\perp}\;\;\;\frac{n\;a^2\;\sqrt{1\;-\;e^2}}{1\;+\;e\;\cos
\nu}\;e\;\;\sin \nu \;\;\;,\;\;\;\;\;\;\;\;\;\;\;
 \label{59}
 \label{477}
 \ea
~\\
 \ba
 \nonumber
 \mubold\;\cdot\;\left(\; \frac{\partial \efbold}{\partial \Omega}\;\times\;{\bf
\vec g}\;-\;\efbold\;\times\;\frac{\partial {\bf \vec g}}{\partial
\Omega}\;
\right)\;=~~~~~~~~~~~~~~~~~~~~~~~~~~~~~~~~~~~~~~~~~~~~~~~~~~~~~~~~~~~~~~~~~~~~
~~~~~~~~~~~~~~~\\
 \label{60}
 \label{478}
 \ea
 \ba
 \nonumber
 \mu_1\;\;\left[\;\;
 \frac{n\;a^2\;\sqrt{1\;-\;e^2}}{1\;+\;e\;\cos\nu}\;\;\sin
 \inc\;\;\left\{\;\;
 \cos \Omega\;\cos\left[\,2\,\left(\omega + \nu \right)\,\right]
 \;-\;\sin \Omega\;\cos \inc\;\sin \left[\,2\,\left(\omega + \nu
 \right)\,\right]\;\;
 \right\}~+~~~~\,~~~~~~~~~~~ \right.\\
 \nonumber\\
 \nonumber\\
 \nonumber  \left.
 \frac{n\;a^2\;\sqrt{1\;-\;e^2}}{1\;+\;e\;\cos\nu}\;\;e\;\;\sin
 \inc\;\;\left\{\;\;
 \cos \Omega\;\cos\left(\nu \,+ \,2\,\omega \right)
 \;-\;2\;\sin \Omega\;\cos \inc\;\sin \left(\omega + \nu
 \right)\;\cos \omega\;\;
 \right\}
 \;\;\right]\;+~~~~~~\\
 \nonumber\\
 \nonumber\\
 \nonumber\\
 \nonumber
 \mu_2\;\;\left[\;\;
 \frac{n\;a^2\;\sqrt{1\;-\;e^2}}{1\;+\;e\;\cos\nu}\;\;\sin
 \inc\;\;\left\{\;\;
 \sin \Omega\;\cos\left[\,2\,\left(\omega + \nu \right)\,\right]
 \;+\;\cos \Omega\;\cos \inc\;\sin \left[\,2\,\left(\omega + \nu
 \right)\,\right]\;\;
 \right\}~+\;\;\;\;\;\;\;\;\;~~~~\,~~~~ \right. \\
 \nonumber\\
 \nonumber\\
 \nonumber  \left.
 \frac{n\;a^2\;\sqrt{1\;-\;e^2}}{1\;+\;e\;\cos\nu}\;\;e\;\;\sin
 \inc\;\;\left\{\;\;
 \sin \Omega\;\cos\left(\nu \,+ \,2\,\omega \right)
 \;+\;2\;\cos \Omega\;\cos \inc\;\sin \left(\omega + \nu
 \right)\;\cos \omega\;\;
 \right\}
 \;\;\right]\;+\\
 \nonumber\\
 \nonumber\\
 \nonumber
 \mu_3\;\left[\;\;\frac{n a^2 \sqrt{1 - e^2}}{1 + e\;\cos\nu}\;\sin^2
 \inc\;\sin \left[\,2\,\left( \omega\,+\,\nu \right) \,\right]
 \,+\,\frac{n a^2 \sqrt{1 - e^2}}{1 + e\;\cos\nu}\;\;2\;e
 \;\left\{\;\;
 -\,\sin \nu \,+ \,\sin^2\inc\;\cos \omega \;\sin \left(\omega + \nu \right)\;
 \; \right\}\;\right]\;\;,
 \ea
~\\
~\\
 \ba
 \nonumber
 \mubold\;\cdot\;\left(\; \frac{\partial \efbold}{\partial \inc}\;\times\;{\bf
 \vec g}\;-\;\efbold\;\times\;\frac{\partial {\bf \vec g}}{\partial
 \inc}\; \right)\;=~~~~~~~~~~~~~~~~~~~~~~~~~~~~~~~~~~~~~~~~~~~~~~~~~~~~~~~~~
 ~~~~~~~~~~~\\
 \label{61}
 \label{479}
 \ea
 \ba
 \nonumber
 \mu_1\;\;\;
 \frac{n a^2 \sqrt{1- e^2}}{1 + e\;\cos\nu}\;\left\{\;
 \left[\;
 \left(\;-\;\sin\Omega\;\cos\inc\;\cos2\omega\;-\;\cos\Omega\;\sin 2\omega\;\right)
 \;\left(\;\cos^2 \nu\;-\;\sin^2 \nu\;\right)\;+~~~~~~~~~~~~~~~~~~~~~~~~~~~~\right.\right.\\
 \nonumber\\
 \nonumber\\
 \nonumber
 \left. 2\;\sin \nu\;\cos\nu\;
 \left(\;
 \sin \Omega\;\cos\inc\;\sin 2\omega\;-\;\cos\Omega\;\cos 2\omega
 \;\right)
 \;\right]\;+ ~~~~~~~~~~~~ \\
 \nonumber\\
 \nonumber\\
 \nonumber
 \left. e\;\,\left[\;
 \left(\;
 -\;\sin \Omega\;\cos\inc\;\cos 2\omega \,-\,\cos\Omega\;\sin
 2\omega\;\right)\;\cos\nu\;
 +\;\left(\;
 \sin\Omega \;\cos\inc\;\sin 2\omega\;-\;2\;\cos \Omega\;\cos^2 \omega \right)\;
 \sin\nu\;\right]\;\right\} ~~~~~~~~~~~~ \\
 \nonumber\\
 \nonumber\\
 \nonumber
 +\;\mu_2\;\;\;
 \frac{n a^2 \sqrt{1- e^2}}{1 + e\;\cos\nu}\;\left\{\;
 \left[\;
 \left(\;\cos\Omega\;\cos\inc\;\cos2\omega\;-\;\sin\Omega\;\sin 2\omega\;\right)
 \;\left(\;\cos^2 \nu\;-\;\sin^2 \nu\;\right)\;+~~~~~~~~~~~~~~~~~~
 ~~~~~~~~\right.\right.\\
 \nonumber\\
 \nonumber\\
 \nonumber
 \left. 2\;\sin \nu\;\cos\nu\;
 \left(\;-\;
 \cos \Omega\;\cos\inc\;\sin 2\omega\;-\;\sin\Omega\;\cos 2\omega
 \;\right)
 \;\right]\;+  ~~~~~~~~~~~~~~~~~~~~~~~~~~~\\
 \nonumber\\
 \nonumber\\
 \nonumber
 \left. e\;\,\left[\;
 \left(\;\cos \Omega\;\cos\inc\;\cos
 2\omega\,-\,\sin\Omega\;\sin2\omega\;\right)\;\cos\nu\,+\,
 \left(\;
-\;2\;\sin\Omega\;\cos^2\omega\,+\,\cos\Omega\;\sin2\omega\;\cos\inc
 \;\right)\;\sin\nu
 \;\right]\;\right\} ~~~~~~~~~~~~~~~\\
 \nonumber\\
 \nonumber\\
 \nonumber
 +\;\mu_3\;\;\;
  \frac{n a^2 \sqrt{1- e^2}}{1 + e\;\cos\nu}\;\left\{\;
 \sin\inc\;\left[\;
 \cos 2\omega \;\left(\;\cos^2\nu\;-\;\sin^2\nu\;\right)\;-\;2\;\sin
 2\omega\;\sin\nu\;\cos\nu
 \;\right]
 \;+~~~~~~~~~~~~~~~~~~~~~~~~~~~
  \right. \\
  \nonumber\\
  \nonumber\\
  \nonumber
  \left.
 e\;\;\left[\; \sin\inc\;\cos 2\omega\;\cos
 \nu\;-\;\sin\inc\;\sin 2\omega\;\sin
 \nu\;\right]\;\right\}  \;\;,~~~~~~~~~~~~~~~~~~~~~~~
 \ea
~\\
 \ba
\mubold\;\cdot\;\left(\; \frac{\partial \efbold}{\partial
M_o}\;\times\;{\bf \vec g}\;-\;\efbold\;\times\;\frac{\partial
{\bf \vec g}}{\partial
M_o}\;\right)\;=\;0\;\;\;,~~~~~~~~~~~~~~~~~~~~~~~~~~~~~~~~~~~~~~~~~~~~
 \label{62}
 \label{480}
 \ea
 ~\\
 \ba
 -\;\dotmubold\;\cdot\;\left(\;\efbold\;\times\;\frac{\partial \efbold}{\partial
 a} \;\right)\;=\;\frac{3}{2}\,\;\dot{\mu}_{\perp}\;\;a\;n\;\left(\;t\;-\;t_o\;\right)\;\sqrt{1\;-\;e^{2}}
 \;\;\;~,~~~~~~~~~~~~~~~~~~~~~~
 \label{481}
 \ea
 ~\\
 \ba
-\;\dotmubold\;\cdot\;\left(\;\efbold\;\times\;\frac{\partial
\efbold}{\partial
 e} \;\right)\;=\;-\;\dot{\mu}_{\perp}\;\;a^2
 \;\;\frac{\left(1-e^{2}\right)}{1+
 e\;\cos \nu}\;\sin \nu\;\;\;.\;\;\;\;
 \;\;\;.\;~~~~~~~~~~~~~~~~~~~~~~~~~~~
 \label{482}
 \ea
  ~\\
 \ba
-\;\dotmubold\;\cdot\;\left(\;\efbold\;\times\;\frac{\partial
\efbold}{\partial
 \omega} \;\right)\;=\;-\;\dot{\mu}_{\perp}\;a^2\;\;\frac{\left(1\;-\;e^2\right)^2}{\left(1\;+\;e\;\cos
\nu\right)^2}\;\;\;\;,~~~~~~~~~~~~~~~~~~~~~~~~~~~~~~~~~~~~
 \label{483}
 \ea
  ~\\
 \ba
 \nonumber
-\;\mubold\;\cdot\;\left(\;\efbold\;\times\;\frac{\partial
\efbold}{\partial
 \Omega} \;\right)\;=
 ~~~~~~~~~~~~~~~~~~~~~~~~~~~~~~~~~~~~~~~~~~~~~~~~~~~~~~~~~~~~~~
 ~~~~~~~~~~~~~~~~~~\\
 \label{484}\\
 \nonumber\\
 \nonumber
 a^2\;\;
 \frac{\left(1\;-\;e^2\right)^2}{\left(1\;+\;e\;\cos
\nu\right)^2}\;\;\left\{\;\;\dot{\mu}_{1}\;
 \;\left[\;
           \cos\Omega\;\cos(\omega +\nu)\;-\;
           \sin \Omega\;\sin(\omega +\nu)\;\cos \inc
  \;\right]\;\sin(\omega + \nu)\;\sin \inc  \right.\\
  \nonumber\\
  \nonumber\\
  \nonumber
+\;\dot{\mu}_2\;
  \left[\;\sin\Omega\;\cos(\omega +\nu)\;+\;
                \cos \Omega\;\sin(\omega +\nu)\;\cos \inc
                   \;\right]\;\sin (\omega + \nu)\;\sin \inc ~~\\
 \nonumber\\
 \nonumber\\
 \nonumber
 \left.
+\;\dot{\mu}_3\;\left[\,\cos^2(\omega + \nu)\,+\,\sin^2(\omega +
 \nu)\;\cos^2\inc \,\right]
 \;\;\right\}~~~~~~~~~~~~~~~~~~~~~~~~~~~~~~~~~
 \ea
  ~\\
 \ba
 \nonumber
-\;\mubold\;\cdot\;\left(\;\efbold\;\times\;\frac{\partial
\efbold}{\partial
 \inc} \;\right)\;=\;
 ~~~~~~~~~~~~~~~~~~~~~~~~~~~~~~~~~~~~~~~~~~~~~~~~~~~~~~~~~~~~~~
 ~~~~~~~~~~~~~~~~~~\\
 \label{485}\\
 \nonumber\\
 \nonumber
 a^2\;\;
 \frac{\left(1\;-\;e^2\right)^2}{\left(1\;+\;e\;\cos
\nu\right)^2}\;\;\left\{\;\;\dot{\mu}_{1}\;
 \;\left[\;
           -\;\cos\Omega\;\sin(\omega +\nu)\;-\;
           \sin \Omega\;\cos(\omega +\nu)\;\cos \inc
  \;\right]\;\sin(\omega + \nu)\;  \right.\\
  \nonumber\\
  \nonumber\\
  \nonumber
+\;\dot{\mu}_2\;
  \left[\;-\;\sin\Omega\;\sin(\omega +\nu)\;+\;
                \cos \Omega\;\cos(\omega +\nu)\;\cos \inc
                   \;\right]\;\sin (\omega + \nu)\;~~\\
 \nonumber\\
 \nonumber\\
 \nonumber
 \left.
+\;\dot{\mu}_3\;\sin(\omega + \nu)\;\cos(\omega + \nu)\;\sin \inc
 \;\;\right\}~~~~~~~~~~~~~~~~~~~~~~~~~~~~~~~~~
 \ea
  ~\\
 \ba
-\;\mubold\;\cdot\;\left(\;\efbold\;\times\;\frac{\partial
\efbold}{\partial
 M_o}
 \;\right)\;=\;-\;\dot{\mu}_{\perp}\;a^2\;\sqrt{1\,-\,e^2}\;\;\;\;,
 ~~~~~~~~~~~~~~~~~~~~~~~~~~~~~~~~~~~~~~~~
 \label{486}
 \ea
 $\mu_1\,,\;\mu_2\,$, and $\,\mu_3\;$ being the Cartesian components of
 the precession rate, as seen in the co-precessing frame, and
$\mu_{\perp}\;$ being the component of the precession rate, aimed
in the direction of the angular momentum; it is given by
(\ref{A33}).

It may seem strange that the right-hand side of (\ref{58}) does
not vanish in the limit of $\;e\;\rightarrow\;0\;\;$. The
absurdity of this will be easily redeemed by the fact that this
term shows up only in the equation for $\;d\omega/dt\;$ and,
therefore, leads to no physical paradoxes in the limit of a
circular orbit. However, for finite values of the eccentricity,
this term contributes to the periapse precession.

Another seemingly calamitous thing is the divergence in
(\ref{482}). This divergence, however, entails no disastrous
physical consequences, because the term (\ref{482}) shows up only
in the planetary equation for $\;dM_o/dt\;$ and simply leads to a
steady shift of the initial condition $\;M_o\;$.

\subsection{The case of a constant precession rate.}

The situation might simplify very considerably if we could also
assume that the precession rate $\;\mubold\;$ stays constant. Then
in equations (\ref{51} - \ref{56}), we would take $\;\mubold\;$
out of the angular brackets and proceed with averaging the
expressions $\;\left(\;(\partial \efbold/\partial C_j)\times
{\bf\vec g}\,-\,\efbold\times(\partial {\bf \vec g}/\partial
C_j)\;\right)\;$ only (while all the terms with $\;\dotmubold\;$
will now vanish). It is, of course, well known that this is
physically wrong, because the planetary precession has a
continuous spectrum of frequencies (some of which are
 commensurate with the orbital frequency of the
 satellite).\footnote{~The case of the Earth rotation and precession is
comprehensively reviewed by Eubanks (1993). The Martian
short-time-scale rotational dynamics is of an equal complexity,
even though Mars lacks oceans and the coupling of its rotation
with the atmospheric motions is weaker than in the case of the
Earth (Defraigne et al 2003, Van Hoolst et al 2000, Dehant et al
2000).} Nevertheless, for the sake of argument let us go on with
this assumption.

The averaging of (\ref{57}) and (\ref{62}) is self-evident. The
averaging of (\ref{58} - \ref{61}) is lengthy and is presented in
the Appendix. All in all, we get, for constant $\;\mubold\;$:
  ~\\
 \ba
 \mubold\,\cdot\,\langle\;\left(\;\frac{\partial \efbold}{\partial a}\,\times\,{\bf \vec
g}\,-\,\efbold\,\times\,\frac{\partial {\bf \vec g}}{\partial
a}\;\right)\;\rangle\,=\,\mubold\,\cdot\,\left(\;\frac{\partial
\efbold}{\partial a}\,\times\,{\bf \vec
g}\,-\,\efbold\,\times\,\frac{\partial {\bf \vec g}}{\partial
a}\;\right) \,=\,\frac{3}{2}\;\;
\mu_{\perp}\;\;\sqrt{\frac{G\;m\;\left(1\;-\;e^2\right)}{a}}\;\;\;,~~
 \label{63}
 \label{487}
 \ea
~\\
  \ba
 \mubold\;\cdot\;\langle\;\left(\; \frac{\partial \efbold}{\partial C_j}\;\times\;{\bf \vec
g}\;-\;\efbold\;\times\;\frac{\partial {\bf \vec g}}{\partial
C_j}\;\right)\;\rangle\;=\;0\;\;\;~,~~~~~~~~~C_j\;=\;e\,,\;\Omega\,,\;\omega\,,\;\inc\,,\;M_o\;\;\;.
~~~~~~~~~~~~~~~~~~~~
 \label{64}
 \label{488}
 \ea
Since the orbital averages (\ref{64}) vanish, then $\;e\;$ will,
along with $\;a\;$, stay constant for as long as our approximation
remains valid. Besides, no trace of $\;\mubold\;$ will be left in
the equations for $\;\Omega\;$ and $\;\inc\;$. This means that, in
the assumed approximation and under the extra assumption of
constant $\;\mubold\;$, the afore-quoted analysis (\ref{10} -
\ref{23}), offered by Goldreich (1965), will remain valid at time
scales which are not too long. At longer scales (of order dozens
of millions of years and longer) one has to take into
consideration the back reaction of the short-period terms upon the
secular ones (Laskar 1990). Beside the latter issue, the problem
with this approximation is that it ignores both the long-term
evolution of the spin axis and the short-term nutations. For these
reasons, this approximation will not be extendable to long periods
of Mars' evolution. This puts forward the bigger question: what
maximal amplitude of obliquity variations could Mars afford, to
keep both its satellites so close to its equatorial plane?


Even in the unphysical case of constant $\;\mubold\;$ the averaged
equation (\ref{56}) for the osculating $\;M_o\;$ differs already
in the first order over $\;\mubold\;$ from equation (\ref{30}) for
the contact $\;M_o\;$. In the realistic case of time-dependent
precession, the averages of terms containing $\;\mubold\;$ do not
vanish (except for $\;\mubold\cdot\left(\;(\partial
\efbold/\partial M_o)\times {\bf\vec g}\,-\,\efbold\times(\partial
{\bf \vec g}/\partial M_o)\;\right)\;$ which is identically nil).
These terms show up in all equations (except in that for $\;a\;$)
and influence the motion. They will be the key to our
understanding the long-term satellite dynamics, including the
secular drift of the orbit plane, caused by the precession
$\;\mubold$.

 \section{An outline of a more accurate analysis.  Resonances
          between the planetary nutation
          and the satellite orbital frequency.}

Precession of any planet contains in itself a continuous spectrum
of circular frequencies involved:\footnote{~A more honest analysis
should take into account also the direct dependence of the
planet's precession rate upon the instantaneous position(s) of its
satellite(s): $\;\mubold\;=\;\mubold
(t\,;\;a\,,\,e\,,\,\Omega\,,\,\omega\,,\,\inc\,,\,M_o)\;$.
However, the back-reaction of the satellites upon the primary is
known to be an effect of a higher order of smallness (Laskar
2004), at least in the case of Mars; and therefore we shall omit
this circumstance by simply assuming that
$\;\mubold\;=\;\mubold(t)\;=\;\mubold\left(\,t(\nu)\,\right)$.}
 \ba
 \mubold(t)\;=\;\int_{-\infty}^{\infty}
 \;\mubold(s)\;e^{-ist}\;ds\;\;\;,
 \label{65}
 \label{489}
 \ea
some modes being more prominent than the others. For our present
purposes, it will be advantageous to express the precession rate
as function of the satellite's true anomaly:
 \ba
 \mubold(\nu)\;=\;\int_{-\infty}^{\infty}
 \;\mubold(W)\;e^{-iW\nu}\;dW\;\;\;,
 \label{490}
 \ea
 $W\,$ being the circular ``frequency" related to the true
 anomaly $\;\nu\;$.  Needless to say,
 $\;\mubold(t)\,\;\mubold(\nu)\,,\;\mubold(s)\,,\;$ and
 $\;\mubold(W)\;$ are four different functions. However, we take
the liberty of using the same notation $\;\mubold(\,.\,.\,.\,)\;$
because the argument will always reveal which of the four
functions we imply.\footnote{~Evidently, $\;\mubold(\nu)\;$ is a
short notation for $\;\mubold(\,t(\nu)\,)\;$. It is also possible
 to demonstrate that, in the limit of vanishing eccentricity,
 $\;\mubold(W)\;\approx\;n\,\mubold(W\,n)\;$ and $\;W\;\approx\;s/n\;$.}
 As ever, it is understood that the physical content is attributed
 to the real parts of $\;\mubold(t)\;$ and $\;\mubold(\nu)\,$.

 If we now plug the real part of (\ref{490}) into (\ref{475} -
 \ref{487}) and carry out averaging in accordance with formula
 (\ref{A14}) of the Appendix, we shall see that the secular parts
 of these $\;\mubold$-dependent terms do not vanish. They reveal
 the influence of the planetary precession upon the satellite
 orbital motion. Especially interesting are the resonant contributions
 provided by the integer $\;W$'s, because these nutation modes are
 commensurate with the orbital motion of the satellite.

For plugging $\;\mubold(\nu)\;$ into the terms (\ref{475} -
\ref{486}), it will be convenient to rewrite (\ref{490}) as
 \ba
 \mubold(\nu)\;=\;\int_{0}^{\infty}\;\left[
 \;\mubold^{(s)}(W)\;\sin({W\nu})\;+\;
 \mubold^{(c)}(W)\;\cos({W\nu})\;\right]\;dW\;\;\;.
 \label{491}
 \ea
Insertion of the latter into (\ref{475} - \ref{486}) will still
result in extremely sophisticated integrals. For example, the term
(\ref{475}) will have the following orbital average:
  \ba
  \nonumber
 \langle\;\;\mubold\;\cdot\;\left(\;\frac{\partial \efbold}{\partial a}\;
 \times\;{\bf \vec g}\;-\;\efbold\;\times\;
 \frac{\partial {\bf \vec g}}{\partial a}\;\right)\;\;\rangle\;=\;
 \frac{3}{2}\;\sqrt{\frac{G\,m\;\left(1\,-\,e^2\right)}{a}} \;\;
 \langle\;\mu_{\perp}\rangle\;=\\
 \label{492}\\
 \nonumber\\
 \nonumber
 \frac{3}{4\pi}\;\sqrt{\frac{G\,m}{a}}
 \;\;\;\left(1\,-\,e^2\right)^2\;\;\int_0^{\infty}dW\;\int_0^{2\pi}d\nu\;\;
  \frac{\mu^{(s)}_{\perp}(W)\;\sin({W\nu})\;+\;
 \mu^{(c)}_{\perp}(W)\;\cos({W\nu})}{\left(1\,+\,e\;\cos\nu  \right)^2}
  \ea
(the averaging rule (\ref{A14}) being employed). Evaluation of the
two integrals emerging in this expression can, in principle, be
carried out in term of the hypergeometric functions, but the
outcome will be hard to work with and hard to interpret
physically. Even worse integrals will show up in the averages of
(\ref{476} - \ref{486}).

To get an idea as to how much the terms (\ref{475} - \ref{486})
contribute to the secular drift of the satellite orbits from the
planet's equator, it may be good to first calculate these term's
averages under the assumption of small eccentricity. Then, for
example, (\ref{492}) will simplify to
  \ba
  \nonumber
 \langle\;\;\mubold\;\cdot\;\left(\;\frac{\partial \efbold}{\partial a}\;
 \times\;{\bf \vec g}\;-\;\efbold\;\times\;
 \frac{\partial {\bf \vec g}}{\partial
 a}\;\right)\;\;\rangle\;\approx\\
 \label{493}
 \ea
 \ba
 \nonumber
 \frac{3}{4\pi}\;\sqrt{\frac{G\,m}{a}}
 \;\;\;\left(1\,-\,e^2\right)^2\;\;\int_0^{\infty}dW\;\int_0^{2\pi}d\nu\;\;
 \left\{\;  \mu^{(s)}_{\perp}(W)\;\sin({W\nu})\;\;
  \left(\;1\;-\;2\;e\;\cos\nu\;+\;3\;e^2\;\cos^2\nu\;+\;.\;.\;.\;\;\right)
  \right. \\
 \nonumber\\
 \nonumber\\
 \nonumber
  \left.+\;
  \mu^{(c)}_{\perp}(W)\;\cos({W\nu})\;
  \left(\;1\;-\;2\;e\;\cos\nu\;+\;3\;e^2\;\cos^2\nu\;+\;.\;.\;.\;\;\right)
 \;\right\}
   \ea
 We immediately see that the above expression consists of two
 parts: the non-resonant one and the resonant one.

 This topic will be addressed in our
 subsequent paper, where we shall try to determine how much time is
 needed for these subtle effects to accumulate enough to cause a
 substantial secular drift of the orbit plane.

\section{Conclusions}

In this article we have prepared an analytical launching pad for
the research of long-term evolution of orbits about a precessing
oblate primary. This paper is the first in a series and is
technical, so we deliberately avoided making whatever quantitative
estimates, leaving those for the next part of our project.

The pivotal question emerging in the context of this research is
whether the orbital planes of near-equatorial satellites will
drift away from the planetary equator in the cause of the planet's
obliquity changes. Several facts have been established in this
regard.

First, the planetary equations for osculating elements of the
satellite do contain terms responsible for such a drift. These
terms contain inputs of the first order and of the second order in
$\;\mubold\;$, and of the first order in $\;\dotmubold\;$, where
$\;\mubold\;$ is the precession rate of the primary.

Second, the first-order (but not the second-order) terms average
out in the case of a constant precession rate, which means that in
this case their effect will accumulate only over extremely long
time scales. We would remind that the sort-period terms of the
planetary equations do exert back-reaction upon the secular ones.
While in the artificial-satellite science, which deals with short
interval of time, the short-period terms often may be omitted, in
long-term astronomical computations (dozens of million years and
higher), the accumulated influence of short-period terms must be
taken into account. A simple explanation of how this should be
done is offered in section 2 of Laskar (1990). Besides, the
contribution of the secular second-order terms will, too,
accumulate over very large time intervals.

Third, the first-order drift terms do \textbf{not} average out in
the case of variable precession. Under these circumstances they
become secular. This means, for example, that the turbulent
history of Mars' obliquity -- history which includes both
long-term changes (Ward 1973, 1974, 1979; Laskar \& Robutel 1993;
Touma \& Wisdom 1994) and short-term nutations (Dehant et al 2000;
Van Hoolst et al 2000; Defraigne et al 2004) -- might have lead to
a secular drift of the initially near-equatorial satellites. If
that were the case, then the current, still near-equatorial,
location of Phobos and Deimos may lead to restrictions upon the
rate and amplitude of the Martian obliquity variations. To render
a judgment on this topic, one should compute how quickly this
drift accumulates. Very likely, this quest will demand heavy-duty
numerics.

~\\

~\\

{\underline{\bf{\Large{Acknowledgments}}}}\\
~\\
I owe my sincere gratitude to
Yurii Batrakov, Peter Goldreich, Pini Gurfil, George Kaplan,
Konstantin Kholshevnikov, Marc Murison, Alessandro Morbidelli,
William Newman, Victor Slabinski, and Jack Wisdom, all of whom
kindly discussed this material with me (and are in no way to blame
for my errors of omission or commission, for which I alone am
responsible). My great thanks are due to my Department Head, John
Bangert, who tolerated the project since its earliest stage, when
its success was yet far from evident.
%
%
This research was supported by NASA grant W-19948.
\\

\pagebreak

 \noindent
{\underline{\bf{\Large{Appendix.}}}} \\
 ~\\
  \noindent
 {\underline{\it{\Large{\bf{The inertial terms emerging in the
 planetary equations\\}}}}}
  ~\\

In this Appendix we write down, as functions of the true anomaly,
the $\;\mubold$- and $\;\dotmubold$-dependent terms in (\ref{41} -
\ref{46}). We also calculate their orbital averages in the case
of a constant precession rate $\;\mubold\,$.\\
~\\

\noindent {{\bf{\large{A.1.~~~The basic formulae\\}}}}

 Formulae (\ref{37} - \ref{42}) contain the two-body unperturbed
 expressions for the position and velocity as functions of
 time and the six Keplerian elements, (\ref{A1}) and (\ref{A7}).
To find their explicit form, one can employ an auxiliary set of
dextral perifocal coordinates $\;\bf{\vec q}\;$, with an origin at
the gravitating centre, and with the first two axes located in the
plane of the orbit:
 \ba
 {\bf{\vec q}}\;=\;\left\{
\;r\;\cos \nu\;,\;\;\;\;r\;\sin \nu\;,\;\;\;0\;\right\}^{\bf\;
T}\;\;=\;\;
 {a}\;\frac{{1\;-\;e^2}}{1\,+\,e\,\cos
\nu}\;\;\left\{ \;\cos \nu\;,\;\;\;\;\sin
\nu\;,\;\;\;0\;\right\}^{\bf\; T} \;\;.
 \label{A15}
 \ea
The corresponding velocities will read:
 \ba
 \nonumber
 {\bf{\dot{\vec q}}}
 \;=\;\left\{
 \;-\,\frac{n\;a\;\sin
\nu}{\sqrt{1\,-\,e^2}}\;,\;\;\;\;\frac{n\;a\;(e\,+\,\cos
\nu)}{\sqrt{1\,-\,e^2}}\;,\;\;\;0\;\right\}^{\bf\; T}
 \;\;=\\
 \nonumber\\
 \label{A16}\\
 \nonumber
 \frac{n\;a}{\sqrt{1\;-\;e^2}}\;\;\left\{\;-\,{\sin \nu}\;,\;\;\;\;{(e\,+\,\cos
 \nu)}\;,\;\;\;0\;\right\}^{\bf\; T}\;\;.\;\;\;\;
 \ea
the radius in (\ref{A15}) being a function of the semiaxis major
$\;a\;$, the eccentricity $\;e\;$, and the true anomaly $\;\nu\;$:
 \ba
 r\;=\;a\;\frac{1\;-\;e^2}{1\;+\;e\;\cos \nu}\;\;,
 \label{A17}
 \ea
the true anomaly $\;\it \nu\;$ itself being a function of $\;a\;$,
$\;e\;$, and of the mean anomaly
$\;M\,\equiv\,M_o\,+\,\int_{t_o}^t n\;dt\;$, where
$\;\;n\,\equiv\,(G\,m)^{1/2}\,a^{-3/2}\;$. Then, in the two-body
setting, the position and velocity, related to some fiducial
inertial frame, will appear as:
 \ba
 \nonumber
{{\erbold}} \;=\;
 \efbold \left(C_1, ... , C_6, \,t
\right)\;= \;{\bf \hat R}(\Omega,\,\inc,\,\omega)\;\,{\bf {\vec
q}}(a,\,e,\,M_o\,,\,t)\;\;\;,\\
 \label{A18} \\
 \nonumber
 {\bf {\dot{\erbold}}}\;=\;
{\bf{\vec g}}\left(C_1, ... , C_6, \,t \right)\;=
 \;{\bf \hat R}(\Omega,\,\inc,\,\omega)\;\,{\bf
{\dot{\vec q}}}(a,\,e,\,M_o\,,\,t)\;\;\;\;,
 \ea
 ${\bf \hat R}(\Omega,\,\inc,\,\omega)\;$ being the matrix of rotation
from the orbital-plane-related axes $\;q\;$ to some fixed
Cartesian axes $\;(x_1,\,x_2,\,x_3)\;$ in the said fiducial
inertial frame wherein the vectors $\;\bf \vec r\;$ and  $\;{\bf
{\dot{\erbold}}}\;$ are defined. The rotation is parameterised by
the three Euler angles: inclination, $\;\inc\;$; the longitude of
the node, $\;\Omega\;$; and the argument of the pericentre,
$\;\omega\,$. Thence, as is well known (see, for example,
Morbidelli (2002), subsection 1.2),
~ \\
 \ba
 \nonumber
 {\left.\;\right.}^{\left.\;\right.}~~
 \cos\Omega\;\cos \omega\,-\,\sin \Omega\;\sin \omega\;\cos \inc
 ~~~~~~
 -\,\cos\Omega\;\sin \omega\,-\,\sin \Omega\;\cos \omega\;\cos \inc
 ~~~~~~~~~
 \sin \Omega\;\sin \inc~~~~
 \nonumber\\
 \nonumber\\
 \nonumber
 {\bf \hat R}\,=\;~~~~
 \sin\Omega\;\cos \omega\,+\,\cos \Omega\;\sin \omega\;\cos \inc
 ~~~~~~
 -\,\sin\Omega\;\sin \omega\,+\,\cos \Omega\;\cos \omega\;\cos \inc
 ~~~~~
 -\,\cos \Omega\;\sin \inc~~~~
 \nonumber\\
 \label{A19}\\
 \nonumber
 {\left.\;\right.}^{\left.\;\right.}~~~~~~~~~~~~~
 \sin \omega\;\sin \inc
 ~~~~~~~~~~~~~~~~~~~~~~~~~~~~~~~~~~~~
 \cos \omega\;\sin \inc ~~~~~~~~~~~~~~~~
 \cos \inc ~~~~~
 \ea
insertion whereof, together with (\ref{A15}), into the first
equation of (\ref{A18}) yields
 \ba
 f_1\;=\;a\;\frac{1\;-\;e^2}{1\;+\;e\;\cos \nu}\;\;\left[\;
 \cos \Omega\;\cos \left(\omega\;+\;\nu  \right)\;-\;\sin
 \Omega\;\sin \left( \omega\;+\;\nu \right)\;\cos \inc
 \;\right] \;\;,
 \label{A20}
 \ea
 \ba
 f_2\;=\;a\;\frac{1\;-\;e^2}{1\;+\;e\;\cos \nu}\;\; \left[ \;
 \sin \Omega\;\cos \left(\omega\;+\;\nu  \right)\;+\;\cos
 \Omega\;\sin \left( \omega\;+\;\nu \right)\;\cos \inc
 \; \right] \;\;,
 \label{A21}
 \ea
 \ba
 f_3\;=\;a\;\frac{1\;-\;e^2}{1\;+\;e\;\cos \nu}\;\;\sin \left( \omega\;+\;\nu \right)\;\sin
 \inc~~\;\;.~~~~~~~~~~~~~~~~~~~~~~~~~~~~~~~~~~~~~~
 \label{A22}
 \ea
 Similarly, substitution of (\ref{A16}) and (\ref{A19}) into the
 second equation of (\ref{A18}) entails:
 \ba
 \nonumber
 g_1\;=\;\frac{n\;a}{\sqrt{1\;-\;e^2}}\;\left[\;
 -\;\cos \Omega\;\sin (\omega\;+\;\nu)\;-\;\sin
 \Omega\;\cos(\omega\;+\;\nu)\;\cos \inc\;+ \right.\\
 \label{A23}\\
 \nonumber
 \left. e\;
 (\;-\;\cos \Omega\;\sin \omega\;-\;\sin \Omega\;\cos \omega\;\cos \inc )
 \;\right]\;\;,
 \ea
 \ba
 \nonumber
 g_2\;=\;\frac{n\;a}{\sqrt{1\;-\;e^2}}\;\left[\;
 -\;\sin \Omega\;\sin (\omega\;+\;\nu)\;+\;\cos
 \Omega\;\cos(\omega\;+\;\nu)\;\cos \inc\;+ \right.\\
 \label{A23}\\
 \nonumber
 \left. e\;
 (\;-\;\sin \Omega\;\sin \omega\;+\;\cos \Omega\;\cos \omega\;\cos \inc )
 \;\right]\;\;,
 \ea
 \ba
 g_3\;=\;\frac{n\;a}{\sqrt{1\;-\;e^2}}\;\sin \inc\;\left[\;\cos (\omega\;+\;\nu)
 \;+\;e\;\cos\omega
 \;\right]\;\;\;,~~~~~~~~~~~~~~~~~~~~~~~~
 \label{A25}
 \ea
the subscripts $\;1\,,\,2\,,\,3\;$ denoting the
$\;x_1\,,\,x_2\,,\,x_3\;$ components in the fiducial inertial
frame wherein (\ref{A18}) is written.\\
 ~\\

\noindent {{\bf{\large{A.2.~~~The averaging rule\\}}}}

 From equations
 \ba
 \cos E\;=\;\frac{e\;+\;\cos \nu}{1\;+\;e\;\cos \nu}\;\;\;\;,\;\;\;\;\;\;\;\;
 \sin E\;=\;\frac{\sqrt{1\;-\;e^2}\;\;\sin \nu}{1\;+\;e\;\cos
 \nu}
 \label{A-1}
 \ea
it follows that
 \ba
 \frac{\partial E}{\partial\nu}\;=\;\frac{\sqrt{1\;-\;e^2}}{1\;+\;e\;\cos
 \nu}\;\;\;.
 \label{A-2}
 \ea
From the first of formulae (\ref{A-1}) and from the Kepler
equation one can derive:
 \ba
 \frac{\partial M}{\partial E}\;=\;\frac{1\;-\;e^2}{1\;+\;e\;\cos\nu}\;\;\;.
 \label{A-3}
 \ea
 Together, (\ref{A-2}) and (\ref{A-3}) entail:
 \ba
 \frac{\partial M}{\partial\nu}\;=\;\frac{\partial M}{\partial E}\;
 \frac{\partial E}{\partial \nu}\;=\;
 \frac{\left(1\;-\;e^2\right)^{3/2}}{\left(1\;+\;e\;\cos\nu\right)^2}
 \label{A-4}
 \ea
 whence
 \ba
 \nonumber
 \frac{1}{2\,\pi}\;\int_{0}^{2\pi}\;dM\;=\;
 \frac{\left(1\,-\,e^2 \right)^{3/2}}{2\,\pi}\;\int_{0}^{2\pi}\;
 \frac{d\nu}{\left(1\;+\;e\;\cos \nu
 \right)^2}\;\;\;.
 \ea
 Calculation of the integral shows that the right-hand side of the
 above equation is equal to unity, which means that the secular
 parts should be calculated through the following averaging rule:
 \ba
 \langle\;\,.\,.\,.\,\;\rangle\;\equiv\;\frac{\left(1\;-\;e^2\right)^{3/2}}{2\;\pi}\;
 \int_{0}^{2\pi}\;.\,.\,.\;\;\;\frac{d\nu}{\left(1\;+\;e\;\cos \nu  \right)^2}
 \label{A14}
 \ea
 In what follows below, we try hard to squeeze the calculations
 as much as possible, but at the same time to leave them
 verifiable for the interested reader.\\
~\\

 \ba
 \nonumber
 \mbox{{\bf{\large{A.3.~~~Calculation of}}}}
\;\;\;\;\mubold\cdot\left(\; \frac{\partial \efbold}{\partial
a}\;\times\;{\bf \vec g}\;-\;\efbold\;\times\;\frac{\partial {\bf
\vec g}}{\partial a}\;
\right)\;\;\;~~~~~~~~~~~~~~~~~~~~~~~~~~~~~~~~~~~~~~~~~~~~~~~~~~~
 \ea\\

As evident from (\ref{A15}) and from the first equation of
(\ref{A18}),
 \ba
 \frac{\partial \efbold}{\partial a}\;=\,
 \left(\frac{\partial \efbold}{\partial a}\right)_{\nu} +\,
 \left(\frac{\partial \efbold}{\partial \nu }\right)_{a}\;
 \left(\frac{\partial \nu}{\partial
 a}\right)_{t,\,e,\,M_o} =\,\frac{\efbold}{a}\,+\,
 \frac{\partial \efbold}{\partial t}\;\frac{\partial t}{\partial
 \nu}\;\left(\frac{\partial \nu}{\partial
 a}\right)_{t,\,e,\,M_o} =\,\frac{\efbold}{a}\,+\,
 {\bf \vec g}\;\frac{\partial t}{\partial
 \nu}\;\left(\frac{\partial \nu}{\partial
 a}\right)_{t,\,e,\,M_o}\;\;\;\;\;
 \label{A26}
 \ea
and, therefore,
 \ba
 \frac{\partial \efbold}{\partial a}\;\times\;{\bf \vec
 g}\;=\;\frac{1}{a}\;\efbold\;\times\;{\bf\vec g}\;\;\;.
 \label{A27}
 \ea
Similarly, from (\ref{A16}) and the second equation of (\ref{A18})
it ensues that
 \ba
 \nonumber
 \frac{\partial {\bf\vec g}}{\partial a}\;=\,
 \left(\frac{\partial {\bf\vec g}}{\partial a}\right)_{\nu} +\,
 \left(\frac{\partial {\bf\vec g}}{\partial \nu }\right)_{a}\;
 \left(\frac{\partial \nu}{\partial
 a}\right)_{t,\,e,\,M_o} \;=~~~~~~~~~~~~~~~~~~~~~~~~~~~~~~~~~~~~~~~~~\\
 \label{A28}\\
 \nonumber\\
 \nonumber
 -\;\frac{{\bf\vec g}}{2\,a}\,+\,
 \frac{\partial {\bf\vec g}}{\partial t}\;\frac{\partial t}{\partial
 \nu}\;\left(\frac{\partial \nu}{\partial
 a}\right)_{t,\,e,\,M_o}\; =\;-\;\frac{{\bf\vec
 g}}{2\,a}\,+\,\left(\;-\;\frac{G\,m}{|\efbold|^3}\;
 {\efbold}\;\right)\;\frac{\partial t}{\partial
 \nu}\;\left(\frac{\partial \nu}{\partial
 a}\right)_{t,\,e,\,M_o}\;\;\;\;\;
 \ea
wherefrom
 \ba
 \efbold\;\times\;\frac{\partial {\bf \vec g}}{\partial
 a}\;=\;-\;\frac{1}{2\,a}\;\efbold\;\times\;{\bf\vec g}\;\;.
 \label{A29}
 \ea
In the undisturbed two-body problem, $\;\efbold\;\times\;{\bf\vec
g}\;$ is the angular momentum (per unit of the reduced mass) and
is equal to
$\;\sqrt{G\,m\;a\;\left(1\;-\;e^2\right)}\;\;{\wbold}\;$, where
the unit vector
 \ba
 {\wbold}\;=\;{\bf\hat{x}}_1\;\sin \inc\;\sin \Omega\;-
 \;{\bf\hat{x}}_2\;\sin
 \inc\;\cos \Omega\;+\;{\bf\hat{x}}_3\;\cos
 \inc\;\;~~~~~~~~~~~~~~~~~~~
 \label{A30}
 \ea
is normal to the instantaneous osculating ellipse, the unit
vectors $\;{\bf\hat{x}}_1,\,{\bf\hat{x}}_2,\,{\bf\hat{x}}_3\;$
making the basis of the co-precessing coordinate system $\,\it
x_1,\,x_2,\,x_3\,$ (the axes $\,\it x_1\,$ and $\,\it x_2\,$
belonging to the planet's equatorial plane).

 Together, (\ref{A27}) and (\ref{A29}) will give:
 \ba
 \nonumber
 \frac{\partial \efbold}{\partial a}\;\times\;{\bf \vec
g}\;-\;\efbold\;\times\;\frac{\partial {\bf \vec g}}{\partial
a}\;=\;\frac{3}{2\,a}\;\efbold\;\times\;{\bf\vec
g}\;=\;\frac{3}{2\,a}\;\sqrt{G\,m\;a\;\left(1\;-\;e^2\right)}\;\;{\wbold}\\
 \label{A31}\\
 \nonumber\\
 \nonumber
 =\;\frac{3}{2}\;\sqrt{\frac{G\;m\;\left(1\;-\;e^2\right)}{a}}\;\;
 \left[
  \;{\bf\hat{x}}_1\;\sin \inc\;\sin \Omega\;-
 \;{\bf\hat{x}}_2\;\sin
 \inc\;\cos \Omega\;+\;{\bf\hat{x}}_3\;\cos
 \inc\;
 \right]
 \ea
and, thereby,
 \ba
 \mubold\;\cdot\;\left(\;\frac{\partial \efbold}{\partial a}\;\times\;{\bf \vec
g}\;-\;\efbold\;\times\;\frac{\partial {\bf \vec g}}{\partial
a}\;\right)\;=\;\frac{3}{2}\;\;
\mu_{\perp}\;\;\sqrt{\frac{G\;m\;\left(1\;-\;e^2\right)}{a}}
 \label{A32}
 \ea
 where
 \ba
 \mu_{\perp}\;\equiv\;\mu_1\;\sin \inc\;\sin \Omega\;-
 \;\mu_2\;\sin \inc\;\cos \Omega\;+\;\mu_3\;\cos \inc\;\;\;\;.
 \label{A33}
 \ea
Since, for constant $\;\mubold\;$, (\ref{A32}) is
$\;\nu$-independent, then in the uniform-precession case it will
coincide with its orbital average.

 ~\\

 \ba
 \nonumber
 \mbox{{\bf{\large{A.4.~~~Calculation of}}}}
\;\;\;\;\mubold\cdot\left(\; \frac{\partial \efbold}{\partial
e}\;\times\;{\bf \vec g}\;-\;\efbold\;\times\;\frac{\partial {\bf
\vec g}}{\partial e}\;
\right)\;\;\;.~~~~~~~~~~~~~~~~~~~~~~~~~~~~~~~~~~~~~~~~~~~~~~~~~~~
 \ea\\

Just as in the preceding subsection, one can write:
 \ba
 \nonumber
 \frac{\partial \efbold}{\partial e}\;=\,
 \left(\frac{\partial \efbold}{\partial e}\right)_{\nu} \,+\,
 \left(\frac{\partial \efbold}{\partial \nu}\right)_{e}\;
 \left(\frac{\partial \nu}{\partial
 e}\right)_{t,\,a,\,M_o} \;=\;
 \left(\frac{\partial \efbold}{\partial e}\right)_{\nu}\;+\;{\bf\vec g}
 \;\frac{\partial t}{\partial
 \nu}\;\left(\frac{\partial \nu}{\partial
 e}\right)_{t,\,a,\,M_o}\\
 \label{A34}\\
 \nonumber\\
 \nonumber
  =\;-\;\frac{2\,e\;+\;\cos \nu\;+\;e^2\;\cos \nu }{
  \left(1\;+\;e\;\cos \nu \right)\;\left(1\;-\;e^2  \right) }\;\efbold\;+\,
 {\bf \vec g}\;\;\frac{\partial t}{\partial
 \nu}\;\left(\frac{\partial \nu}{\partial
 e}\right)_{t,\,a,\,M_o}\;\;\;\;\;~~~~~~~~~~~~~~~~~~~~~~~~~~~~~~~~~
 \ea
whence
 \ba
 \frac{\partial \efbold}{\partial e}\;\times\;{\bf \vec
g}\;=\;-\;\frac{2\,e\;+\;\cos \nu\;+\;e^2\;\cos \nu }{
  \left(1\;+\;e\;\cos \nu \right)\;\left(1\;-\;e^2  \right)
  }\;\efbold\;\times\;{\bf\vec g}\;\;\;.
 \label{A35}
 \ea
With (\ref{A16}) taken into account, the derivative of the
two-body velocity will take the form:
 \ba
 \nonumber
 \frac{\partial \bf \vec g}{\partial e}\;=\,
 \left(\frac{\partial \bf \vec g}{\partial e}\right)_{\nu} \,+\,
 \left(\frac{\partial \bf \vec g}{\partial \nu}\right)_{e}\;
 \left(\frac{\partial \nu}{\partial
 e}\right)_{t,\,a,\,M_o} \;=\;\frac{e}{1\;-\;e^2}\;{\bf\vec
 g}\;+\;{\bf\vec b}\;+\;\frac{\partial \bf \vec g}{\partial t}\;\frac{\partial t}{\partial
 \nu}\;\left(\frac{\partial \nu}{\partial
 e}\right)_{t,\,a,\,M_o}~=\\
 \label{A36}\\
 \nonumber
 \frac{e}{1\;-\;e^2}\;{\bf\vec
 g}\;+\;{\bf\vec
 b}\;+\;\left(\;-\;\frac{G\;m}{|\efbold|^3}\;\efbold
 \right)
 \;\frac{\partial t}{\partial
 \nu}\;\left(\frac{\partial \nu}{\partial
 e}\right)_{t,\,a,\,M_o}~~~~~~
 \ea
where
 \ba
 {\bf\vec
 b}\,\equiv~~~~~~~~~~~~~~~~~~~~~~~~~~~~~~~~~~~~~~~~~
 ~~~~~~~~~~~~~~~~~~~~~~~~~~~~~~~~~~~~
 \label{A37}
 \ea
 \ba
 \nonumber
 \frac{n\;a}{\sqrt{1-e^2}}\;
 \left\{\,
 -\,\cos\Omega \;\sin\omega \,-\, \sin \Omega \;\cos \omega \; \cos \inc
 \;\;,\;\;\;
 -\,\sin\Omega \;\sin\omega \,+\, \cos \Omega \;\cos \omega \; \cos \inc
 \;\;,\;\;\;\;
 \sin \inc\;\cos \omega
   \;\right\}^{\;{\bf{T}}}\;.
 \label{A38}
 \ea
 From this we easily get, with aid of (\ref{A30}):
 \ba
 \nonumber
 \efbold\;\times\;{\bf\vec b}\;=\;\frac{n\;a^2\;\sqrt{1\,-\,e^2}}{1\,+\,e\,\cos
 \nu}\;\;
 \left\{\;
 \sin \Omega\;\sin \inc\;\cos \nu \;\;\;,\;\;\;\; -\;\cos
 \Omega\;\sin \inc\;\cos \nu \;\;\;,\;\;\;\; \cos \inc\;\cos \nu
 \;\right\}^{\;\bf T}~=~~~~~\\
 \label{A39}\\
 \nonumber
 \frac{n\;a^2\;\sqrt{1\,-\,e^2}}{1\,+\,e\,\cos
 \nu}\;\;\cos \nu\;\;
 \left\{\;
 \sin \inc\;\cos \Omega \;\;\;,\;\;\;\; -\;\sin \inc\;\cos \Omega \;\;\;,
 \;\;\;\; \cos \inc
 \;\right\}^{\;\bf T}\; = \;{\wbold} ~~\frac{n\;a^2\;\sqrt{1\,-\,e^2}}{1\,+\,e\,\cos
 \nu}\;\;\cos \nu\;\;~.~~
 \ea
 ~\\
 Together, (\ref{A35}), (\ref{A36}), and (\ref{A38}) yield:
 \ba
 \nonumber
 \frac{\partial \efbold}{\partial e}\;\times\;{\bf \vec
g}\;-\;\efbold\;\times\;\frac{\partial {\bf \vec g}}{\partial
e}\;=\;\left[\;
 -\;\frac{2\;e\;+\;\cos \nu\;+\;e^2\;\cos \nu}{\left(1\;+\;e\;\cos \nu \right)\,
 \left(1\;-\;e^2\right)}\;-\;\frac{e}{1\;-\;e^2}
\;\right]\;\efbold\;\times\;{\bf\vec
g}\;-\;\efbold\;\times\;{\bf\vec b}\;=\\
 \nonumber\\
 \nonumber\\
 \nonumber
 -\;\,\wbold\;\,\frac{3\,e\,+\,\cos \nu\,+\,2\,e^2\,\cos \nu}{\left(1\,+\,e\,\cos \nu \right)\,
 \left(1\,-\,e^2\right)}\;\;\sqrt{G\,m\,a\,\left(1\,-\,e^2\right)}\;\,-\;\,{\wbold} ~~\frac{n\;a^2\;\sqrt{1\,-\,e^2}}{1\,+\,e\,\cos
 \nu}\;\;\cos \nu\;=
~~~~~~~~~~~~~
 \\
 \nonumber\\
 \label{A40}\\
 \nonumber
 -\;\,\wbold\;\,\frac{n\,a^2\;\left(3\,e\,+\,2\,\cos \nu\,+\,e^2\,\cos \nu  \right)}{\left(1\,+\,e\,\cos
 \nu\right)\;\sqrt{1\;-\;e^2}}
 ~~~,~~~~~~~~~~~~~~~~~~~~~~~~~~~~~~~~~~~~~~~~~~~~~~~~~~~~~~~~~~~~~~~~~~~
 \ea
whence
 \ba
 \mubold\cdot\left(\frac{\partial \efbold}{\partial e}\;\times\;{\bf \vec
 g}\;-\;\efbold\;\times\;\frac{\partial {\bf \vec g}}{\partial
 e}\right)\;=\;
  -\;\,\mu_{\perp}\;\,\frac{n\,a^2\;\left(3\,e\,+\,2\,\cos \nu\,+\,e^2\,\cos \nu  \right)}{\left(1\,+\,e\,\cos
 \nu\right)\;\sqrt{1\;-\;e^2}}
 ~~~.~~~~~~~~~~~~~~~~~~~~~~~~~~~~~~~~~~~~~~~~~~~~~~~~~~~~~~~~~~~~~~~~~~~
 \label{4120}
 \ea
 With aid of integrals
 \ba
 \int_0^{2\pi}\;\frac{d\nu}{\left(1\,+\,e\,\cos\nu\right)^3}\;=\;\pi\;\frac{2\;+\;e^2}{\left(1\;-\;e^2\right)^{5/2}}
 \label{integral1}
 \ea
 and
  \ba
 \int_0^{2\pi}\;\frac{\cos\nu\;\;d\nu}{\left(1\,+\,e\,\cos\nu\right)^3}\;=\;\pi\;\frac{-\;3\;e}{\left(1\;-\;e^2\right)^{5/2}}
 \label{integral2}
 \ea
 it is easy to demonstrate that, under the assumption of constant
 $\;\mubold\;$, the orbital average of (\ref{4120}) is nil.
 ~\\

 \ba
 \nonumber
 \mbox{{\bf{\large{A.5.~~~Calculation of}}}}
\;\;\;\;\mubold\cdot\left(\; \frac{\partial \efbold}{\partial
\omega}\;\times\;{\bf \vec g}\;-\;\efbold\;\times\;\frac{\partial
{\bf \vec g}}{\partial \omega}\;
\right)\;\;\;.~~~~~~~~~~~~~~~~~~~~~~~~~~~~~~~~~~~~~~~~~~~~~~~~~~~
 \ea\\

A straightforward calculation, based on formulae (\ref{A15}),
(\ref{A16}), (\ref{A18}), and (\ref{A19}), gives:
 \ba
 \left(\; \frac{\partial \efbold}{\partial \omega}\;\times\;{\bf
\vec g}\;\right)_1\;=\;\frac{\partial f_2}{\partial
\omega}\;g_3\;-\;\frac{\partial f_3}{\partial
\omega}\;g_2\;=\;-\;\frac{n\;a^2\;\sqrt{1\;-\;e^2}}{1\;+\;e\;\cos
\nu}\;e\;\sin \Omega\;\sin \inc\;\sin \nu \;\;\;,\;\;\;
 \label{A41}
 \ea
 \ba
 \left(\; \frac{\partial \efbold}{\partial \omega}\;\times\;{\bf
\vec g}\;\right)_2\;=\;\frac{\partial f_3}{\partial
\omega}\;g_1\;-\;\frac{\partial f_1}{\partial
\omega}\;g_3\;\;=\;\;\frac{n\;a^2\;\sqrt{1\;-\;e^2}}{1\;+\;e\;\cos
\nu}\;e\;\cos \Omega\;\sin \inc\;\sin \nu \;\;\;,\;\;\;
 \label{A42}
 \ea
 \ba
 \left(\; \frac{\partial \efbold}{\partial \omega}\;\times\;{\bf
\vec g}\; \right)_3\;=\;\frac{\partial f_1}{\partial
\omega}\;g_2\;-\;\frac{\partial f_2}{\partial
\omega}\;g_1\;=\;-\;\frac{n\;a^2\;\sqrt{1\;-\;e^2}}{1\;+\;e\;\cos
\nu}\;\;e\;\;\cos \inc\;\sin \nu \;\;\;.\;\;\;
 \label{A43}
 \ea
The two-body angular momentum cannot depend on the argument of the
pericentre,
 \ba
 \frac{\partial}{\partial \omega}\;\left(\;\efbold\;\times\;{\bf\vec g}\;\right)\;=\;0\;\;\;,
 \label{A44}
 \ea
i.e.,
 \ba
 \left(\;\efbold\;\times\;\frac{\partial \bf \vec g}{\partial \omega} \;\right)_j
 \;=\;-\;\left(\; \frac{\partial \efbold}{\partial \omega}\;\times\;{\bf
\vec g}\; \right)_j\;\;\;,
 \label{A45}
 \ea
whereby
 \ba
 \left(\; \frac{\partial \efbold}{\partial \omega}\;\times\;{\bf
\vec g}\;-\;\efbold\;\times\;\frac{\partial {\bf \vec g}}{\partial
\omega}\;
\right)_1\;=\;-\;2\;\;\frac{n\;a^2\;\sqrt{1\;-\;e^2}}{1\;+\;e\;\cos
\nu}\;e\;\sin \Omega\;\sin \inc\;\sin \nu \;\;\;,\;\;\;
 \label{A46}
 \ea
 \ba
 \left(\; \frac{\partial \efbold}{\partial \omega}\;\times\;{\bf
\vec g}\;-\;\efbold\;\times\;\frac{\partial {\bf \vec g}}{\partial
\omega}\;
\right)_2\;\;=\;\;2\;\;\frac{n\;a^2\;\sqrt{1\;-\;e^2}}{1\;+\;e\;\cos
\nu}\;e\;\cos \Omega\;\sin \inc\;\sin \nu \;\;\;,\;\;\;\;\;
 \label{A47}
 \ea
 \ba
 \left(\; \frac{\partial \efbold}{\partial \omega}\;\times\;{\bf
\vec g}\;-\;\efbold\;\times\;\frac{\partial {\bf \vec g}}{\partial
\omega}\;
\right)_3\;=\;-\;2\;\;\frac{n\;a^2\;\sqrt{1\;-\;e^2}}{1\;+\;e\;\cos
\nu}\;e\;\;\cos \inc\;\sin \nu \;\;\;.\;\;\;\;\;\;\;\;\;\;\;
 \label{A48}
 \ea
Altogether, these formulae may be written down as:
 \ba
 \left(\; \frac{\partial \efbold}{\partial \omega}\;\times\;{\bf
\vec g}\;-\;\efbold\;\times\;\frac{\partial {\bf \vec g}}{\partial
\omega}\;
\right)\;=\;-\;2\;\;{\wbold}\;\;\;\frac{n\;a^2\;\sqrt{1\;-\;e^2}}{1\;+\;e\;\cos
\nu}\;e\;\;\sin \nu \;\;\;,\;\;\;\;\;\;\;\;\;\;\;
 \label{Aomega}
 \ea
$\wbold\;$ being the unit vector pointing in the direction of the
angular momentum. It is given by (\ref{A30}). Finally,
 \ba
 \mubold\cdot\left(\; \frac{\partial \efbold}{\partial \omega}\;\times\;{\bf
\vec g}\;-\;\efbold\;\times\;\frac{\partial {\bf \vec g}}{\partial
\omega}\;
\right)\;=\;-\;2\;\;\mu_{\perp}\;\;\;\frac{n\;a^2\;\sqrt{1\;-\;e^2}}{1\;+\;e\;\cos
\nu}\;e\;\;\sin \nu \;\;\;.\;\;\;\;\;\;\;\;\;\;\;
 \label{4133}
 \ea
 Since $\;\sin\nu\;$ is odd and $\;\left(1\;+\;e\;\cos\nu
 \right)^{-3}\;$ is even, it is obvious that, for a constant
 $\;\mubold\;$, the orbital average of (\ref{4133}) will vanish.

~\\

 \ba
 \nonumber
 \mbox{{\bf{\large{A.6.~~~Calculation of}}}}
\;\;\;\;\mubold\cdot\left(\; \frac{\partial \efbold}{\partial
\Omega}\;\times\;{\bf \vec g}\;-\;\efbold\;\times\;\frac{\partial
{\bf \vec g}}{\partial \Omega}\;
\right)\;\;\;.~~~~~~~~~~~~~~~~~~~~~~~~~~~~~~~~~~~~~~~~~~~~~~~~~~~
 \ea\\

Once again, formulae  (\ref{A15}), (\ref{A16}), (\ref{A18}), and
(\ref{A19}) yield:
 \ba
 \nonumber
\left(\;\frac{\partial \efbold}{\partial \Omega}\;\times\;{\bf
\vec g}\;\right)_1\;=\;\frac{\partial f_2}{\partial
\Omega}\;g_3\;-\;\frac{\partial f_3}{\partial
\Omega}\;g_2\;=~~~~~~~~~~~~~~~~~~~~~~~~~~~~~~~~~~~~~~~~~~~~~~~~~~~~~~~~~~~~\\
 \label{A49}\\
 \nonumber
{\left. \; \right.}^{\left. \; \right.}~~~~
\frac{n\;a^2\;\sqrt{1\;-\;e^2}}{1\;+\;e\;\cos\nu}\;\left[\; \cos
\Omega\;\cos (\omega\;+\;\nu)\;-\;\sin \Omega\;\sin
(\omega\;+\;\nu)\;\cos \inc
\;\right]\;\cos\left(\omega\;+\;\nu\right)\;\sin\inc~~~\\
 \nonumber\\
 \nonumber\\
 \nonumber
 +\;\;\frac{n\;a^2\;\sqrt{1\;-\;e^2}}{1\;+\;e\;\cos\;nu}\;\;e\;\;\left[\;
 \cos
\Omega\;\cos (\omega\;+\;\nu)\;-\;\sin \Omega\;\sin
(\omega\;+\;\nu)\;\cos \inc \;\right]\;\cos
\omega\;\sin\inc~~~,~~~~
 \ea
 ~\\
  \ba
 \nonumber
\left(\;\frac{\partial \efbold}{\partial \Omega}\;\times\;{\bf
\vec g}\;\right)_2\;=\;\frac{\partial f_3}{\partial
\Omega}\;g_1\;-\;\frac{\partial f_1}{\partial
\Omega}\;g_3\;=~~~~~~~~~~~~~~~~~~~~~~~~~~~~~~~~~~~~~~~~~~~~~~~~~~~~~~~~~~~~\\
 \label{A50}\\
 \nonumber
{\left. \; \right.}^{\left. \; \right.}~~~~
\frac{n\;a^2\;\sqrt{1\;-\;e^2}}{1\;+\;e\;\cos\nu}\;\left[\; \sin
\Omega\;\cos (\omega\;+\;\nu)\;+\;\cos \Omega\;\sin
(\omega\;+\;\nu)\;\cos \inc
\;\right]\;\cos\left(\omega\;+\;\nu\right)\;\sin\inc~~~\\
 \nonumber\\
 \nonumber\\
 \nonumber
 +\;\;\frac{n\;a^2\;\sqrt{1\;-\;e^2}}{1\;+\;e\;\cos\nu}\;\;e\;\;\left[\;
 \sin
\Omega\;\cos (\omega\;+\;\nu)\;+\;\cos \Omega\;\sin
(\omega\;+\;\nu)\;\cos \inc \;\right]\;\cos
\omega\;\sin\inc~~~,~~~~
 \ea
 ~\\
 \ba
\left(\;\frac{\partial \efbold}{\partial \Omega}\;\times\;{\bf
\vec g}\;\right)_3\;=\;\frac{\partial f_1}{\partial
\Omega}\;g_2\;-\;\frac{\partial f_2}{\partial
\Omega}\;g_1\;=~\,~~~~~~~~~~~~~~~~~~~~~~~~~~~~~~~~~~~~~~~~~~~~~~~~~~~~~~~~~~~
 \label{A51}
 \ea
 \ba
 \nonumber
 \frac{n a^2 \sqrt{1- e^2}}{1 + e\;\cos\nu}\;
 \left\{\;
 \left[\;-\, \sin \Omega\;\cos (\omega + \right.\right.
 \nu)~~~~~~~~~~~~~~~~~~~~~~~~~~~~~~~~~~~~~~~~~~~~~~~~~~~~~~~~~~~~~~~~~~~~~~~~~~~~~~~~~~~~~
 \\
 \nonumber\\
 \nonumber\\
 \nonumber
 \left.
 -\;\cos \Omega\;\sin
(\omega + \nu)\;\cos \inc \;\right]\;\;\left[\;-\,\sin\Omega\;
 \sin\left(\omega + \nu\right)\;+\;\cos\Omega\;\cos\inc\;\cos(\omega + \nu)
 \right]\,-~~~
  \\
  \nonumber\\
  \nonumber\\
  \nonumber
 \left.
 \left[\;\cos \Omega\;\cos (\omega + \nu)\;-\;\sin \Omega\;\sin
(\omega + \nu)\;\cos \inc \;\right]\;\;\left[\;-\;\cos\Omega\;
 \sin\left(\omega + \nu\right)\;-\;\sin\Omega\;\cos\inc\;\cos(\omega + \nu)
 \right]
 \; \right\}
~~~\\
 \nonumber\\
 \nonumber\\
 \nonumber
+\;\frac{n\;a^2\;\sqrt{1\;-\;e^2}}{1\;+\;e\;\cos\nu}\;\;e\;\;
 \left\{\;
 \left[\;-\; \sin \Omega\;\cos (\omega\;+\;\nu)\;-~~~~~~~~~~~~~~~~~~~~~~~~~~~~~~~~~~~~~~~~~~~~~
 ~~~~~~~~~~~~~~~~~~~~~~ \right. \right. \\
 \nonumber\\
 \nonumber\\
 \nonumber
 \left. \cos \Omega\;\sin
(\omega\;+\;\nu)\;\cos \inc \;\right]\;\;\left[\;-\;\sin\Omega\;
 \sin\omega\;+\;\cos\Omega\;\cos\inc\;\cos\omega\;
 \right]\;~~~~~~~~~~~~~~~~~~\\
 \nonumber\\
 \nonumber\\
 \nonumber
 \left.
 -\;\left[\;\cos \Omega\;\cos (\omega\;+\;\nu)\;-\;\sin \Omega\;\sin
(\omega\;+\;\nu)\;\cos \inc \;\right]\;\;\left[\;-\;\cos\Omega\;
 \sin\omega\;-\;\sin\Omega\;\cos\inc\;\cos\omega\;
 \right]
 \; \right\}\;=
 ~~~~~~~~
 \ea
 ~\\
 \ba
 \nonumber
 \frac{n a^2 \sqrt{1- e^2}}{1 + e\;\cos\nu}\;\left\{\;
 \sin (\omega + \nu)\;\cos (\omega + \nu)\;\sin^2\inc\;+\;
 \;e\;\;\left[\; \sin \omega\;\cos (\omega + \nu) \;-\;\sin(\omega +
 \nu)\;\cos \omega\;\cos^2\inc\;\right]\;\right\}\;\;,
 \ea
~\\
 \ba
\left(\;\efbold\;\times\;\frac{\partial \bf\vec g}{\partial
\Omega}\;\right)_1\;=\;f_2\;\frac{ \partial {\bf\vec
g}_3}{\partial \Omega}\;-\;f_3\;\frac{\partial {\bf\vec
g}_2}{\partial
\Omega}\;=~~~~~~~~~~~~~~~~~~~~~~~~~~~~~~~~~~~~~~~~~~~~~~~~~~~~~~~~~~~~~~~~~~~
 \label{A52}
 \ea
 \ba
 \nonumber\\
 \nonumber
{\left. \; \right.}^{\left. \; \right.}~~~~
\frac{n\;a^2\;\sqrt{1\;-\;e^2}}{1\;+\;e\;\cos\nu}\;\;\;\left[\;-\;\sin(\omega
+ \nu)\;\sin \inc \;\right]\;\;\;\left[\;-\;\cos \Omega\;\sin
(\omega\;+\;\nu)\;-\;\sin \Omega\;\cos (\omega\;+\;\nu)\;\cos \inc
\;\right]\;~~~~~~~~~\\
 \nonumber\\
 \nonumber\\
 \nonumber
 +\;\;\frac{n\;a^2\;\sqrt{1\;-\;e^2}}{1\;+\;e\;\cos\nu}\;\;\;e\;\;\;
 \left[\;
 -\;\sin(\omega + \nu)\;\sin \inc \;\right]\;\;\;
 \left[\;-\;\cos\Omega\;\sin \omega\;-\;\sin \Omega\;\cos \omega\;\cos \inc
 \;\right]~~~~~~~~~~~~~~~~~~~~~~~~
 \ea
 ~\\
 \ba
 \nonumber
 =\;\frac{n\;a^2\;\sqrt{1\;-\;e^2}}{1\;+\;e\;\cos\nu}\;\;\sin
 \inc\;\;\left[\;
 \sin^2\left(\omega + \nu \right)\;\cos \Omega \;+\;\sin\left(\omega + \nu \right)\;
 \cos \left(\omega + \nu \right)\;\sin\Omega\;\cos \inc
 \;\right]~~~~~~\\
 \nonumber\\
 \nonumber\\
 \nonumber
 {\left. \; \right.}^{\left. \; \right.}~~~+\;\;\frac{n\;a^2\;\sqrt{1\;-\;e^2}}{1\;+\;e\;\cos\;nu}\;\;\;e\;\;\sin \inc\;\;\;
 \left[\;
 \sin\left(\omega + \nu \right)\;\sin \omega\;\cos \Omega \;+\;\sin\left(\omega + \nu \right)\;
 \cos \omega \;\sin\Omega\;\cos \inc
 \;\right]\;\;\;,
 \ea
 ~\\
 \ba
\left(\;\efbold\;\times\;\frac{\partial \bf\vec g}{\partial
\Omega}\;\right)_2\;=\;f_3\;\frac{ \partial {\bf\vec
g}_1}{\partial \Omega}\;-\;f_1\;\frac{\partial {\bf\vec
g}_3}{\partial
\Omega}\;=~~~~~~~~~~~~~~~~~~~~~~~~~~~~~~~~~~~~~~~~~~~~~~~~~~~~~~~~~~~~~~~~~~~
 \label{A53}
 \ea
 \ba
 \nonumber\\
 \nonumber
{\left. \; \right.}^{\left. \; \right.}~~~~
\frac{n\;a^2\;\sqrt{1\;-\;e^2}}{1\;+\;e\;\cos\nu}\;\;\;\sin(\omega
+ \nu)\;\sin \inc \;\;\;\left[\;\sin \Omega\;\sin
(\omega\;+\;\nu)\;-\;\cos\Omega\;\cos (\omega\;+\;\nu)\;\cos \inc
\;\right]\;~~~~~~~~~\\
 \nonumber\\
 \nonumber\\
 \nonumber
 +\;\;\frac{n\;a^2\;\sqrt{1\;-\;e^2}}{1\;+\;e\;\cos\nu}\;\;\;e\;\;\;
 \sin(\omega + \nu)\;\sin \inc \;\;\;
 \left[\;\sin\Omega\;\sin \omega\;-\;\cos \Omega\;\cos \omega\;\cos \inc
 \;\right]~~~,~~~~~~~~~~~~~~~~~~~~
 \ea
 ~\\
 \ba
\left(\;\frac{\partial \efbold}{\partial \Omega}\;\times\;{\bf
\vec g}\;\right)_3\;=\;\frac{\partial f_1}{\partial
\Omega}\;g_2\;-\;\frac{\partial f_2}{\partial
\Omega}\;g_1\;=~\,~~~~~~~~~~~~~~~~~~~~~~~~~~~~~~~~~~~~~~~~~~~~~~~~~~~~~~~~~~~
 \label{A54}
 \ea
 \ba
 \nonumber
 \frac{n a^2 \sqrt{1- e^2}}{1 + e\;\cos\nu}\;
 \left\{\;
 \left[\; \cos \Omega\;\cos (\omega +
 \nu)~~~~~~~~~~~~~~~~~~~~~~~~~~~~~~~~~~~~~~~~~~~~~~~~~~~~~~~~~~~~~
 ~~~~~~~~~~~~~~~~~~~~~~~~ \right.\right.
 \\
 \nonumber\\
 \nonumber\\
 \nonumber
 \left.
 -\;\sin \Omega\;\sin
(\omega + \nu)\;\cos \inc \;\right]\;\;\left[\;-\,\cos\Omega\;
 \sin\left(\omega + \nu\right)\;-\;\sin\Omega\;\cos\inc\;\cos(\omega + \nu)
 \right]\,-~~~
  \\
  \nonumber\\
  \nonumber\\
  \nonumber
 \left.
 \left[\;\sin \Omega\;\cos (\omega + \nu)\;+\;\cos \Omega\;\sin
(\omega + \nu)\;\cos \inc \;\right]\;\;\left[\;\sin\Omega\;
 \sin\left(\omega + \nu\right)\;-\;\cos\Omega\;\cos\inc\;\cos(\omega + \nu)
 \right]
 \; \right\}
~~~\\
 \nonumber\\
 \nonumber\\
 \nonumber
+\;\frac{n\;a^2\;\sqrt{1\;-\;e^2}}{1\;+\;e\;\cos\nu}\;\;e\;\;
 \left\{\;
 \left[\;\cos \Omega\;\cos (\omega\;+\;\nu)\;-~~~~~~~~~~~~~~~~~~~~~~~~~~~~~~~~~~~~~~~~~~~~~
 ~~~~~~~~~~~~~~~~~~~~~~ \right.\right.\\
 \nonumber\\
 \nonumber\\
 \nonumber
 \left.\sin \Omega\;\sin
(\omega\;+\;\nu)\;\cos \inc \;\right]\;\;\left[\;-\;\cos\Omega\;
 \sin\omega\;-\;\sin\Omega\;\cos\inc\;\cos\omega\;
 \right]\;~~~~~~~~~~~~~~~~~~\\
 \nonumber\\
 \nonumber\\
 \nonumber
 \left.
 -\;\left[\;\sin \Omega\;\cos (\omega\;+\;\nu)\;+\;\cos \Omega\;\sin
(\omega\;+\;\nu)\;\cos \inc \;\right]\;\;\left[\;\sin\Omega\;
 \sin\omega\;-\;\cos\Omega\;\cos\inc\;\cos\omega\;
 \right]
 \; \right\}\;=
 ~~~~~~~~
 \ea
 ~\\
 \ba
 \nonumber
 \frac{n a^2 \sqrt{1- e^2}}{1 + e\;\cos\nu}\;\left\{\;
 -\,\sin (\omega + \nu)\;\cos (\omega + \nu)\;\sin^2\inc\;+\;
 e\;\left[\;-\, \sin \omega\;\cos (\omega + \nu) \,+\,\sin(\omega +
 \nu)\;\cos \omega\;\cos^2\inc\;\right]\;\right\}\;.
 \ea
~\\
In (\ref{A49} - \ref{A50}) and in (\ref{A52} - \ref{A53}) we
benefitted from $\;\it f_3\;$ being independent of $\;\Omega\;$.
Now, by subtracting, accordingly, (\ref{A52}) from (\ref{A49}),
and (\ref{A53}) from (\ref{A50}), and (\ref{A54}) from
(\ref{A51}), we arrive, after some intermediate algebra, at the
following three expressions:
 \ba
 \left(\; \frac{\partial \efbold}{\partial \Omega}\;\times\;{\bf
\vec g}\;-\;\efbold\;\times\;\frac{\partial {\bf \vec g}}{\partial
\Omega}\;
\right)_1\;=~~~~~~~~~~~~~~~~~~~~~~~~~~~~~~~~~~~~~~~~~~~~~~~
~~~~~~~~~~~~~~~~~~~~~~~~~~
 \label{A55}\\
 \nonumber\\
 \nonumber\\
 \nonumber
 \frac{n\;a^2\;\sqrt{1\;-\;e^2}}{1\;+\;e\;\cos\nu}\;\;\sin
 \inc\;\;\left\{\;\;
 \cos \Omega\;\cos\left[\,2\,\left(\omega + \nu \right)\,\right]
 \;-\;\sin \Omega\;\cos \inc\;\sin \left[\,2\,\left(\omega + \nu
 \right)\,\right]\;\;
 \right\}~~~~~\,~~~~\\
 \nonumber\\
 \nonumber\\
 \nonumber
 +\;\;\frac{n\;a^2\;\sqrt{1\;-\;e^2}}{1\;+\;e\;\cos\nu}\;\;e\;\;\sin
 \inc\;\;\left\{\;\;
 \cos \Omega\;\cos\left(\nu \,+ \,2\,\omega \right)
 \;-\;2\;\sin \Omega\;\cos \inc\;\sin \left(\omega + \nu
 \right)\;\cos \omega\;\;
 \right\}\;\;\;,
 \ea
~\\
 \ba
 \left(\; \frac{\partial \efbold}{\partial \Omega}\;\times\;{\bf
\vec g}\;-\;\efbold\;\times\;\frac{\partial {\bf \vec g}}{\partial
\Omega}\;
\right)_2\;=~~~~~~~~~~~~~~~~~~~~~~~~~~~~~~~~~~~~~~~~~~~~~~~
~~~~~~~~~~~~~~~~~~~~~~~~~~
 \label{A56}\\
 \nonumber\\
 \nonumber\\
 \nonumber
 \frac{n\;a^2\;\sqrt{1\;-\;e^2}}{1\;+\;e\;\cos\nu}\;\;\sin
 \inc\;\;\left\{\;\;
 \sin \Omega\;\cos\left[\,2\,\left(\omega + \nu \right)\,\right]
 \;+\;\cos \Omega\;\cos \inc\;\sin \left[\,2\,\left(\omega + \nu
 \right)\,\right]\;\;
 \right\}~~~~~\,~~~~\\
 \nonumber\\
 \nonumber\\
 \nonumber
 +\;\;\frac{n\;a^2\;\sqrt{1\;-\;e^2}}{1\;+\;e\;\cos\nu}\;\;e\;\;\sin
 \inc\;\;\left\{\;\;
 \sin \Omega\;\cos\left(\nu \,+ \,2\,\omega \right)
 \;+\;2\;\cos \Omega\;\cos \inc\;\sin \left(\omega + \nu
 \right)\;\cos \omega\;\;
 \right\}\;\;\;,
 \ea
~\\
 \ba
 \left(\; \frac{\partial \efbold}{\partial \Omega}\;\times\;{\bf
\vec g}\;-\;\efbold\;\times\;\frac{\partial {\bf \vec g}}{\partial
\Omega}\;
\right)_3\;=~~~~~~~~~~~~~~~~~~~~~~~~~~~~~~~~~~~~~~~~~~~~~~~
~~~~~~~~~~~~~~~~~~~~~~~~~~~~~
 \label{A57}
 \label{4142}
 \ea
 \ba
 \nonumber\\
 \nonumber
 \frac{n\;a^2\;\sqrt{1\;-\;e^2}}{1\;+\;e\;\cos\nu}\;\;\sin^2
 \inc\;\;\sin \left[\,2\,\left( \omega\,+\,\nu \right) \,\right]~~~~~\,~~~~
 ~~~~~~~~~~~~~~~~~~~~~~~~~~~~~~~~~~~~~~~~~~~~~~~~\\
 \nonumber\\
 \nonumber\\
 \nonumber
 +\;\;\frac{n\;a^2\;\sqrt{1\;-\;e^2}}{1\;+\;e\;\cos\nu}\;\;e
 \;\;\left\{\;\;
 2\;\sin \omega\;\cos\left(\nu \,+ \,\omega \right)
 \;-\;2\;\sin\left(\omega + \nu \right)\;\cos \omega\;\cos^2 \inc\;\;
 \right\}\;=\;\;~~~~\\
 \nonumber\\
 \nonumber\\
 \nonumber
 \frac{n a^2 \sqrt{1 - e^2}}{1 + e\;\cos\nu}\;\sin^2
 \inc\;\sin \left[\,2\,\left( \omega\,+\,\nu \right) \,\right]
 \,+\,\frac{n a^2 \sqrt{1 - e^2}}{1 + e\;\cos\nu}\;\;2\;e
 \;\;\left\{\;\;
 -\,\sin \nu \,+ \,\sin^2\inc\;\cos \omega \;\sin \left(\omega + \nu \right)\;
 \; \right\}\;\;.
 \ea
Hence, the overall expression for
$\;\;\mubold\cdot\left(\,({\partial \efbold}/{\partial
\Omega})\times {\bf\vec g}\,-\,\efbold\times ({\partial \bf\vec
g}/{\partial \Omega})\,\right)\;$ will be given by (\ref{478}). To
average this expression, in the simple case of a constant
$\;\mubold\;$, it will be sufficient to separately average the
above three expressions (\ref{A55} - \ref{A57}) and then to
multiply the averages by $\;\mu_1\,,\;\mu_2\,$ and $\,\mu_3\;$,
correspondingly, and to add everything up. To carry out this
procedure, it will be convenient first to regroup terms in
(\ref{A55} - \ref{A57}) and to through out the inputs proportional
to the odd functions $\;\sin\nu\;$ and $\;\sin 2\nu\;$, because
these inputs will vanish after averaging via formula (\ref{A14}).
Thus we get:
 \ba
 {\langle} \;\left(\; \frac{\partial \efbold}{\partial \Omega}\;\times\;{\bf
\vec g}\;-\;\efbold\;\times\;\frac{\partial {\bf \vec g}}{\partial
\Omega}\;
\right)_1\;\rangle\;=~~~~~~~~~~~~~~~~~~~~~~~~~~~~~~~~~~~~~~~~~~~~~~~
~~~~~~~~~~~~~~~~~~~~~~~~~~
 \label{4143}\\
 \nonumber\\
 \nonumber\\
 \nonumber
\langle\;\;
\frac{n\;a^2\;\sqrt{1\;-\;e^2}}{1\;+\;e\;\cos\nu}\;\;\sin
 \inc\;\;\left\{\;\;
 \cos \Omega\;\cos2\omega\;\cos 2\nu
 \;-\;\cos \Omega\;\sin 2\omega\;\sin 2\nu ~~~~~~~~~~~~~~~~~~~~~ \right.\\
 \nonumber\\
 \nonumber\\
 \nonumber
 \left.
 -\;\sin \Omega\;\cos \inc\;\sin 2\omega\;\cos 2\nu\;
 +\;\sin \Omega\;\cos \inc\;\cos 2\omega\;\sin 2\nu
 \;\;\right\}~~\\
 \nonumber\\
 \nonumber\\
 \nonumber
 +\;\;\frac{n\;a^2\;\sqrt{1\;-\;e^2}}{1\;+\;e\;\cos\nu}\;\;e\;\;\sin
 \inc\;\;\left\{\;\;
 \cos \Omega\; \cos 2\omega\;\cos\nu\;
 -\;\cos \Omega\;\sin 2\omega\;\sin\nu ~~~~~~~~~~~~~~~~~~~~~~~~ \right.\\
 \nonumber\\
 \nonumber\\
 \nonumber
 \left.
 -\;\sin \Omega\;\cos \inc\;\sin 2\omega \;\cos \nu
 +\;2\;\sin \Omega\;\cos\inc\;\cos^2\omega\;\sin\nu
 \;\right\}\;\;\rangle\;=\;\;
 \ea
 ~\\
 \ba
 \nonumber
\langle\;\;
\frac{n\;a^2\;\sqrt{1\;-\;e^2}}{1\;+\;e\;\cos\nu}\;\;\sin
 \inc\;\;\left(\;\;
 \cos \Omega\;\cos2\omega\;-\;\sin \Omega\;\cos \inc\;\sin 2\omega
 \;\;\right)\;\left(\;\cos 2\nu\;+\;e\;\cos\nu \;\right)\;\;\rangle~~=~~
 \ea
 ~\\
 \ba
 \nonumber
 \frac{\left(1-e^2\right)^{3/2}}{2\;\pi}\;n\,a^2\;\sqrt{1-e^2}\;\sin\inc\;\,
 \left(\;
 \cos \Omega\;\cos2\omega\;-\;\sin \Omega\;\cos \inc\;\sin 2\omega
 \;\right)\,
 \int_0^{2\pi}\frac{\cos 2\nu\,+\,e\,\cos\nu}{\left(1\,+\,e\,\cos\nu
 \right)^3}\,d\nu\,=\,0\;\;,
 \ea
~\\
 \ba
 \langle\;\left(\; \frac{\partial \efbold}{\partial \Omega}\;\times\;{\bf
\vec g}\;-\;\efbold\;\times\;\frac{\partial {\bf \vec g}}{\partial
\Omega}\;
\right)_2\;\rangle\;=~~~~~~~~~~~~~~~~~~~~~~~~~~~~~~~~~~~~~~~~~~~~~~~
~~~~~~~~~~~~~~~~~~~~~~~~~~
 \label{4144}\\
 \nonumber\\
 \nonumber\\
 \nonumber
 \langle\;\;
 \frac{n\;a^2\;\sqrt{1\;-\;e^2}}{1\;+\;e\;\cos\nu}\;\;\sin
 \inc\;\;\left\{\;\;
 \sin \Omega\;\cos 2\omega\;\cos 2\nu\;-\;\sin\Omega\;\sin
 2\omega\;\sin 2\nu ~~~~~~~~~~~~~~~~~~~~~~~~~ \right. \\
 \nonumber\\
 \nonumber\\
 \nonumber
 \left.
 +\;\cos \Omega\;\cos \inc\;\sin 2\omega\;\cos 2\nu\;
 +\;\cos \Omega\;\cos \inc\;\cos 2\omega\;\sin 2\nu
 \;\;
 \right\}~~~~~\,~~~~\\
 \nonumber\\
 \nonumber\\
 \nonumber
 +\;\;\frac{n\;a^2\;\sqrt{1\;-\;e^2}}{1\;+\;e\;\cos\nu}\;\;e\;\;\sin
 \inc\;\;\left\{\;\;
 \sin \Omega\;\cos 2\omega \;\cos\nu\;
 -\;\sin\Omega\;\sin 2\omega \;\sin\nu ~~~~~~~~~~~~~~~~~~~~~~~~ \right. \\
 \nonumber\\
 \nonumber\\
 \nonumber
 \left.
 +\;\cos \Omega\;\cos \inc\;\sin 2\omega \;\cos \nu\;
 +\;2\;\cos \Omega\;\cos \inc\;\cos^2\omega\;\sin\nu
 \right\}\;\;\rangle\;=\;\;
 \ea
~\\
 \ba
 \nonumber
 \langle\;\; \frac{n\;a^2\;\sqrt{1\;-\;e^2}}{1\;+\;e\;\cos\nu}\;\;\sin
 \inc\;\;\left(
\sin\Omega\;\cos 2\omega\;+\;\cos\Omega\;\cos\inc\;\sin 2\omega
 \right)\;\left(\cos 2\nu\;+\;e\;\cos\nu \right)\;\;\rangle\;=\\
 \nonumber\\
 \nonumber\\
 \nonumber
 \frac{\left(1-e^2\right)^{3/2}}{2\;\pi}\;n\,a^2\;\sqrt{1-e^2}\;\sin\inc\;\,
 \left(\;
 \sin \Omega\;\cos2\omega\;+\;\cos \Omega\;\cos \inc\;\sin 2\omega
 \;\right)\,
 \int_0^{2\pi}\frac{\cos 2\nu\,+\,e\,\cos\nu}{\left(1\,+\,e\,\cos\nu
 \right)^3}\,d\nu\,=\,0\;\;,
 \ea
 \ba
 \langle\;\left(\; \frac{\partial \efbold}{\partial \Omega}\;\times\;{\bf
\vec g}\;-\;\efbold\;\times\;\frac{\partial {\bf \vec g}}{\partial
\Omega}\;
\right)_3\;\rangle\;=~~~~~~~~~~~~~~~~~~~~~~~~~~~~~~~~~~~~~~~~~~~~~~~
~~~~~~~~~~~~~~~~~~~~~~~~~~~~~
 \label{4145}
 \ea
 \ba
 \nonumber\\
 \nonumber
 \langle\;\;
 \frac{n a^2 \sqrt{1 - e^2}}{1 + e\;\cos\nu}\;\sin^2
 \inc\;\left(\;\sin 2\omega\;\cos 2\nu\;+\;\cos 2\omega\;\sin 2\nu\;\right)
 \,+ ~~~~~~~~~~~~~~~ \\
 \nonumber\\
 \nonumber\\
 \nonumber
 \frac{n a^2 \sqrt{1 - e^2}}{1 + e\;\cos\nu}\;\;2\;e
 \;\left(\;
 -\,\sin \nu \,+ \,\sin^2\inc\;\cos \omega \;\sin \omega \;\cos \nu
 \;+\;\sin^2\inc\;\cos^2 \omega \;\sin \nu
 \right)\;\;\rangle\;=\\
 \nonumber\\
 \nonumber\\
 \nonumber
 \langle\;\;
  \frac{n a^2 \sqrt{1 - e^2}}{1 + e\;\cos\nu}\;\sin^2
 \inc\;\sin 2 \omega\;\left(\cos 2 \nu\;+\;e\;\cos\nu  \right)
 \;\;\rangle\;=~~~~~~~~~~~~~~~~\\
 \nonumber\\
 \nonumber\\
 \nonumber
  \frac{\left(1-e^2\right)^{3/2}}{2\;\pi}\;n\,a^2\;\sqrt{1-e^2}\;\sin^2\inc\;\,
  \sin2 \omega\,
 \int_0^{2\pi}\frac{\cos 2\nu\,+\,e\,\cos\nu}{\left(1\,+\,e\,\cos\nu
 \right)^3}\,d\nu\,=\,0\;\;.
 \ea
We see that the averages of all three Cartesian components of
$\;\left(\,({\partial \efbold}/{\partial \Omega})\times {\bf\vec
g}\,-\,\efbold\times (\partial{\bf \vec
g}/{\partial\Omega})\,\right)\;$ vanish. In the developments
(\ref{4143} - \ref{4145}) the following integrals were used:
 \ba
 \int_{0}^{2\pi}\frac{\cos \nu}{\left(1\,+\,e\;\cos\nu
 \right)^3}\;=\;-\;\sqrt{\frac{1\,-\,e}{1\,+\,e}}\;\frac{3\;\pi\;e}{\left(1\,-\,e \right)^3
 \,\left(1\,+\,e \right)^2}
 \label{integral3}
 \ea
and
 \ba
 \int_{0}^{2\pi}\frac{\cos 2\nu}{\left(1\,+\,e\;\cos\nu
 \right)^3}\;=\;\sqrt{\frac{1\,-\,e}{1\,+\,e}}\;\frac{3\;\pi\;e^2}{\left(1\,-\,e \right)^3
 \,\left(1\,+\,e \right)^2}\;\;.
 \label{integral4}
 \ea

~\\

 \ba
 \nonumber
 \mbox{{\bf{\large{A.7.~~~Calculation of}}}}
\;\;\;\;\mubold\cdot\left(\; \frac{\partial \efbold}{\partial
\inc}\;\times\;{\bf \vec g}\;-\;\efbold\;\times\;\frac{\partial
{\bf \vec g}}{\partial \inc}\;
\right)\;\;\;.~~~~~~~~~~~~~~~~~~~~~~~~~~~~~~~~~~~~~~~~~~~~~~~~~~~
 \ea\\

Just like the preceding subsections, this one is based on formulae
(\ref{A15}), (\ref{A16}), (\ref{A18}), and (\ref{A19}):

 \ba
\left(\;\frac{\partial \efbold}{\partial \inc}\;\times\;{\bf \vec
g}\;\right)_1\;=\;\frac{\partial f_2}{\partial
\inc}\;g_3\;-\;\frac{\partial f_3}{\partial
\inc}\;g_2\;=~~~~~~~~~~~~~~~~~~~~~~~~~~~~~~~~~~~~~~~~~~~~~~~~~~~~~~~~~~~~
 \label{A58}
 \ea
  ~\\
 \ba
 \nonumber
 \frac{n a^2 \sqrt{1- e^2}}{1 + e\;\cos\nu}\;
 \left\{\;
 \left[\;-\;\sin (\omega + \nu)\;\sin \inc\;\cos \Omega \;\right]\;\;\left[\;\sin\inc\;
 \cos\left(\omega + \nu\right)\;
 \right]\,- \right. ~~~~~~~~~~~~~~~~~~~~~~~~~~~~~~~~~~~~~~~~~~~
  \\
  \nonumber\\
  \nonumber\\
  \nonumber
 \left.
 \left[\;\cos\inc\;\sin(\omega +
 \nu)\;\right]\;\;\left[\;-\;\sin\Omega\;
 \sin\left(\omega + \nu\right)\;+\;\cos\Omega\;\cos(\omega + \nu)\;\cos \inc \right]
 \; \right\}
~~~\\
 \nonumber\\
 \nonumber\\
 \nonumber
+\;\frac{n\;a^2\;\sqrt{1\;-\;e^2}}{1\;+\;e\;\cos\nu}\;\;e\;\;
 \left\{\;
 \left[\;-\;\sin (\omega + \nu)\;\sin \inc\;\cos \Omega \;\right]\;\;\left[\;\sin\inc\;
 \cos\omega \;
 \right]\,- \right. ~~~~~~~~~~~~~~~~~~~~~~~~~~~~~~~~~~~~~~~~~~~
  \\
  \nonumber\\
  \nonumber\\
  \nonumber
 \left.
 \left[\;\cos\inc\;\sin(\omega +
 \nu)\;\right]\;\;\left[\;-\;\sin\Omega\;
 \sin\omega\;+\;\cos\Omega\;\cos\omega\;\cos \inc \right]
 \; \right\}\;=
 ~~~~~~~~
 \ea
 ~\\
 \ba
 \nonumber
 \frac{n a^2 \sqrt{1- e^2}}{1 + e\;\cos\nu}\;\left\{\;
 \left[\;\sin\Omega\;
 \sin^2 (\omega + \nu)\;\cos\inc\;-\;\cos
 \Omega\;\sin\left(\omega + \nu \right)\;\cos\left(\omega +
 \nu\right)
 \;\right]\;+ \right. \\
 \nonumber\\
 \nonumber\\
 \nonumber
 \left. \;e\;\;\sin\left(\omega + \nu \right)\left[\; \sin \Omega\;\sin \omega \;\cos \inc  \;-\;\cos\Omega\;
 \cos \omega\;\right]\;\right\}\;\;,
 \ea
 ~\\
 \ba
\left(\;\frac{\partial \efbold}{\partial \inc}\;\times\;{\bf \vec
g}\;\right)_2\;=\;\frac{\partial f_3}{\partial
\inc}\;g_1\;-\;\frac{\partial f_1}{\partial
\inc}\;g_3\;=~~~~~~~~~~~~~~~~~~~~~~~~~~~~~~~~~~~~~~~~~~~~~~~~~~~~~~~~~~~~
 \label{A59}
 \ea
  ~\\
 \ba
 \nonumber
 \frac{n a^2 \sqrt{1- e^2}}{1 + e\;\cos\nu}\;
 \left\{\;\;
 \left[\;\;\sin (\omega + \nu)\;\cos \inc\;\;\right]\;\;
 \left[\;-\;\cos \Omega \;\sin\left(\omega +
 \nu\right)\;-\;\sin\Omega\;\cos\inc\;\cos\left(\omega + \nu   \right)
 \;\;\right]\,- \right.~~~~~~
  \\
  \nonumber\\
  \nonumber\\
  \nonumber
 \left.
 \left[\;\sin\Omega\;\sin\inc\;\sin(\omega +
 \nu)\;\right]\;\;\left[\;\sin\inc\;
 \cos\left(\omega + \nu\right)\;\right]
 \; \right\}
~~~\\
 \nonumber\\
 \nonumber\\
 \nonumber
+\;\frac{n\;a^2\;\sqrt{1\;-\;e^2}}{1\;+\;e\;\cos\;nu}\;\;e\;\;
 \left\{\;
 \left[\;\sin (\omega + \nu)\;\cos \inc\;\right]\;\;\left[\;-\;\cos \Omega
 \;\sin\omega\;-\;\sin\Omega\;\cos\omega \;\cos\inc
 \;\right]\,- \right. ~~~~~~~~~~~~~~~~~
  \\
  \nonumber\\
  \nonumber\\
  \nonumber
 \left.
 \left[\;\sin\Omega\;\sin(\omega +
 \nu)\;\sin\inc\;\right]\;\;\left[\;\sin\inc\;
 \cos\omega\; \right]
 \; \right\}\;= ~~~~~~~~
 \ea
 ~\\
 \ba
 \nonumber
 \frac{n a^2 \sqrt{1- e^2}}{1 + e\;\cos\nu}\;\left\{\;
 \left[\;-\;\sin\Omega\;
 \sin (\omega + \nu)\;\cos (\omega + \nu)\;-\;\cos
 \Omega\;\sin^2\left(\omega + \nu \right)\;\cos\inc
 \;\right]\;+ \right. \\
 \nonumber\\
 \nonumber\\
 \nonumber
 \left. \;e\;\;\left[\;-\; \sin \Omega\;\sin\left(\omega + \nu \right)\;\cos \omega  \;-\;\cos\Omega\;
 \cos\inc\;\sin \omega\;\sin\left(\omega + \nu \right)\;\right]\;\right\}\;\;,
 \ea
 ~\\
 \ba
\left(\;\frac{\partial \efbold}{\partial \inc}\;\times\;{\bf \vec
g}\;\right)_3\;=\;\frac{\partial f_1}{\partial
\inc}\;g_2\;-\;\frac{\partial f_2}{\partial
\inc}\;g_1\;=~~~~~~~~~~~~~~~~~~~~~~~~~~~~~~~~~~~~~~~~~~~~~~~~~~~~~~~~~~~~
 \label{A60}
 \ea
  ~\\
 \ba
 \nonumber
 \frac{n a^2 \sqrt{1- e^2}}{1 + e\;\cos\nu}\;
 \left\{\;
 \left[\;\sin (\omega + \nu)\;\sin \inc\;\sin \Omega \;\right]\;\;\left[\;-\;\sin\Omega\;
 \sin\left(\omega + \nu\right)\;+\;\cos\Omega\;\cos\inc\;\cos (\omega + \nu)
 \;\right]\,- \right. ~~~~~
  \\
  \nonumber\\
  \nonumber\\
  \nonumber
 \left.
 \left[\;-\;\cos\Omega\;\sin(\omega +
 \nu)\;\sin\inc\;\right]\;\;\left[\;-\;\cos\Omega\;
 \sin\left(\omega + \nu\right)\;-\;\sin\Omega\;\cos(\omega + \nu)\;\cos \inc \;\right]
 \; \right\}
~~~\\
 \nonumber\\
 \nonumber\\
 \nonumber
+\;\frac{n\;a^2\;\sqrt{1\;-\;e^2}}{1\;+\;e\;\cos\nu}\;\;e\;\;
 \left\{\;
 \left[\;\sin (\omega + \nu)\;\sin \inc\;\sin \Omega \;\right]\;\;\left[\;-\;\sin\Omega\;
 \sin\omega \;+\;\cos\Omega\;\cos \inc\;\cos\omega
 \;\right]\,- \right. ~~~~~~~~~~~
  \\
  \nonumber\\
  \nonumber\\
  \nonumber
 \left.
 \left[\;-\;\cos\Omega\;\sin(\omega +
 \nu)\;\sin\inc\;\right]\;\;\left[\;-\;\cos\Omega\;
 \sin\omega\;-\;\sin\Omega\;\cos\omega\;\cos \inc \; \right]
 \; \right\}\;=
 ~~~~~~~~
 \ea
 ~\\
 \ba
 \nonumber
 \frac{n a^2 \sqrt{1- e^2}}{1 + e\;\cos\nu}\;\left\{\;
 \;-\;\sin^2 (\omega + \nu)\;\sin\inc\;\;-\;
\;e\;\;\sin\left(\omega + \nu \right)\; \sin \omega\;\sin  \inc
\;\;\right\}\;\;,
 \ea
 ~\\
 \ba
\left(\;\efbold\;\times\;\frac{\partial {\bf \vec g}}{\partial
\inc}\; \right)_1\;=\;f_2\;\frac{\partial g_3}{\partial
\inc}\;-\;f_3\;\frac{\partial g_2}{\partial
\inc}\;=~~~~~~~~~~~~~~~~~~~~~~~~~~~~~~~~~~~~~~~~~~~
~~~~~~~~~~~~~~~~~~~~
 \label{A61}
 \ea
~\\
 \ba
 \nonumber
 \frac{n\;a^2\;\sqrt{1\;-\;e^2}}{1\;+\;e\;\cos\nu}\;\left\{\;
 \left[\;\sin\Omega\;\cos(\omega + \nu)\;+\;\cos\Omega\;\sin(\omega + \nu)\;\cos \inc \right]\;
 \;\cos(\omega + \nu)\;\cos \inc\;- \right.\\
 \nonumber\\
 \nonumber\\
 \nonumber
 \left. \sin(\omega + \nu)\;\sin \inc\;\;\left[\;-\;\cos \Omega \;\cos(\omega +
 \nu)\;\sin\inc
 \;\right]
 \;\right\}\\
 \nonumber\\
 \nonumber\\
 \nonumber
  +\;\frac{n\;a^2\;\sqrt{1\;-\;e^2}}{1\;+\;e\;\cos\nu}\;\;e\;\;\left\{\;
 \left[\;\sin\Omega\;\cos(\omega + \nu)\;+\;\cos\Omega\;\sin(\omega + \nu)\;
 \cos \inc \right]\;
 \;\cos\omega\;\cos \inc\;-~~ \right.\\
 \nonumber\\
 \nonumber\\
 \nonumber
 \left. \sin(\omega + \nu)\;\sin \inc\;\;\left[\;-\;\cos \Omega \;\cos\omega\;\sin\inc
 \;\right]
 \;\right\}~=\\
  \nonumber\\
 \nonumber\\
 \nonumber
  \frac{n\;a^2\;\sqrt{1\;-\;e^2}}{1\;+\;e\;\cos\nu}\;
  \left\{\;
  \left[\;
  \cos \Omega\;\sin(\omega + \nu)\;\cos(\omega +
  \nu)\;+\;\sin\Omega\;\cos^2(\omega + \nu)\;\cos \inc
  \;\right]\;+ \right.\\
 \nonumber\\
 \nonumber\\
 \nonumber
\left.  e\;\;
  \left[\;
 \cos \Omega\;\sin(\omega + \nu)\;\cos \omega\;+\;\sin\Omega\;\cos(\omega +
 \nu)\;\cos \omega\;\cos\inc
  \;\right]
  \;\right\}\;\;,
 \ea
 ~\\
 \ba
\left(\;\efbold\;\times\;\frac{\partial {\bf \vec g}}{\partial
\inc}\; \right)_2\;=\;f_3\;\frac{\partial g_1}{\partial
\inc}\;-\;f_1\;\frac{\partial g_3}{\partial
\inc}\;=~~~~~~~~~~~~~~~~~~~~~~~~~~~~~~~~~~~~~~~~~~~~
~~~~~~~~~~~~~~~~~~
 \label{A62}
 \ea
~\\
 \ba
 \nonumber
 \frac{n\;a^2\;\sqrt{1\;-\;e^2}}{1\;+\;e\;\cos\nu}\;\left\{\;
 \;\sin(\omega + \nu)\;\sin\inc\;\;\sin\Omega\;\sin \inc\;\cos(\omega + \nu)
 \;- \right.~~~~~~~~~~~~~~~~~~~~~~~~~~~~~~\\
 \nonumber\\
 \nonumber\\
 \nonumber
 \left.
 \left[\;\cos\Omega\;\cos(\omega + \nu)\;-\;\sin\Omega\;\sin(\omega + \nu)\;
 \cos \inc \right]\;
 \;\cos(\omega + \nu)\;\cos \inc\; \right\}\\
 \nonumber\\
 \nonumber\\
 \nonumber
  +\;\frac{n\;a^2\;\sqrt{1\;-\;e^2}}{1\;+\;e\;\cos\nu}\;\;e\;\;\left\{\;
 \;\sin(\omega + \nu)\;\sin\inc\;\;\sin\Omega\;\sin \inc\;\cos\omega
 \;- \right.~~~~~~~~~~~~~~~~~~~~~~~~~~~~~~\\
 \nonumber\\
 \nonumber\\
 \nonumber
 \left.
 \left[\;\cos\Omega\;\cos(\omega + \nu)\;-\;\sin\Omega\;\sin(\omega + \nu)\;
 \cos \inc \right]\;
 \;\cos\omega\;\cos \inc\; \right\}
 ~=\\
  \nonumber\\
 \nonumber\\
 \nonumber
  \frac{n\;a^2\;\sqrt{1\;-\;e^2}}{1\;+\;e\;\cos\nu}\;
  \left\{\;
  \left[\;
  \sin \Omega\;\sin(\omega + \nu)\;\cos(\omega +
  \nu)\;-\;\cos\Omega\;\cos^2(\omega + \nu)\;\cos \inc
  \;\right]\;+ \right.\\
 \nonumber\\
 \nonumber\\
 \nonumber
\left.  e\;\;
  \left[\;
 \sin \Omega\;\sin(\omega + \nu)\;\cos \omega\;-\;\cos\Omega\;\cos(\omega +
 \nu)\;\cos \omega\;\cos\inc
  \;\right]
  \;\right\}\;\;,
 \ea
 ~\\
 \ba
 \left(\;\efbold\;\times\;\frac{\partial {\bf \vec g}}{\partial
 \inc}\; \right)_3\;=\;f_1\;\frac{\partial g_2}{\partial
 \inc}\;-\;f_2\;\frac{\partial g_1}{\partial
 \inc}\;=~~~~~~~~~~~~~~~~~~~~~~~~~~~~~~~~~~~~~~~~~~~
 ~~~~~~~~~~~~~~~~~~
 \label{A63}
 \ea
 ~\\
 \ba
 \nonumber
 \frac{na^2\,\sqrt{1-e^2}}{1+e\;\cos\nu}\;\left\{\;
 \left[\;
 \cos\Omega\;\cos(\omega + \nu)\,-\,\sin\Omega\;\sin(\omega + \nu)\;\cos\inc\;
 \;\right]\;\;\left[\;-\,
 \cos\Omega\;\sin \inc\;\cos(\omega + \nu)
 \;\right]\;- \right.\\
 \nonumber\\
 \nonumber\\
 \nonumber
 \left.
 \left[\;\sin\Omega\;\cos(\omega + \nu)\;+\;\cos\Omega\;\sin(\omega + \nu)\;
 \cos \inc \right]\;\sin\Omega\;\cos(\omega + \nu)\;\cos \inc\; \right\}\\
 \nonumber\\
 \nonumber\\
 \nonumber
 +\;\frac{na^2\,\sqrt{1-e^2}}{1+e\;\cos\nu}\;\;e\;\;\left\{\;
 \left[\;
 \cos\Omega\;\cos(\omega + \nu)\,-\,\sin\Omega\;\sin(\omega + \nu)\;\cos\inc\;
 \;\right]\;\;\left[\;-
 \,\cos\Omega\;\sin \inc\;\cos\omega
 \;\right]\;- \right.\\
 \nonumber\\
 \nonumber\\
 \nonumber
 \left.
 \left[\;\sin\Omega\;\cos(\omega + \nu)\;+\;\cos\Omega\;\sin(\omega + \nu)\;
 \cos \inc \right]\;\sin\Omega\;\cos\omega \;\cos \inc
 \;\right\}~=\\
 \nonumber\\
 \nonumber\\
 \nonumber
 \frac{n\;a^2\;\sqrt{1\;-\;e^2}}{1\;+\;e\;\cos\nu}\;
 \left\{\;
 \;-\;\sin\inc\;\cos^2(\omega +
 \nu)\;-\;e\;\sin\inc\;\cos(\omega + \nu)\;\cos \omega
 \;\right\}~~~.~~~~~~~~~~~~~~~~~~~~~~~~
 \ea
 ~\\
 Now, by putting together, accordingly, (\ref{A58}) with (\ref{A61}), (\ref{A59}) with
 (\ref{A62}), and (\ref{A60}) with (\ref{A63}), we arrive, after some intermediate
 algebra, to the following expressions for the three Cartesian components:
 \ba
 \left(\; \frac{\partial \efbold}{\partial \inc}\;\times\;{\bf
 \vec g}\;-\;\efbold\;\times\;\frac{\partial {\bf \vec g}}{\partial
 \inc}\; \right)_1\;=\;
 \label{A64}
 \ea
 ~\\
 \ba
 \nonumber
 \frac{n a^2 \sqrt{1- e^2}}{1 + e\;\cos\nu}\;\left\{\;
 \left[\;\sin\Omega\;\cos\inc\;\left(\;
 \sin^2 (\omega + \nu)\;-\;\cos^2 (\omega + \nu)\;\right)
 -\;2\;\cos
 \Omega\;\sin\left(\omega + \nu \right)\;\cos\left(\omega +
 \nu\right)
 \;\right]\;+ \right. \\
 \nonumber\\
 \nonumber\\
 \nonumber
 \left. \;e\;\;\left[\; \sin \Omega\;\cos\inc\;\left(\;-\;\cos\left(\omega + \nu \right)\;\cos\omega\;
 +\;\sin\left(\omega + \nu \right)\;\sin\omega\;\right)\;-\;2\;\cos \Omega\;\sin\left(\omega + \nu \right)\;
 \cos\omega\;\right]\;\right\}\;=
 \ea
 ~\\
 \ba
 \nonumber
 \frac{n a^2 \sqrt{1- e^2}}{1 + e\;\cos\nu}\;\left\{\;
 \left[\;
 \left(\;-\;\sin\Omega\;\cos\inc\;\cos2\omega\;-\;\cos\Omega\;\sin 2\omega\;\right)
 \;\left(\;\cos^2 \nu\;-\;\sin^2 \nu\;\right)\;+~~~~~~~~~~~~~~~\right.\right.\\
 \nonumber\\
 \nonumber\\
 \nonumber
 \left. 2\;\sin \nu\;\cos\nu\;
 \left(\;
 \sin \Omega\;\cos\inc\;\sin 2\omega\;-\;\cos\Omega\;\cos 2\omega
 \;\right)
 \;\right]\;+ ~~~~~~~ \\
 \nonumber\\
 \nonumber\\
 \nonumber
 \left. e\;\,\left[\;
 \left(\;
 -\;\sin \Omega\;\cos\inc\;\cos 2\omega \,-\,\cos\Omega\;\sin
 2\omega\;\right)\;\cos\nu\;
 +\;\left(\;
 \sin\Omega \;\cos\inc\;\sin 2\omega\;-\;2\;\cos \Omega\;\cos^2 \omega \right)\;
 \sin\nu\;\right]\;\right\}\;\;,
 \ea
~\\
  \ba
 \left(\; \frac{\partial \efbold}{\partial \inc}\;\times\;{\bf
\vec g}\;-\;\efbold\;\times\;\frac{\partial {\bf \vec g}}{\partial
\inc}\; \right)_2\;=\;
 \label{A65}
 \ea
  ~\\
 \ba
 \nonumber
 \frac{n a^2 \sqrt{1- e^2}}{1 + e\;\cos\nu}\;\left\{\;
 \left[\;\cos\Omega\;\cos\inc\;\left(\;
 \cos^2 (\omega + \nu)\;-\;\sin^2 (\omega + \nu)\;\right)
 -\;2\;\sin
 \Omega\;\sin\left(\omega + \nu \right)\;\cos\left(\omega +
 \nu\right)
 \;\right]\;+ \right. \\
 \nonumber\\
 \nonumber\\
 \nonumber
 \left. \;e\;\;\left[\; \cos \Omega\;\cos\inc\;\left(\;\cos\left(\omega + \nu \right)\;\cos\omega\;
 -\;\sin\left(\omega + \nu \right)\;\sin\omega\;\right)\;-\;
 2\;\sin \Omega\;\sin\left(\omega + \nu \right)\;
 \cos\omega\;\right]\;\right\}\;=
 \ea
 ~\\
 \ba
 \nonumber
 \frac{n a^2 \sqrt{1- e^2}}{1 + e\;\cos\nu}\;\left\{\;
 \left[\;
 \left(\;\cos\Omega\;\cos\inc\;\cos2\omega\;-\;\sin\Omega\;\sin 2\omega\;\right)
 \;\left(\;\cos^2 \nu\;-\;\sin^2 \nu\;\right)\;+~~~~~~~~~~~~~~~~~~\right.\right.\\
 \nonumber\\
 \nonumber\\
 \nonumber
 \left. 2\;\sin \nu\;\cos\nu\;
 \left(\;-\;
 \cos \Omega\;\cos\inc\;\sin 2\omega\;-\;\sin\Omega\;\cos 2\omega
 \;\right)
 \;\right]\;+  ~~~~~~~\\
 \nonumber\\
 \nonumber\\
 \nonumber
 \left. e\;\,\left[\;
 \left(\;\cos \Omega\;\cos\inc\;\cos
 2\omega\,-\,\sin\Omega\;\sin2\omega\;\right)\;\cos\nu\,+\,
 \left(\;
-\;2\;\sin\Omega\;\cos^2\omega\,+\,\cos\Omega\;\sin2\omega\;\cos\inc
 \;\right)\;\sin\nu
 \;\right]\;\right\}\;\;,
 \ea
~\\
 \ba
 \left(\; \frac{\partial \efbold}{\partial \inc}\;\times\;{\bf
\vec g}\;-\;\efbold\;\times\;\frac{\partial {\bf \vec g}}{\partial
\inc}\; \right)_3\;=\;
 \label{A66}
 \ea
 ~\\
 \ba
 \nonumber
 \frac{n a^2 \sqrt{1- e^2}}{1 + e\;\cos\nu}\;\left\{\;
 \sin\inc\;\left[\;
 \cos^2 (\omega + \nu)\;-\;\sin^2 (\omega + \nu)\;\right]
 \;+~~~~~~~~~~~~~~~~~~~~~~~~~~~~~~~~~~~~~~~~~~~~~~~~~~~~~~~~~~~~~~~
  \right. \\
  \nonumber\\
  \nonumber\\
  \nonumber
  \left.
 e\;\;\left[\; \sin\inc\;\cos \omega\;\cos(\omega +
 \nu)\;-\;\sin\inc\;\sin \omega\;\sin (\omega + \nu)\;\right]\;\right\}\;=~~~~~~~~~\\
 \nonumber\\
 \nonumber\\
 \nonumber
 \frac{n a^2 \sqrt{1- e^2}}{1 + e\;\cos\nu}\;\left\{\;
 \sin\inc\;\left[\;
 \cos 2\omega \;\left(\;\cos^2\nu\;-\;\sin^2\nu\;\right)\;-\;2\;\sin
 2\omega\;\sin\nu\;\cos\nu
 \;\right]
 \;+~~~~~~~~~~~~~~~~~~~~~~~~~~~~~~~~~
  \right. \\
  \nonumber\\
  \nonumber\\
  \nonumber
  \left.
 e\;\;\left[\; \sin\inc\;\cos 2\omega\;\cos
 \nu\;-\;\sin\inc\;\sin 2\omega\;\sin
 \nu\;\right]\;\right\}\;\;.~~~~~~~~~~~~~~~~~~~~~~~~~
 \ea
Averaging of the afore calculated three Cartesian components of
$\;\left(\,({\partial \efbold}/{\partial \inc})\times {\bf\vec
g}\,-\,\efbold\times (\partial{\bf \vec
g}/{\partial\inc})\,\right)\;$ almost exactly coincides with
averaging of $\;\left(\,({\partial \efbold}/{\partial
\Omega})\times {\bf\vec g}\,-\,\efbold\times (\partial{\bf \vec
g}/{\partial\Omega})\,\right)\;$ presented in the preceding
subsection, and the result is the same, nil. Indeed, as can be
easily seen from (\ref{A64} - \ref{A66}), after we through out the
(vanishing after averaging) inputs proportional to the odd
functions $\;\sin\nu\;$ and $\;\sin 2\nu\;$, each of the three
expressions (\ref{A64} - \ref{A66}) will be proportional to
$\;(\cos 2 \nu\;+\;e\;\cos\nu)/(1\;+\;e\;\cos\nu)\;$. Averaging of
this expression via formula (\ref{A14}) will vanish, as can be
easily seen from (\ref{integral3} - \ref{integral4}).

 ~\\

 \ba
 \nonumber
 \mbox{{\bf{\large{A.8.~~~Calculation of}}}}
\;\;\;\;\mubold\cdot\left(\; \frac{\partial \efbold}{\partial
M_o}\;\times\;{\bf \vec g}\;-\;\efbold\;\times\;\frac{\partial
{\bf \vec g}}{\partial M_o}\;
\right)\;\;\;.~~~~~~~~~~~~~~~~~~~~~~~~~~~~~~~~~~~~~~~~~~~~~~~~~~~
 \ea\\

This calculation will be the easiest: as
 \ba
 \frac{\partial \efbold}{\partial M_o}\;=\;\frac{\partial \efbold}{\partial t}\;
 \frac{\partial t}{\partial M_o}\;=\;{\bf\vec g}\;\frac{\partial t}{\partial M_o}
 \;\;\;\;\;\;\;\;\;\;\;\;\;\;\;\;\;\;\;\;
 \label{A67}
 \ea
and
  \ba
 \frac{\partial \bf \vec g}{\partial M_o}\;=\;\frac{\partial \bf \vec g}{\partial t}\;
 \frac{\partial t}{\partial M_o}\;=\;\left(\;-\;\frac{G\;m}{|\efbold|^3}{\efbold}\;\right)
 \;\frac{\partial t}{\partial M_o}\;\;\;\;,
 \label{A68}
 \ea
then, obviously,
 \ba
 \mubold\cdot\left(\frac{\partial \efbold}{\partial M_o}\;\times\;{\bf
\vec g}\;-\;\efbold\;\times\;\frac{\partial {\bf \vec g}}{\partial
M_o}\right)\;=\;\mubold\cdot\left(0\;-\;0\right)\;=\;0\;\;\;.
 \label{A69}
 \ea

 ~\\

 \ba
 \nonumber
 \mbox{{\bf{\large{A.9.~~~Calculation of}}}}
 \;\;\;\;\dotmubold\cdot\left(\;-\;\efbold\;\times\;
 \frac{\partial {\efbold}}{\partial a}\;
 \right)\;\;\;.~~~~~~~~~~~~~~~~~~~~~~~~~
 ~~~~~~~~~~~~~~~~~~~~~~~~~~
 \ea\\

With aid of (\ref{A26}) and (\ref{50}) we establish:
 \ba
 \nonumber
 -\;\efbold\,\times\,\frac{\partial {\efbold}}{\partial
 a}\,=\,-
 \;\efbold\,\times\,{\bf\vec g}\;\left(\frac{\partial t}{\partial E}\right)_{a,\,e,\,M_o}
 \left(\frac{\partial E}{\partial
 a}\right)_{t,\,e,\,M_o} =~~~~~~~~~~~~~~~~~~~~~~~~~~~~~~~~~~~~~~~~~~\\
 \label{A70}\\
 \nonumber\\
 \nonumber
 -\;{\bf{\vec w}}\;
 \sqrt{Gma\,\left(1-e^2\right)}\;\;\frac{1}{n}\;\frac{1-e^{2}}{1+
 e\;\cos \nu}\;\left(\frac{\partial E}{\partial
 a}\right)_{t,\,e,\,M_o}\;=\;-\;{\bf{\vec w}}\;\;a^2\;\frac{\left(1-e^{2}\right)^{3/2}}{1+
 e\;\cos \nu}\;\left(\frac{\partial E}{\partial
 a}\right)_{t,\,e,\,M_o}\;\;\;\;
 \ea
 where we used the formula
 \ba
 \left(\,\frac{\partial t}{\partial
 E}\,\right)_{a,\,e,\,M_o}\;=\;\frac{1\;-\;e\;\cos E}{n}\;=\;
 \frac{1}{n}\;\;\frac{1\,-\,e^2}{1\,+\,e\;\cos \nu}~~~~~~~~~~~~
 \label{A71}
 \ea
 that follows from the Kepler equation and from the first equation of (\ref{A-1}).

 It remains to compute $\;{\partial E}/{\partial a} \;$. From the Kepler
 equation
 \ba
 E\;-\;e\sin E\;=\;M_o\;+\;n\;\left(\;t\;-\;t_o\;\right)
 \label{A72}
 \ea
 it follows that
 \ba
 dE\;\left(\,1\;-\;e\;\cos E\,\right)\;=\;-\;\frac{3}{2}\;\;n\;\;a^{-1}\;\;
 (t\;-\;t_o)\;da
 \label{A73}
 \ea
 wherefrom
  \ba
  \nonumber
  \left(\frac{\partial E}{\partial a}\right)_{t,\,e,\,M_o}\;=
  \;-\;\frac{3}{2}\;\;a^{-1}\;\;
  \frac{\;n\;\left(\;t\;-\;t_o\;\right)}{1\;-\;e\;\cos E}\;=\;
  \;-\;\frac{3}{2}\;\;a^{-1}\;\;
  \frac{\;E\;-\;e\;\sin E\;-\;M_o\;}{1\;-\;e\;\cos E} \\
  \nonumber\\
  \label{A74}\\
  \nonumber
  =\;-\;\frac{3}{2}\;\;n\;\;a^{-1}\;\;\frac{1\;+\;\cos
  \nu}{1\;-\;e^2}\;\;\left(\;t\;-\;t_o\;\right)\;\;\;.\;\;\;\;
  ~~~~~~~~~~~~~~~~~
  \ea
 Insertion of the above into (\ref{A70}) will give:
 \ba
  \nonumber
 -\;\efbold\,\times\,\frac{\partial {\efbold}}{\partial
 a}\,=\;\frac{3}{2}\,\;{\bf{\vec
 w}}\;\;a\;n\;\left(\;t\;-\;t_o\;\right)\;\sqrt{1\;-\;e^{2}}
 \;\;\;,
 \label{A75}
 \ea
expression linear in $\;t\;-\;t_o\;$.


~\\

 \ba
 \nonumber
 \mbox{{\bf{\large{A.10.~~~Calculation of}}}}
 \;\;\;\;\dotmubold\cdot\left(\;-\;\efbold\;\times\;
 \frac{\partial {\efbold}}{\partial e}\;
 \right)\;\;\;.~~~~~~~~~~~~~~~~~~~~~~~~~
 ~~~~~~~~~~~~~~~~~~~~~~~~~~
 \ea\\

Repeating the line of reasoning presented in the preceding
calculation, we derive from (\ref{A34}), (\ref{45}), and
(\ref{A71}) the following:
 \ba
 -\;\efbold\,\times\,\frac{\partial {\efbold}}{\partial
 e}\,=\,-\;\efbold\,\times\,{\bf\vec g}\;\;
\left( \frac{\partial t}{\partial E}\right)_{a,\,e,\,M_o}
 \;\left(\frac{\partial E}{\partial
 e}\right)_{t,\,e,\,M_o}
  =\,-\;{\bf{\vec w}}\;\;a^2
 \;\;\frac{\left(1-e^{2}\right)^{3/2}}{1+
 e\;\cos \nu}\;\left(\frac{\partial E}{\partial
 e}\right)_{t,\,a,\,M_o}\;\;\;.\;\;\;\;
 \label{A76}
 \ea
 To compute $\;{\partial E}/{\partial e}\;$, one can start with
 the Kepler equation,
 \ba
 E\;-\;e\;\sin E\;=\;M_o\;+\;n\;\left(t\;-\;t_o\right)
 \label{A77}
 \ea
which yields, for $\;t,\;a,\;M_o\;$ being fixed:
 \ba
 dE\;-\;e\;\cos E\;dE\;-\;\sin E\;de\;=\;0\;\;\;,
 \label{A78}
 \ea
insertion of (\ref{A-1}) wherein will entail
 \ba
 \left(\frac{\partial E}{\partial e}\right)_{t,\,a,\,M_o}\;=\;
 \frac{\sin E}{1\;-\;e\;\cos E}\;=\;\frac{\sin
 \nu}{\sqrt{1\;-\;e^2}}\;\;\;.
 \label{A79}
 \ea
All in all,
  \ba
 -\;\efbold\,\times\,\frac{\partial {\efbold}}{\partial
 e}\,=\;
-\;{\bf{\vec w}}\;\;a^2
 \;\;\frac{\left(1-e^{2}\right)}{1+
 e\;\cos \nu}\;\sin \nu\;\;\;.\;\;\;\;
 \;\;\;.\;
 \label{A80}
 \ea

~\\

 \ba
 \nonumber
 \mbox{{\bf{\large{A.11~~~Calculation of}}}}
 \;\;\;\;\dotmubold\cdot\left(\;-\;\efbold\;\times\;
 \frac{\partial {\efbold}}{\partial \omega}\;
 \right)\;\;\;.~~~~~~~~~~~~~~~~~~~~~~~~~
 ~~~~~~~~~~~~~~~~~~~~~~~~~~
 \ea\\

Evidently, this vector is perpendicular to the instantaneous
orbital plane and is $\;\omega$-independent. A direct computation,
based upon (\ref{A20} - \ref{A22}), will lead us to:
 \ba
 \nonumber
 \left(\;\frac{\partial \efbold}{\partial \omega}\;\times\;\efbold\;\right)_1\;=\;
 \frac{\partial f_2}{\partial \omega}\;f_3\;-\;\frac{\partial f_3}{\partial
 \omega}\;f_2\;=~~~~~~~~~~~~~~~~~\\
 \label{A81}\\
 \nonumber
 r^2\;\left[\;-\;\sin\Omega\;\sin(\omega +\nu)\;+\;
  \cos \Omega\;\cos(\omega +\nu)\;\cos \inc \;\right]\;\sin(\omega + \nu)\;\sin \inc\;-\\
  \nonumber\\
  \nonumber
 r^2\;\cos (\omega + \nu)\;\sin \inc\;\left[\; \sin\Omega\;\cos(\omega +\nu)\;+\;
  \cos \Omega\;\sin(\omega +\nu)\;\cos \inc \;\right]\;=\;\;\;\;\\
  \nonumber\\
  \nonumber
a^2\;\;\frac{\left(1\;-\;e^2\right)^2}{\left(1\;+\;e\;\cos
\nu\right)^2}\;\;\left[\;-\;\sin \Omega\;\sin \inc
\;\right]\;\;\;,
 \ea
~\\
 \ba
 \nonumber
 \left(\;\frac{\partial \efbold}{\partial \omega}\;\times\;\efbold\;\right)_2\;=\;
 \frac{\partial f_3}{\partial \omega}\;f_1\;-\;\frac{\partial f_1}{\partial
 \omega}\;f_3\;=~~~~~~~~~~~~~~~~~\\
 \label{A82}\\
 \nonumber\\
 \nonumber
 r^2\;\cos(\omega + \nu)\;\sin \inc\;\left[\;\cos\Omega\;\cos(\omega +\nu)\;-\;
  \sin \Omega\;\sin(\omega +\nu)\;\cos \inc \;\right]\;-~~~~~~~\\
  \nonumber\\
  \nonumber
 r^2\;\left[\;-\; \cos\Omega\;\sin(\omega +\nu)\;-\;
  \sin \Omega\;\cos(\omega +\nu)\;\cos \inc \;\right]\;\sin (\omega + \nu)\;\sin \inc\;=\;\;\;\;\\
  \nonumber\\
  \nonumber
a^2\;\;\frac{\left(1\;-\;e^2\right)^2}{\left(1\;+\;e\;\cos
\nu\right)^2}\;\;\left[\;\cos \Omega\;\sin \inc \;\right]\;\;\;,
 \ea
~\\
 \ba
 \nonumber
 \left(\;\frac{\partial \efbold}{\partial \omega}\;\times\;\efbold\;\right)_3\;=\;
 \frac{\partial f_1}{\partial \omega}\;f_2\;-\;\frac{\partial f_2}{\partial
 \omega}\;f_1\;=\\
 \label{A83}
 \ea
 \ba
 \nonumber
 r^2\;\left[\;-\; \cos\Omega\;\sin(\omega +\nu)\;-\;
  \sin \Omega\;\cos(\omega +\nu)\;\cos \inc \;\right]\;\left[\; \sin\Omega\;\cos(\omega +\nu)\;+\;
  \cos \Omega\;\sin(\omega +\nu)\;\cos \inc \;\right]\;-\\
  \nonumber\\
  \nonumber
  r^2\;\left[\;-\;\sin\Omega\;\sin(\omega +\nu)\;+\;
  \cos \Omega\;\cos(\omega +\nu)\;\cos \inc \;\right]\;\left[\;\cos\Omega\;\cos(\omega +\nu)\;-\;
  \sin \Omega\;\sin(\omega +\nu)\;\cos \inc \;\right]\;\;\;\\
  \nonumber\\
  \nonumber
 =\;-\;r^2\;\cos \inc\;=\;-\;a^2\;\;\frac{\left(1\;-\;e^2\right)^2}{\left(1\;+\;e\;\cos
\nu\right)^2}\;\cos
\inc~~~.~~~~~~~~~~~~~~~~~~~~~~~~~~~~~~~~~~~~~~~~~~~~~~~~~~~
 \ea
Briefly,
 \ba
 -\;\left(\;\efbold\;\times\;\frac{\partial \efbold}{\partial
 \omega}\;\right)\;=\;-\;{\bf{\vec{w}}}\;\;a^2\;\;\frac{\left(1\;-\;e^2\right)^2}{\left(1\;+\;e\;\cos
\nu\right)^2}\;\;\;\;.~~~~~~~~
 \label{briefly}
 \ea

~\\

 \ba
 \nonumber
 \mbox{{\bf{\large{A.12.~~~Calculation of}}}}
 \;\;\;\;\dotmubold\cdot\left(\;-\;\efbold\;\times\;
 \frac{\partial {\efbold}}{\partial \Omega}\;
 \right)\;\;\;.~~~~~~~~~~~~~~~~~~~~~~~~~
 ~~~~~~~~~~~~~~~~~~~~~~~~~~
 \ea\\

 \ba
 \nonumber
 \left(\;\frac{\partial \efbold}{\partial \Omega}\;\times\;\efbold\;\right)_1\;=\;
 \frac{\partial f_2}{\partial \Omega}\;f_3\;-\;\frac{\partial f_3}{\partial
 \Omega}\;f_2\;=~~~~~~~~~~~~~~~~~\\
 \label{A84}\\
 \nonumber
 r^2\;\left[\;
           \cos\Omega\;\cos(\omega +\nu)\;-\;
           \sin \Omega\;\sin(\omega +\nu)\;\cos \inc
  \;\right]\;\sin(\omega + \nu)\;\sin \inc\;=\;\;\;\;~~~~~~~~~~~~~\\
  \nonumber\\
  \nonumber
a^2\;\;\frac{\left(1\;-\;e^2\right)^2}{\left(1\;+\;e\;\cos
\nu\right)^2}\;\;\left[\;
           \cos\Omega\;\cos(\omega +\nu)\;-\;
           \sin \Omega\;\sin(\omega +\nu)\;\cos \inc
  \;\right]\;\sin(\omega + \nu)\;\sin \inc\;\;\;\;,
 \ea
~\\
 \ba
 \nonumber
 \left(\;\frac{\partial \efbold}{\partial \Omega}\;\times\;\efbold\;\right)_2\;=\;
 \frac{\partial f_3}{\partial \Omega}\;f_1\;-\;\frac{\partial f_1}{\partial
 \Omega}\;f_3\;=~~~~~~~~~~~~~~~~~\\
 \label{A85}\\
 \nonumber\\
 \nonumber
 -\;r^2\;\left[\;
                -\; \sin\Omega\;\cos(\omega +\nu)\;-\;
                \cos \Omega\;\sin(\omega +\nu)\;\cos \inc
  \;\right]\;\sin (\omega + \nu)\;\sin \inc\;=\;\;\;\;~~~~~~~~~~~~\\
  \nonumber\\
  \nonumber
a^2\;\;\frac{\left(1\;-\;e^2\right)^2}{\left(1\;+\;e\;\cos
\nu\right)^2}\;\;\left[\;
                 \sin\Omega\;\cos(\omega +\nu)\;+\;
                \cos \Omega\;\sin(\omega +\nu)\;\cos \inc
                   \;\right]\;\sin (\omega + \nu)\;\sin \inc\;\;\;,
 \ea
~\\
 \ba
 \nonumber
 \left(\;\frac{\partial \efbold}{\partial \Omega}\;\times\;\efbold\;\right)_3\;=\;
 \frac{\partial f_1}{\partial \Omega}\;f_2\;-\;\frac{\partial f_2}{\partial
 \Omega}\;f_1\;=\\
 \label{A86}
 \ea
 \ba
 \nonumber
 r^2\;\left[\;
           -\;\sin\Omega\;\cos(\omega +\nu)\;-\;
           \cos \Omega\;\sin(\omega +\nu)\;\cos \inc
  \;\right]\;\left[\;
           \sin\Omega\;\cos(\omega +\nu)\;+\;
           \cos \Omega\;\sin(\omega +\nu)\;\cos \inc
  \;\right]\;-\\
  \nonumber\\
  \nonumber
  r^2\;\left[\;
           \cos\Omega\;\cos(\omega +\nu)\;-\;
           \sin \Omega\;\sin(\omega +\nu)\;\cos \inc
  \;\right]\;\left[\;
           \cos\Omega\;\cos(\omega +\nu)\;-\;
           \sin \Omega\;\sin(\omega +\nu)\;\cos \inc
  \;\right]\;=\;\;\;\\
  \nonumber\\
  \nonumber
 r^2\,\left[\,-\,\cos^2(\omega + \nu)\,-\,\sin^2(\omega +
 \nu)\;\cos^2\inc
 \,\right]\,=
 -\;a^2\;\frac{\left(1 - e^2\right)^2}{\left(1 + e\;\cos
 \nu\right)^2}\;\left[\,\cos^2(\omega + \nu)\,+\,\sin^2(\omega +
 \nu)\;\cos^2\inc \,\right]~~.
 \ea

~\\

 \ba
 \nonumber
 \mbox{{\bf{\large{A.13.~~~Calculation of}}}}
 \;\;\;\;\mubold\cdot\left(\;-\;\efbold\;\times\;
 \frac{\partial {\efbold}}{\partial \inc}\;
 \right)\;\;\;.~~~~~~~~~~~~~~~~~~~~~~~~~
 ~~~~~~~~~~~~~~~~~~~~~~~~~~
 \ea\\

 \ba
 \nonumber
 \left(\;\frac{\partial \efbold}{\partial \inc}\;\times\;\efbold\;\right)_1\;=\;
 \frac{\partial f_2}{\partial \inc}\;f_3\;-\;\frac{\partial f_3}{\partial
 \inc}\;f_2\;=~~~~~~~~~~~~~~~~~\\
 \label{A87}\\
 \nonumber\\
 \nonumber
 r^2\; \left\{ \;
 \left[ \;
           -\;\cos \Omega \; \sin ( \omega + \nu ) \; \sin \inc
  \; \right] \; \sin ( \omega + \nu )\;\sin \inc\;-
  ~~~~~ ~~~~~~~~~~~~~~~~~~~~~~~~~~~~~~~~~~~~~~
  \right. \\
  \nonumber\\
  \nonumber
  \left.
  \sin(\omega + \nu)\;\cos \inc\;
  \left[\;\sin\Omega\;\cos(\omega + \nu)\;+\;\cos\Omega\;\sin(\omega + \nu)\;\cos \inc \;\right]
  \;\right\}\;=\;\;\;\;~~~~~~~~~~~~~
  \nonumber\\
  \nonumber\\
  \nonumber
a^2\;\;\frac{\left(1\;-\;e^2\right)^2}{\left(1\;+\;e\;\cos
\nu\right)^2}\;\;\left[\;
           -\;\cos\Omega\;\sin(\omega +\nu)\;-\;
           \sin \Omega\;\cos(\omega +\nu)\;\cos \inc
  \;\right]\;\sin(\omega + \nu)\;\;\;\;\;,
 \ea
~\\
 \ba
 \nonumber
 \left(\;\frac{\partial \efbold}{\partial \inc}\;\times\;\efbold\;\right)_2\;=\;
 \frac{\partial f_3}{\partial \inc}\;f_1\;-\;\frac{\partial f_1}{\partial
 \inc}\;f_3\;=~~~~~~~~~~~~~~~~~\\
 \label{A88}\\
 \nonumber\\
 \nonumber
 r^2\;\left\{\;
 \sin(\omega + \nu)\;\cos\inc\;\left[
                \;\cos\Omega\;\cos(\omega +\nu)\;-\;
                \sin \Omega\;\sin(\omega +\nu)\;\cos \inc
  \;\right]\;- ~~~~~~~~~~~~~~~~~~  \right. \\
  \nonumber\\
  \nonumber\\
  \nonumber
  \left.
  \sin\Omega\;\sin(\omega + \nu)\;\sin \inc
  \; \sin (\omega + \nu)\;\sin \inc\;
  \;\right\}
  =\;\;\;\;~~~~~~~~~~~~~~~\\
  \nonumber\\
  \nonumber\\
  \nonumber
  a^2\;\;\frac{\left(1\;-\;e^2\right)^2}{\left(1\;+\;e\;\cos \nu\right)^2}\;\;
 \left[\;
          -\;\sin\Omega\;\sin(\omega +\nu)\;+\;
                    \cos \Omega\;\cos(\omega +\nu)\;\cos \inc
                   \;\right]\;\sin (\omega + \nu)\;\;\;\;,
 \ea
~\\
 \ba
 \nonumber
 \left(\;\frac{\partial \efbold}{\partial \inc}\;\times\;\efbold\;\right)_3\;=\;
 \frac{\partial f_1}{\partial \inc}\;f_2\;-\;\frac{\partial f_2}{\partial
 \inc}\;f_1\;=\\
 \label{A89}
 \ea
 \ba
 \nonumber
 r^2\;\left[\;
           \sin\Omega\;\sin(\omega +\nu)\;\sin\inc
  \;\right]\;\left[\;
           \sin\Omega\;\cos(\omega +\nu)\;+\;
           \cos \Omega\;\sin(\omega +\nu)\;\cos \inc
  \;\right]\;-~~~~~~~~~~~~~~~~~~~~\\
  \nonumber\\
  \nonumber
  r^2\;\left[\;-\;
           \cos \Omega\;\sin(\omega +\nu)\;\sin \inc
  \;\right]\;\left[\;
           \cos\Omega\;\cos(\omega +\nu)\;-\;
           \sin \Omega\;\sin(\omega +\nu)\;\cos \inc
  \;\right]\;=\;\;\;~~~~~~~~~~~~~\\
  \nonumber\\
  \nonumber
 r^2\;\sin(\omega + \nu)\;\cos(\omega + \nu)\;\sin \inc \,=
 a^2\;\frac{\left(1 - e^2\right)^2}{\left(1 + e\;\cos
 \nu\right)^2}\;\,\sin(\omega + \nu)\;\cos(\omega + \nu)\;\sin \inc
 ~~~.~~~~~~~~~~
 \ea

~\\

 \ba
 \nonumber
 \mbox{{\bf{\large{A.14.~~~Calculation of}}}}
 \;\;\;\;\mubold\cdot\left(\;-\;\efbold\;\times\;
 \frac{\partial {\efbold}}{\partial M_o}\;
 \right)\;\;\;.~~~~~~~~~~~~~~~~~~~~~~~~~
 ~~~~~~~~~~~~~~~~~~~~~~~~~~
 \ea\\

The Kepler equation (\ref{A76}) entails that
 \ba
 \frac{\partial {\efbold}}{\partial M_o}\;=\;
 \frac{1}{n}\;\;\;\frac{\partial {\efbold}}{\partial \,
 t}\;=\;\frac{a^{3/2}}{\sqrt{G\,m}}\;\;{\bf\vec g}
 \label{A90}
 \ea
and, therefore, with aid of (\ref{45}) we obtain:
 \ba
 -\;\efbold\;\times\;\frac{\partial {\efbold}}{\partial
 M_o}\;=\;-\;{\bf{\vec w}}\;\sqrt{G\,m\,a\,\left(1\,-\,e^2  \right)}\;
 \frac{a^{3/2}}{\sqrt{G\,m}}\;=\;
 -\;{\bf{\vec w}}\;a^2\;\;\sqrt{1\,-\,e^2}\;\;\;.\;\;\;\;\;
 \label{A91}
 \ea

~\\

 \ba
 \nonumber
 \mbox{{\bf{\large{A.15.~~~The terms}}}}
 \;\;\;\left(\;\mubold\;\times\;\efbold\;\right)\;\cdot\;
 \frac{\partial }{\partial C_j}\;\left(\;\mubold\;\times\;{\efbold}\;
 \right)\;\;\;.~~~~~~~~~~~~~~~~~~~~~~~~~~~~~~~~~~~~~~~~~~~~~~~~~~~
 \ea\\

In the course of numerical computation, it is convenient to keep
such terms as parts of the Hamiltonian (\ref{34}). In case these
terms are to be dealt with analytically, one will need the
following general formulae valid for any
$\;C_j\;,\;\;\;j\;=\;a\,,\;e\,,\;\Omega\,,\;\omega\,,\;\inc\,,\;M_o\;$:
 \ba
 \nonumber
 \left(\mubold\,\times\,\efbold\right)\,\cdot\,
 \frac{\partial }{\partial
 C_j}\,\left(\mubold\,\times\,{\efbold}\right)\,=\,
 \left(\mubold\,\times\,\efbold\right)\,\cdot\,\left(\frac{\partial \mubold}{\partial C_j}
 \,\times\,\efbold\right)\,+\,
 \left(\mubold\,\times\,\efbold\right)\,\cdot\,
 \left(\mubold\,\times\,\frac{\partial \efbold}{\partial
 C_j}\right)~~~~~~~~~~~\\
 \nonumber\\
 \label{A92}\\
 \nonumber
 \,\approx\,
 \left(\mubold\,\times\,\efbold\right)\,\cdot\,
 \left(\mubold\,\times\,\frac{\partial \efbold}{\partial
 C_j}\right)\;\;,\;\;\;~~~~~~~~~~~~~~~~~~~~~~~~~~~~~~~~~~~~~~~~~~~~~~~~~~~~~~~~~~~~~~~~~~~~~~~~~~
 \ea
the neglect of the term with $\;\partial \mubold / \partial C_j\;$
being legitimate, because this input is of a even higher order
than the other one.\footnote{~The physical meaning of $\;\partial
\mubold /\partial C_j\;$ is the influence of the satellite upon
the precession rate of the primary. As demonstrated by Laskar
(2004), this effect is of a higher order of smallness.} A direct
algebraic calculation then yields, under the said approximation:
 \ba
 \nonumber
 \left(\mubold\,\times\,\efbold\right)\,\cdot\,\frac{\partial }{\partial
 C_j}\,\left(\mubold\,\times\,{\efbold}\right)
 =\;\left(\;\mu_2^2\;+\;\mu_3^2\;\right)\;f_1\;\frac{\partial f_1}{\partial
 C_j}\;+\;
 \left(\;\mu_3^2\;+\;\mu_1^2\;\right)\;f_2\;\frac{\partial f_2}{\partial
 C_j}\;+\;
 \left(\;\mu_1^2\;+\;\mu_2^2\;\right)\;f_3\;\frac{\partial f_3}{\partial
 C_j}~\\
 \nonumber\\
 \label{A93}\\
 \nonumber
 -\;
 \mu_1\;\mu_2\;\left(\;f_1\;\frac{\partial f_2}{\partial C_j}\;+\;
 \frac{\partial f_1}{\partial C_j}\;f_2\;\right)\;-\;
 \mu_2\;\mu_3\;\left(\;f_2\;\frac{\partial f_3}{\partial C_j}\;+\;
 \frac{\partial f_2}{\partial C_j}\;f_3\;\right)\;-\;
 \mu_3\;\mu_1\;\left(\;f_3\;\frac{\partial f_1}{\partial C_j}\;+\;
 \frac{\partial f_1}{\partial C_j}\;f_3\;\right)\;~,
 \ea
expression for $\;f_j\;$ given above by (\ref{A20} - \ref{A22})

 \pagebreak

\end{document}